\documentclass[11pt,a4paper]{article}
\pdfoutput=1
\usepackage{jstyle}
\usepackage{amsmath, amsfonts, amssymb}
\usepackage{graphicx, hyperref, color, verbatim}
\usepackage[dvipsnames]{xcolor}
\usepackage{float}
\usepackage{caption}
\usepackage{subcaption}

\title{Scattering bound states in AdS}
\author[a,b]{Wen-Jie Ma,}
\author[c]{Xinan Zhou}
\affiliation[a]{Beijing Institute of Mathematical Sciences and Applications (BIMSA), Beijing 101408, China.}
\affiliation[b]{Yau Mathematical Sciences Center, Tsinghua University, Beijing 100084, China.}
\affiliation[c]{Kavli Institute for Theoretical Sciences, University of Chinese Academy of Sciences, Beijing 100190, China.}
\abstract{We initiate the study of bound state scattering in AdS space at the level of Witten diagrams. For concreteness, we focus on the case with only scalar fields and analyze several basic diagrams which more general diagrams reduce to. We obtain closed form expressions for their Mellin amplitudes with arbitrary conformal dimensions, which exhibit interesting behavior. In particular, we observe that certain tree-level bound state Witten diagrams have the same structure as loop diagrams in AdS.}
\emailAdd{wenjia.ma@bimsa.cn \\ \hskip 42pt  
xinan.zhou@ucas.ac.cn}

\begin{document}
\maketitle
\tableofcontents

\newpage

\section{Introduction}

Holographic correlators play a central role in checking and exploiting the AdS/CFT correspondence. Thanks to the recent breakthroughs of the bootstrap methods, holographic four-point functions of $\frac{1}{2}$-BPS operators with arbitrary Kaluza-Klein masses have been systematically computed at tree level in a plethora of string theory/M-theory models \cite{Rastelli:2016nze,Rastelli:2017udc,Rastelli:2019gtj,Alday:2020lbp,Alday:2020dtb,Alday:2021odx}.\footnote{See also \cite{Bissi:2022mrs} for a review of these results.} These bootstrap methods rely only on symmetries and basic consistency conditions, and circumvent the enormous difficulties related to the traditional method which stalled progress in this field for many years. Note that according to the standard recipe of AdS/CFT, Kaluza-Klein modes of the AdS supergravity fields are dual to ``single-trace'' operators in the CFT.\footnote{To be precise, the single-trace operators are the leading part of the dual operator. There are also higher-trace operators which are suppressed by inverse powers of the central charge \cite{Arutyunov:1999en,Arutyunov:2000ima,Rastelli:2017udc,Aprile:2018efk,Aprile:2019rep,Alday:2019nin,Aprile:2020uxk}.} In the bulk, they are mapped to states which are ``single-particle''. However, in the dual CFT there are also ``double-trace'' (or more generally, ``multi-trace'') operators which are normal ordered products of single-trace operators. Correlation functions involving such operators can be viewed in the bulk as scattering processes where some of the scattering states are multiple-particle ``bound states''. In principle, such correlators are already contained in the set of all ``single-trace'' correlators because we can produce ``double-trace'' operators from taking the OPE limit. In practice, however, computing these bound state correlators via such a detour through higher-point functions seems rather inefficient. Already computing five-point functions is a highly nontrivial task even equipped with bootstrap techniques \cite{Goncalves:2019znr,Alday:2022lkk}, and going beyond that to higher multiplicities presents serious challenges for the current technology.  Therefore, it will be of great interest to develop a more straightforward approach that allows us to directly apply the bootstrap strategy to such correlators with bound state operators. 

In this paper, we make progress in this direction by initiating a study of the underlying Witten diagrams. This is necessary because the properties of these diagrams  related to bound state scattering processes have not been explored in the literature.\footnote{Some bound state Witten diagrams can be trivially obtained. These are the ones which can be written as a product of single-particle Witten diagrams. They have appeared in, {\it e.g.}, \cite{Giombi:2018qox,Antunes:2021abs}.} In particular, there is currently no knowledge of their analytic structure in Mellin space, which will become important if we want to adapt the bootstrap methods of \cite{Rastelli:2016nze,Rastelli:2017udc,Alday:2020lbp,Alday:2020dtb,Alday:2021odx} to this case. Another motivation for looking into these diagrams comes from the recent work \cite{Ceplak:2021wzz}. The series of papers \cite{Giusto:2019pxc,Giusto:2018ovt,Galliani:2017jlg,Bombini:2017sge,Giusto:2020neo} developed an alternative approach to the bootstrap methods to compute holographic correlators in the  $AdS_3\times S^3$ background. This approach starts from a ``heavy-heavy-light-light'' (HHLL) limit of the four-point function. The correlator in this limit can be computed semi-classically as the fluctuation dual to the light operators in a supergravity background created by the heavy operators. By taking a formal limit where the heavy operators become light, the HHLL correlator can produce four-point functions with all light operators. Extending this method, \cite{Ceplak:2021wzz} managed to compute all light four-point correlators at tree level with two single-particle states and two $n$-particle bound states. Interestingly, \cite{Ceplak:2021wzz} found that for $n\geq 2$ the correlators necessarily contain higher order polylogarithms while in the single-particle case at most dilogarithms appear. Curiously, these higher order polylogarithms also show up in loop-level correlators of single-particle operators \cite{Aprile:2017bgs,Aprile:2017qoy,Aprile:2019rep,Bissi:2020wtv,Bissi:2020woe,Huang:2021xws,Drummond:2022dxw}. This seems to imply that certain tree-level diagrams with external bound states might share structural similarity with AdS loop diagrams.\footnote{It is important to note that the supergravity calculation of \cite{Ceplak:2021wzz} is semi-classical. Therefore, the contributing Witten diagrams are tree-level diagrams.} In this paper, we will provide strong evidence that there is indeed such a connection.

\begin{figure}
  \centering
\begin{subfigure}{0.32\textwidth}
 \centering
  \includegraphics[width=0.7\linewidth]{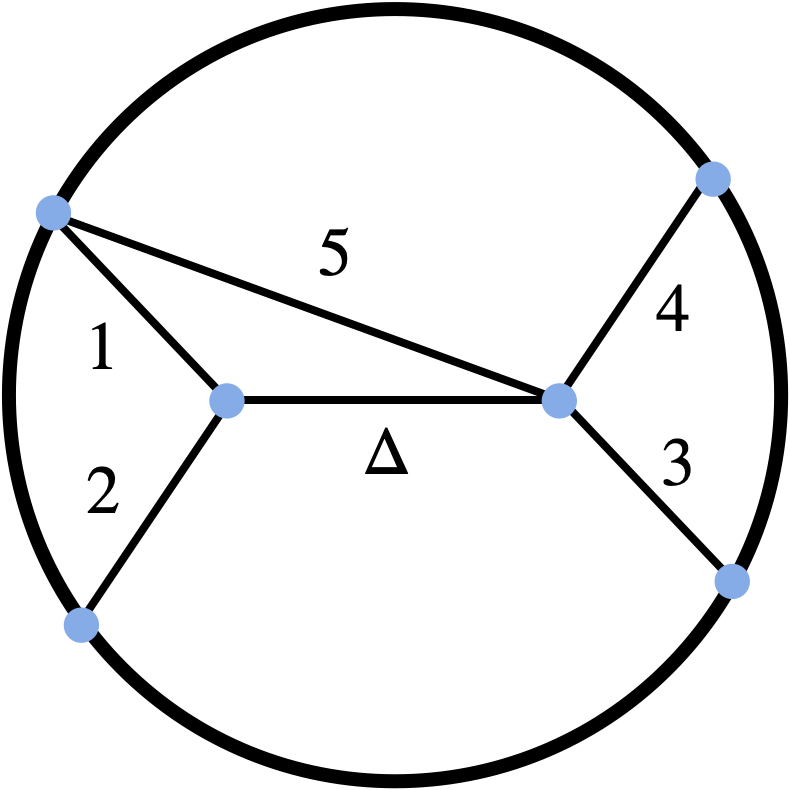}
  \caption{One bound state}
  \label{subfig:bccc}
\end{subfigure}
\begin{subfigure}{0.32\textwidth}
  \centering
  \includegraphics[width=0.7\linewidth]{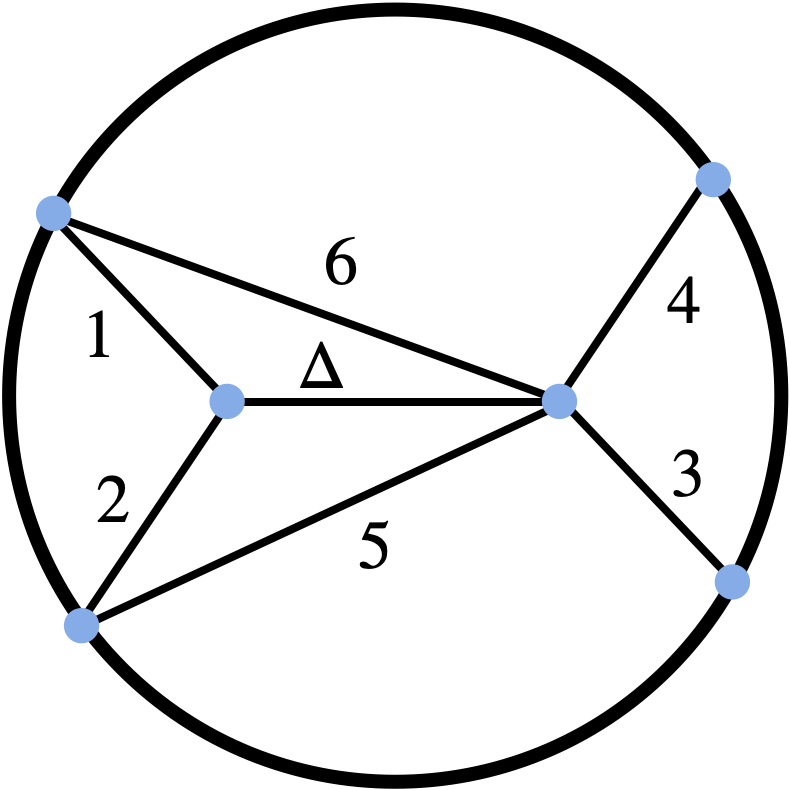}
  \caption{Two bound states (type I)}
  \label{subfig:bbcctypeI}
\end{subfigure}
\begin{subfigure}{0.32\textwidth}
  \centering
  \includegraphics[width=0.7\linewidth]{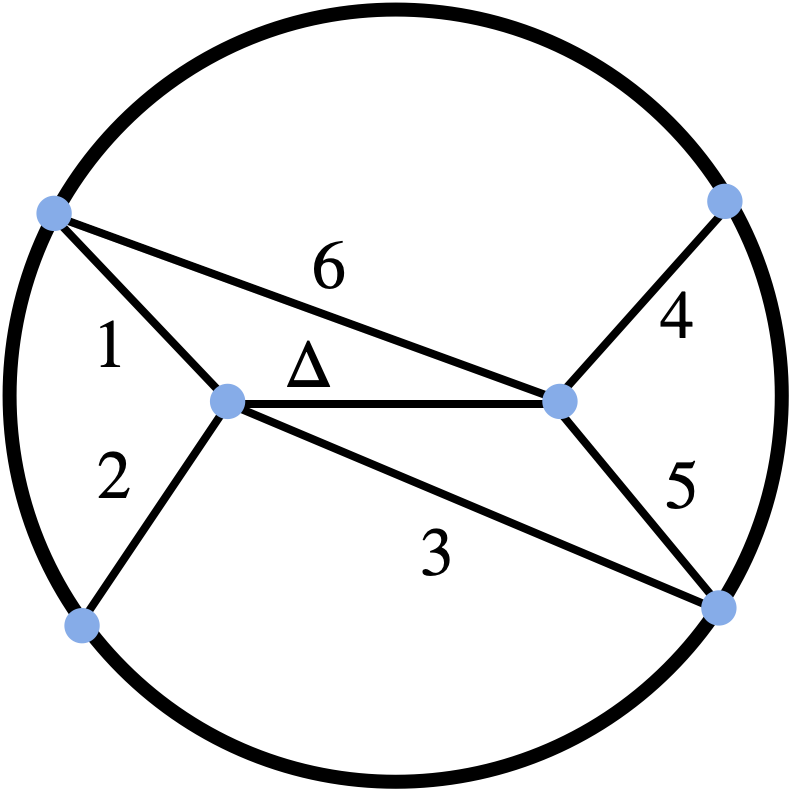}
  \caption{Two bound states (type II)}
  \label{subfig:bbcctypeII}
\end{subfigure}
\caption{Three basic of Witten diagrams with one or two bound states and only one bulk-to-bulk propagator. Here the external bound states are the boundary points from which two lines emanate. }
\label{fig:bsdiags}
\end{figure}

As a first step towards a more systematic exploration, we will limit ourselves to studying diagrams of scalar fields in this paper. More precisely, we will mostly focus on the three diagrams depicted in Fig. \ref{fig:bsdiags} which contain up to two bound states and only one bulk-to-bulk propagator. The bound states are of the ``bi-particle'' type and correspond to double-trace operators in the CFT. This might seem a very small set of diagrams. However, we will explain how a vast array of bound state tree-level diagrams with more bulk-to-bulk propagators can be reduced to these basic diagrams. Moreover, even with just these three diagrams, we find that there is already a rich spectrum of behavior. We will see that the first two diagrams are similar in structure to tree-level exchange diagrams with single-particle external states while the last diagram resembles a one-loop diagram in AdS.

Due to the technical nature of this paper, we offer below a brief summary of the sections to help the reader to navigate it, and highlight some of our main results.

In Section \ref{Sec:Preliminaries}, we review the basic technical ingredients which will be used in this paper. This includes the Mellin representation which recasts correlators in a form similar to flat-space amplitudes and manifests their analytic structure. We will also review two important properties of AdS propagators. One is an integrated vertex identity that allows us to integrate out an internal line of the diagram connected to two external lines via a cubic vertex. The other is the so-called split representation of the bulk-to-bulk propagator. 

These two properties of the bulk-to-bulk propagator are used in Section \ref{Sec:Wcccc} in a warm-up example where we compute the tree-level exchange Witten diagram with single-particle states as its external states. We reproduce the well known result in the literature and the Mellin amplitude takes the following form
\begin{equation}
\mathcal{M}_{\circ\circ\circ\circ}=\sum_{m=0}^\infty \frac{C^{(0)}_m}{s-\Delta-2m}\;,
\end{equation}
where $C^{(0)}_m$ are constants and $\Delta$ is the conformal dimension of the exchanged scalar field. We used $\circ$ to denote an external single-particle operator while later we will also use $\bullet$ to denote a two-particle bound state.

We begin to consider diagrams with bound states in Section \ref{Sec:Wbccc}. We will compute Fig. \ref{subfig:bccc} using two methods. The first method is based on the integrated vertex identity, and generalizes the single-particle case considered in the warm-up section. The second method computes the diagram by taking a coincidence limit of a five-point single-particle diagram. Both approaches lead to the same answer for the Mellin amplitude which has the following schematic form
\begin{equation}
\mathcal{M}_{\bullet\circ\circ\circ}=\sum_{m=0}^\infty \frac{C^{(1)}_m}{s-\Delta-\Delta_5-2m}\;.
\end{equation}
Here $C^{(1)}_m$ are constants and $\Delta_5$ is the conformal dimension of the additional scalar line that makes the single-particle exchange diagram the bound state diagram Fig. \ref{subfig:bccc}. From the expression, we find that the Mellin amplitude is quite similar to the exchange amplitude $\mathcal{M}_{\circ\circ\circ\circ}$, except that the poles are now at shifted locations.

In Section \ref{Sec:WbbcctypeI} we study the diagram in Fig. \ref{subfig:bbcctypeI} which has two bound states and was dubbed ``Type I''. The method based on   the integrated vertex identity can also be applied to this case and leads to the following Mellin amplitude
\begin{equation}
\mathcal{M}^{\rm I}_{\bullet\bullet\circ\circ}=\sum_{m=0}^\infty \frac{C^{(2)}_m}{s-\Delta-\Delta_5-\Delta_6-2m}\;.
\end{equation}
Here the numerators $C^{(2)}_m$ do not depend on Mandelstam variables and $\Delta_5$, $\Delta_6$ are the conformal dimensions of the two additional scalar lines. The amplitude again has a ``tree-like''  analytic structure. We will also confirm this result by reproducing it from taking a coincidence limit of a six-point function. 

We consider the ``Type II'' two bound state diagram (Fig. \ref{subfig:bbcctypeII}) in Section \ref{Sec:WbbcctypeII}, which turns out to have a drastically different structure in Mellin space. The method using the integrated vertex identity no longer applies here because the all the vertices are quartic. However, we can still compute this diagram by taking the coincidence limit. We find that its Mellin amplitude has the form of a sum of simultaneous poles 
\begin{equation}
\mathcal{M}^{\rm II}_{\bullet\bullet\circ\circ}=\sum_{m,n=0}^\infty \frac{\widetilde{C}^{(2)}_{mn}}{(s-\Delta-\Delta_3-\Delta_6-2m)(t-\Delta-\Delta_1-\Delta_5-2n)}\;.
\end{equation}
Remarkably, this is the same structure of one-loop correlators found in $AdS_5\times S^5$ IIB supergravity and $AdS_5\times S^3$ SYM \cite{Alday:2018kkw,Alday:2019nin,Alday:2021ajh}. This connection is further sharpened in Section \ref{Sec:Wbbbccand1loop} where we look at a family of examples of Fig. \ref{subfig:bbcctypeII} with special conformal dimensions. We will show that in the flat-space limit the Mellin amplitudes of these diagrams reduce to the massless one-loop box diagram in flat space. 

Finally, in Section \ref{Sec:morediagrams} we discuss how we can use the three diagrams in Fig. \ref{fig:bsdiags} as building blocks to obtain other more complicated diagrams. We conclude in Section \ref{Sec:outlook} with an outline of future research directions. The paper also contains several appendices where we relegate additional technical details and collect useful formulae. 
\section{Preliminaries}\label{Sec:Preliminaries}
\subsection{Mellin representation}

To discuss holographic correlators, a convenient language is the Mellin representation \cite{Mack:2009mi,Penedones:2010ue}. In this formalism, holographic correlators in general display simple analytic structure. In particular, tree-level correlators with external single-particle states have Mellin amplitudes similar to flat-space scattering amplitudes. An $n$-point function of scalar operators is represented as a multi-fold inverse Mellin transformation
\begin{equation}\label{defMellinnpt}
\langle \mathcal{O}_1(x_1)\ldots \mathcal{O}_n(x_n)\rangle=\int [d\gamma_{ij}] \bigg(\prod_{i<j} (x_{ij}^2)^{-\gamma_{ij}}\Gamma[\gamma_{ij}]\bigg) \mathcal{M}(\gamma_{ij})\;.
\end{equation}
Here we have defined $x_{ij}^2\equiv (\vec{x}_i-\vec{x}_j)^2$, and we can set 
\begin{equation}\label{MMcond1}
\gamma_{ij}=\gamma_{ji}\;,\quad \gamma_{ii}=-\Delta_i\;.
\end{equation}
Conformal invariance requires
\begin{equation}\label{MMcond2}
\sum_{j=1}^n\gamma_{ij}=0\;.
\end{equation} 
The variables $\gamma_{ij}$ then satisfy the same set of constraints as the flat-space Mandelstam variables, except that the external squared masses are now replaced by the conformal dimensions $m_i^2=\Delta_i$. The function $\mathcal{M}(\gamma_{ij})$ is defined to be the Mellin amplitude which contains all the nontrivial dynamic information. Let us also write down the case of $n=4$ explicitly. We can write the correlator as 
\begin{equation}
\langle\mathcal{O}_1(x_1)\ldots \mathcal{O}_4(x_4)\rangle=\frac{1}{(x_{12}^2)^{\frac{\Delta_1+\Delta_2}{2}}(x_{34}^2)^{\frac{\Delta_3+\Delta_4}{2}}}\left(\frac{x_{14}^2}{x_{24}^2}\right)^a\left(\frac{x_{14}^2}{x_{13}^2}\right)^b \mathcal{G}(U,V)\;,
\end{equation}
where $a=\frac{1}{2}(\Delta_2-\Delta_1)$, $b=\frac{1}{2}(\Delta_3-\Delta_4)$, and 
\begin{equation}\label{eq:ConformalCrossRatios}
U=\frac{x_{12}^2x_{34}^2}{x_{13}^2x_{24}^2}\;,\quad V=\frac{x_{14}^2x_{23}^2}{x_{13}^2x_{24}^2}
\end{equation}
are the conformal cross ratios. The function $\mathcal{G}(U,V)$ is represented by
\begin{equation}\label{defMellin4pt}
\begin{split}
\mathcal{G}(U,V)=&\int_{-i\infty}^{i\infty}\frac{dsdt}{(4\pi i)^2}U^{\frac{s}{2}}V^{\frac{t}{2}-\frac{\Delta_2+\Delta_3}{2}}\mathcal{M}(s,t)\,\Gamma[\tfrac{\Delta_1+\Delta_2-s}{2}]\Gamma[\tfrac{\Delta_3+\Delta_4-s}{2}]\\
&\quad\quad\quad\times \Gamma[\tfrac{\Delta_1+\Delta_4-t}{2}]\Gamma[\tfrac{\Delta_2+\Delta_3-t}{2}] \Gamma[\tfrac{\Delta_1+\Delta_3-u}{2}]\Gamma[\tfrac{\Delta_2+\Delta_4-u}{2}]\;,
\end{split}
\end{equation}
and the Mandelstam variables satisfy $s+t+u=\sum_{i=1}^4\Delta_i$.

\subsection{Basics of AdS diagrams}
The Witten diagrams which we will consider in this paper are built from AdS propagators, following rules that are similar to the flat-space position space Feynman rules. These propagators are Green's functions in AdS and can further be divided into bulk-to-bulk or bulk-to-boundary depending on the points of insertions. To write down these propagators it is useful to introduce the so-called embedding space formalism. In this formalism, a point $x^\mu$ in $\mathbb{R}^{d}$ is represented by a null ray $P^A$ in a $d+2$ dimensional embedding space $\mathbb{R}^{d+1,1}$
\begin{equation}
P^AP_A=0\;,\quad P^A\sim \lambda P^A\;.
\end{equation}
The nonlinear conformal transformations are linearized in the embedding space as rotations in $\mathbb{R}^{d+1,1}$. To make connection with the coordinates $x^\mu$, we can fix the rescaling degree of freedom and parameterize the null ray as 
\begin{equation}
P^A=\bigg(\frac{1+x^2}{2},\frac{1-x^2}{2},\vec{x}\bigg)\;,
\end{equation}
where the signature is $(-,+,+,\ldots,+)$. The distance in  $\mathbb{R}^{d}$ is represented in the embedding space as
\begin{equation}
x_{ij}^2=-2P_i\cdot P_j\equiv P_{ij}\;.
\end{equation}
The AdS space can also be conveniently represented by the embedding space. A point with Poincar\'e coordinates $z^\mu=(z_0,\vec{z})$ becomes a point $Z$ in $\mathbb{R}^{d+1,1}$
\begin{equation}
Z^A=\frac{1}{z_0}\bigg(\frac{1+z_0^2+\vec{z}^2}{2},\frac{1-z_0^2-\vec{z}^2}{2},\vec{z}\bigg)\;.
\end{equation}
Using this formalism, the bulk-to-boundary propagator of a scalar field with dimension $\Delta$ reads 
\begin{equation}
G_{B\partial}^\Delta(P,Z)=\frac{1}{(-2Z\cdot P)^\Delta}\;.
\end{equation}
The scalar bulk-to-bulk propagator is given by
\begin{equation}
G_{BB}^\Delta(Z,W)=\tilde{C}_\Delta (2u^{-1})^\Delta {}_2F_1\bigg(\Delta,\Delta-\frac{d}{2}+\frac{1}{2};2\Delta-d+1;-2u^{-1}\bigg)\;,
\end{equation}
where
\begin{equation}
\tilde{C}_\Delta=\frac{\Gamma[\Delta]\Gamma[\Delta-\frac{d}{2}+\frac{1}{2}]}{(4\pi)^{\frac{d+1}{2}}\Gamma[2\Delta-d+1]}\;,
\end{equation}
and 
\begin{equation}
u=\frac{(z_0-w_0)^2+(\vec{z}-\vec{w})^2}{2z_0w_0}=\frac{(Z-W)^2}{2}\;.
\end{equation}
The bulk-to-bulk propagator satisfies the equation of motion
\begin{equation}\label{EOMGBB}
(\square_Z-\Delta(\Delta-d))G_{BB}^\Delta(Z,W)=\delta(Z,W)\;.
\end{equation}

\begin{figure}
  \centering
\begin{subfigure}{0.45\textwidth}
 \centering
  \includegraphics[width=0.5\linewidth]{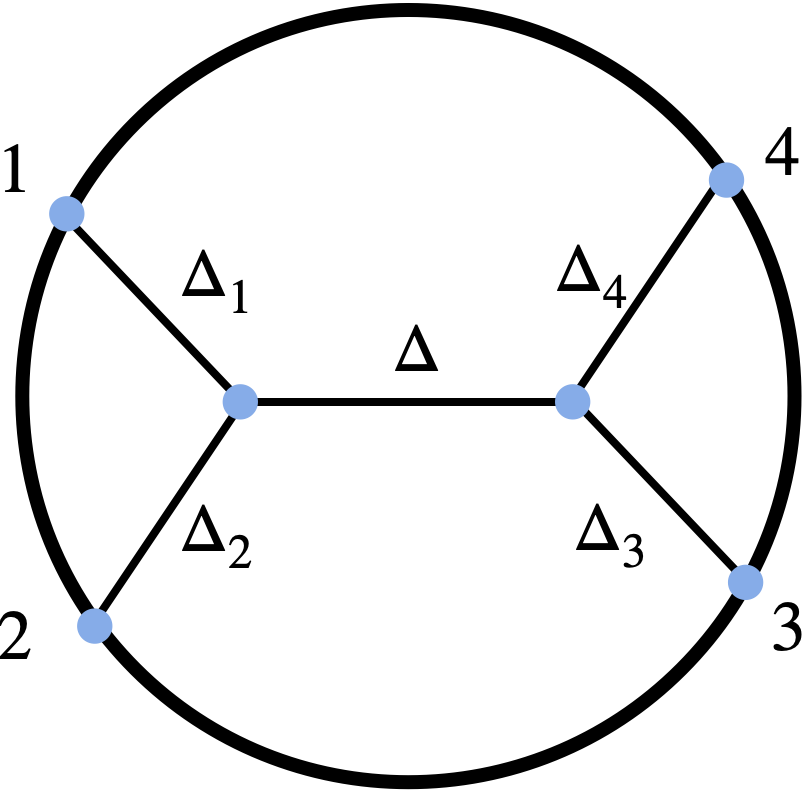}
  \caption{Exchange Witten diagram}
  \label{subfig:cccc}
\end{subfigure}
\begin{subfigure}{0.45\textwidth}
  \centering
  \includegraphics[width=0.5\linewidth]{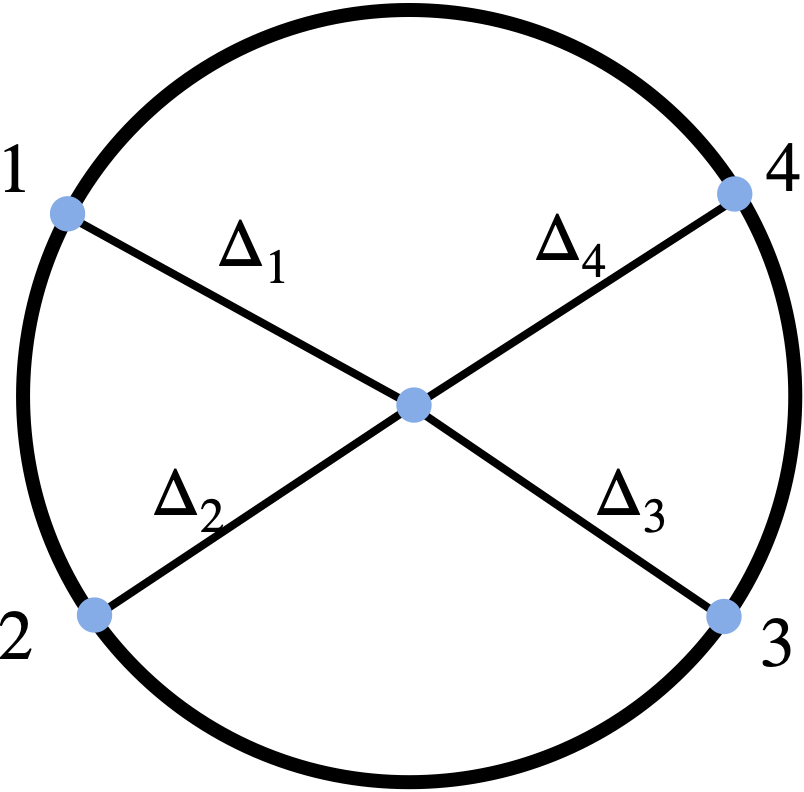}
  \caption{Contact Witten diagram}
  \label{subfig:contact}
\end{subfigure}
\caption{The simplest four-point Witten diagrams at tree level.}
\label{fig:basic4pttrees}
\end{figure}

The two simplest tree-level diagrams we can construct from these propagators are the exchange Witten diagram (Fig. \ref{subfig:cccc})
\begin{equation}
\label{eq:Wcccc}W_{\circ\circ\circ\circ}(P_i)=\int_{\rm AdS} dZ_1dZ_2 G_{B\partial}^{\Delta_1}(P_1,Z_1)G_{B\partial}^{\Delta_2}(P_2,Z_1)G_{BB}^\Delta(Z_1,Z_2)G_{B\partial}^{\Delta_3}(P_3,Z_2)G_{B\partial}^{\Delta_4}(P_4,Z_2)\;,
\end{equation}
and the contact Witten diagram (Fig. \ref{subfig:contact})
\begin{equation}
W_{\rm contact}(P_i)=\int_{\rm AdS} dZG_{B\partial}^{\Delta_1}(P_1,Z)G_{B\partial}^{\Delta_2}(P_2,Z)G_{B\partial}^{\Delta_3}(P_3,Z)G_{B\partial}^{\Delta_4}(P_4,Z)\;.
\end{equation}
Here we have used the notation $W_{\circ\circ\circ\circ}$, where the symbol $\circ$ denotes a single-particle state. This is in anticipation of later discussions of diagrams with external bound states which are denoted by $\bullet$. The contact Witten diagram $W_{\rm contact}$ is commonly known in the literature as the $D$-function and is denoted by $D_{\Delta_1\Delta_2\Delta_3\Delta_4}$. Before we proceed, let us point out two useful properties of these propagators which we will use in this paper. 

\vspace{0.5cm}
\noindent{\bf The integrated vertex identity}
\vspace{0.3cm}

\begin{figure}[h]
\centering
\includegraphics[width=0.75\textwidth]{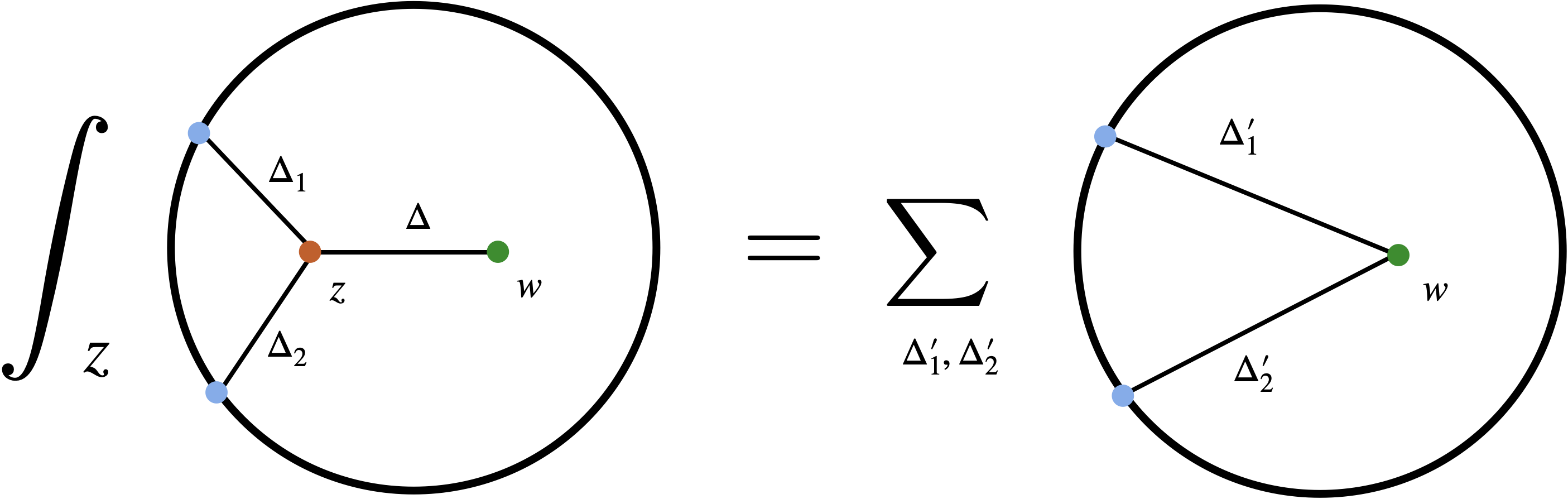}
\caption{An illustration of the integrated vertex identities. After integrating out the scalar bulk-to-bulk propagator, the part of the diagram with a cubic vertex can be written as a sum of contact vertices.}
    \label{fig:ivi}
\end{figure}

\noindent The first useful property is an identity about the following three-point integral 
\begin{equation}
I(P_1,P_2;W)=\int_{\rm AdS} dZ G_{B\partial}^{\Delta_1}(P_1,Z)G_{B\partial}^{\Delta_2}(P_2,Z)G_{BB}^\Delta(Z,W)\;,
\end{equation}
which involves two bulk-to-boundary propagators and one bulk-to-bulk propagator. The bulk-to-bulk propagator can be integrated out and the integral reduces to a sum of products of bulk-to-boundary propagators \cite{DHoker:1999mqo,Zhou:2018sfz}
\begin{equation}\label{ividentity}
\begin{split}
{}&I(P_1,P_2,Z)=\sum_{i=0}^\infty (-2P_1\cdot P_2)^i T_i G_{B\partial}^{\Delta_1+i}(P_1,Z)G_{B\partial}^{\Delta_2+i}(P_2,Z)\\
{}&\quad\quad+\sum_{i=0}^\infty (-2P_1\cdot P_2)^{\frac{\Delta-\Delta_1-\Delta_2+2i}{2}} Q_i G_{B\partial}^{\frac{\Delta+\Delta_1-\Delta_2}{2}+i}(P_1,Z)G_{B\partial}^{\frac{\Delta-\Delta_1+\Delta_2}{2}+i}(P_2,Z)\;,
\end{split}
\end{equation}
where 
\begin{equation}
T_i=\frac{(\Delta_1)_i (\Delta_2)_i}{(\Delta -\Delta_1-\Delta_2) (-d+\Delta +\Delta_1+\Delta_2) \left(\frac{-\Delta +\Delta_1+\Delta_2+2}{2}\right)_i \left(\frac{-d+\Delta +\Delta_1+\Delta_2+2}{2}\right)_i}\;,
\end{equation}
and
\begin{equation}
\begin{split}
Q_i={}&\frac{(-1)^i \Gamma[\frac{d-2 i-2\Delta}{2}]\sin[\frac{\pi  (d-2 \Delta )}{2}]\Gamma[\frac{-d+\Delta +\Delta_1+\Delta_2}{2}]\Gamma[\frac{-\Delta -\Delta_1+\Delta_2+2}{2}] }{4 \pi  \Gamma [i+1]\Gamma [\Delta_1] \Gamma [\Delta_2]}\\
{}&\times \frac{\Gamma [\frac{\Delta -\Delta_1+\Delta_2}{2}] \Gamma [\frac{\Delta +\Delta_1-\Delta_2}{2}] \Gamma[\frac{-\Delta +\Delta_1+\Delta_2}{2}]\Gamma[\frac{-\Delta +\Delta_1-\Delta_2+2}{2}]  }{\Gamma[\frac{-\Delta +\Delta_1-\Delta_2-2 i+2}{2}]\Gamma[\frac{-\Delta -\Delta_1+\Delta_2-2 i+2}{2}]}\;.
\end{split}
\end{equation}
We will refer to this identity as the {\it integrated vertex identity} and it is diagrammatically depicted in Fig. \ref{fig:ivi}. Using this identity we can, for example, write the exchange Witten diagram (Fig. \ref{subfig:cccc}) as the sum of infinitely many contact Witten diagrams (Fig. \ref{subfig:contact}).

\vspace{0.5cm}
\noindent{\bf The split representation}
\vspace{0.3cm}

\begin{figure}[h]
\centering
\includegraphics[width=0.75\textwidth]{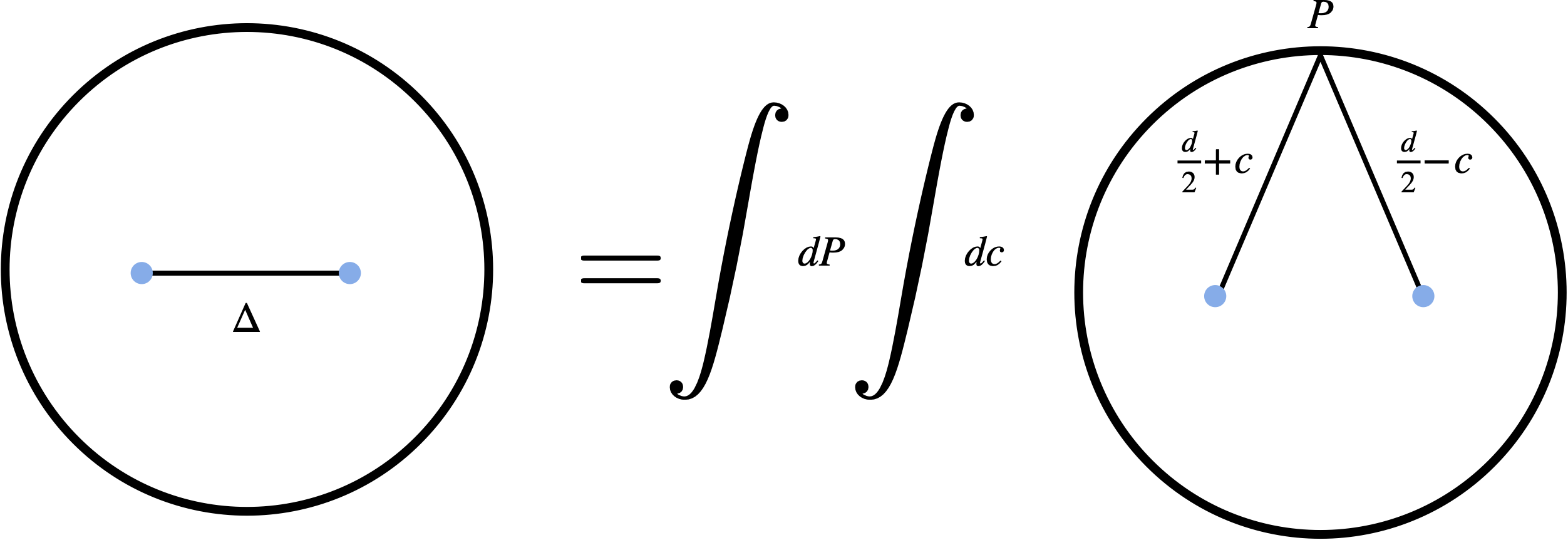}
\caption{An illustration of the split representation. The bulk-to-bulk propagator can be written as a product of bulk-to-boundary propagators integrated over the boundary point and the spectral parameter.}
    \label{fig:sr}
\end{figure}

\noindent The second useful property is the so-called split representation for the bulk-to-bulk propagator \cite{Costa:2014kfa} and is illustrated in Fig. \ref{fig:sr}. The bulk-to-bulk to propagator can be written as a product of a pair of bulk-to-boundary propagators with dimension $\frac{d}{2}+c$ and $\frac{d}{2}-c$ along the principal series, and is further integrated over the boundary point and the parameter $c$. More precisely, we have
\begin{align}
\label{eq:split}G_{BB}(Y,Z)=\int dP\int_{-i\infty}^{i\infty}\frac{dc}{2\pi i}\frac{2}{(\Delta-h)^2-c^2}\frac{\Gamma[h+c]\Gamma[h-c]}{\Gamma[c]\Gamma[-c]}G^{h+c}_{B\partial}(Y,P)G_{B\partial}^{h-c}(Z,P)\;,
\end{align}
where we have defined 
\begin{align*}
h=\frac{d}{2}\;.
\end{align*}

\section{Warm-up: No bound states}\label{Sec:Wcccc}
Let us warm up in this section with the simple case of an exchange Witten diagram (\ref{eq:Wcccc}) where all the external operators are dual to single-particle states. This is a standard example in the literature and the answer has been known for a long time. Our purpose of revisiting this example is to demonstrate the techniques reviewed in Section \ref{Sec:Preliminaries}, which will later be applied to more complicated examples.

\subsection{Using the integrated vertex identity}\label{Subsec:Wccccivi}
Let us first compute this diagram using the integrated vertex identity. Using (\ref{ividentity}), we can write the exchange Witten diagram (\ref{eq:Wcccc}) as 
\begin{equation}\label{WccccasDfun}
\begin{split}
W_{\circ\circ\circ\circ}(x_i)={}&\sum_{i=0}^\infty (x_{12}^2)^i T_i D_{\Delta_1+i,\Delta_2+i,\Delta_3,\Delta_4}\\
{}&+\sum_{i=0}^\infty (x_{12}^2)^{\frac{\Delta-\Delta_1-\Delta_2+2i}{2}} Q_i D_{\frac{\Delta+\Delta_1-\Delta_2}{2}+i,\frac{\Delta-\Delta_1+\Delta_2}{2}+i,\Delta_3,\Delta_4}\;.
\end{split}
\end{equation}
This form of the answer is not particularly illuminating, therefore we now translate it into Mellin space. The Mellin amplitude of a $D$-function is just a constant \cite{Penedones:2010ue}
 \begin{equation}\label{defDfunMellin}
 D_{\Delta_1\ldots \Delta_k}(x_i)=\int [d\delta_{ij}]\prod_{i<j} (x_{ij}^2)^{-\delta_{ij}}\Gamma[\delta_{ij}] \mathcal{M}_{\Delta_1,\ldots,\Delta_k}\;,
 \end{equation}
where 
\begin{equation}
\mathcal{M}_{\Delta_1,\ldots,\Delta_k}=\frac{\frac{\pi^{\frac{d}{2}}}{2}\Gamma[\frac{\sum_i\Delta_i-d}{2}]}{\prod_i \Gamma[\Delta_i]}\;.
\end{equation}
Simple manipulations then give the Mellin amplitude for the following type of functions
\begin{equation}\label{Dnfunctions}
D^{\{n_{ij}\}}_{\Delta_1\ldots \Delta_k}(x_i)\equiv \prod_{i<j} (x_{ij}^2)^{n_{ij}}D_{\Delta^n_1\ldots \Delta^n_k}(x_i)\;,
\end{equation}
which appear in the RHS of (\ref{WccccasDfun}) with $n=4$. Here we require the parameters $n_{ij}$ and $\Delta_i^n$ to satisfy
\begin{equation}
n_{ij}=n_{ji}\;,\quad n_{ii}=0\;,\quad \sum_j n_{ij}=\Delta_i^n-\Delta_i\;,
\end{equation}
so that the external dimensions of each $D^{\{n_{ij}\}}_{\Delta_1\ldots \Delta_k}(x_i)$ are still $\Delta_i$. Using the Mellin representation (\ref{defDfunMellin}) on the RHS of (\ref{Dnfunctions}), we find that, after appropriately shifting the variables, the Mellin representation of $D^{\{n_{ij}\}}_{\Delta_1\ldots \Delta_k}(x_i)$ is 
\begin{equation}
D^{\{n_{ij}\}}_{\Delta_1\ldots \Delta_k}(x_i)=\int [d\delta_{ij}]\prod_{i<j} (x_{ij}^2)^{-\delta_{ij}}\Gamma[\delta_{ij}] \mathcal{M}^{n_{ij}}_{\Delta_1,\ldots,\Delta_k}(\delta_{ij})\;,
\end{equation}
where
\begin{equation}\label{MellinofDn}
\mathcal{M}^{n_{ij}}_{\Delta_1,\ldots,\Delta_k}(\delta_{ij})=\prod_{i<j}\frac{\Gamma[\delta_{ij}+n_{ij}]}{\Gamma[\delta_{ij}]}\mathcal{M}_{\Delta^n_1,\ldots,\Delta^n_k}\;.
\end{equation}
Specifying to our current case, we arrive at the following expression for the Mellin amplitude of the exchange Witten diagram
\begin{equation}\label{MellinMccccinterm}
\begin{split}
\mathcal{M}_{\circ\circ\circ\circ}(s,t)={}&\sum_{i=0}^\infty T_i \frac{\Gamma[\frac{\Delta_1+\Delta_2-s}{2}+i]}{\Gamma[\frac{\Delta_1+\Delta_2-s}{2}]}\mathcal{M}_{\Delta_1+i,\Delta_2+i,\Delta_3,\Delta_4}\\
{}&+\sum_{i=0}^\infty Q_i \frac{\Gamma[\frac{\Delta-s}{2}+i]}{\Gamma[\frac{\Delta_1+\Delta_2-s}{2}]}\mathcal{M}_{\frac{\Delta+\Delta_1-\Delta_2}{2}+i,\frac{\Delta-\Delta_1+\Delta_2}{2}+i,\Delta_3,\Delta_4}\;.
\end{split}
\end{equation}
On the other hand, it is known that the Mellin amplitude has the following analytic structure 
\begin{equation}\label{MellinMcccc}
\mathcal{M}_{\circ\circ\circ\circ}(s,t)=\sum_{m=0}^\infty \frac{C^{(0)}_m}{s-\Delta-2m}\;.
\end{equation}
This structure is anticipated from the large $N$ expansion analysis \cite{Penedones:2010ue} and can be rigorously derived by using the Casimir equation (equation of motion identity) in Mellin space.\footnote{More precisely, the identity is given by 
\begin{equation}
\big({\rm Cas}-\Delta(\Delta-d)\big)W_{\circ\circ\circ\circ}=W_{\rm contact}\;,
\end{equation}
where ${\rm Cas}=-\frac{1}{2}(L_1^{AB}+L_2^{AB})(L_{1,AB}+L_{2,AB})$ is the bi-particle quadratic conformal Casimir built from the conformal generators $L^{AB}_{1,2}$ acting on operators 1 and 2. This identity follows from (\ref{EOMGBB}) which is the equation of motion for the AdS scalar field, and translates into a difference equation for the Mellin amplitude in the Mellin space. For more details, see for instance Appendix C of the review \cite{Bissi:2022mrs}.
} Therefore, we can just focus on the poles at $s=\Delta+2m$ in (\ref{MellinMccccinterm}), and we get 
\begin{equation}\label{C0}
C^{(0)}_m=-\frac{\pi^{\frac{d}{2}}\Gamma[\frac{\Delta+\Delta_1+\Delta_2-d}{2}]\Gamma[\frac{\Delta+\Delta_3+\Delta_4-d}{2}]}{4\Gamma[\Delta_1]\Gamma[\Delta_2]\Gamma[\Delta_3]\Gamma[\Delta_4]\Gamma[1-\frac{d}{2}+\Delta]}\frac{(\frac{\Delta-\Delta_1-\Delta_2+2}{2})_m(\frac{\Delta-\Delta_3-\Delta_4+2}{2})_m}{m!(-\frac{d}{2}+\Delta+1)_m}\;.
\end{equation}

\subsection{Using the split representation}\label{SplitM4}

In this subsection, we will use split representation to compute the Mellin amplitudes of $W_{\circ\circ\circ\circ}$. Using the split representation for bulk-to-bulk propagator $G^{\Delta}_{BB}(Y,Z)$ in the definition (\ref{eq:Wcccc}), $W_{\circ\circ\circ\circ}$ can be written as
\begin{align}
W_{\circ\circ\circ\circ}=\frac{1}{2\pi^d}\int_{-i\infty}^{i\infty}\frac{dc}{2\pi i}\frac{1}{(\Delta-h)^2-c^2}\frac{\Gamma(h+c)\Gamma(h-c)}{\Gamma(c)\Gamma(-c)}\int dP_0 W^{L}_{\circ\circ\circ}W^R_{\circ\circ\circ}\;,
\end{align}
where
\begin{align}
W^L_{\circ\circ\circ}=\int dZ\bigg(\prod_{i=1}^2G^{\Delta_i}_{B\partial}(Z,P_i)\bigg)G^{h+c}_{B\partial}(Z,P_0)\;,
\end{align}
and
\begin{align}
W^R_{\circ\circ\circ}=\int dY\bigg(\prod_{i=3}^4G^{\Delta_i}_{B\partial}(Y,P_i)\bigg)G^{h-c}_{B\partial}(Y,P_0)\;.
\end{align}
The left and right three-point amplitudes can be recast in the Mellin representation through\footnote{This is not necessary for the current case because the Mellin-Mandelstam variables are completely fixed by (\ref{MMcond1}) and (\ref{MMcond2}). However, this representation will become nontrivial and useful later when we will follow the same procedure to compute other exchange diagrams where the vertices are quartic or higher.}
\begin{align}
\int dZ\prod_{i=1}^nG^{\Delta_i}_{B\partial}(Z,P_i)=\frac{\pi^h}{2}\Gamma\left[\frac{\sum_{i=1}^n\Delta_i-d}{2}\right]\prod_{i=1}^n\frac{1}{\Gamma[\Delta_i]}\int[d\gamma]\prod_{i<j}^n\Gamma[\gamma_{ij}]P_{ij}^{-\gamma_{ij}}\;,
\end{align}
leading to 
\begin{align}
W_{\circ\circ\circ\circ}=&\frac{1}{8}\prod_{i=1}^4\frac{1}{\Gamma[\Delta_i]}\int\frac{dc}{2\pi i}\frac{1}{(\Delta-h)^2-c^2}\frac{\Gamma[\frac{\Delta_1+\Delta_2-h+c}{2}]\Gamma[\frac{\Delta_3+\Delta_4-h-c}{2}]}{\Gamma[c]\Gamma[-c]}\\\nonumber
&\times\int[d\tilde{\gamma}]_L[dl]_L\Gamma[\tilde{\gamma}_{12}]P_{12}^{-\tilde{\gamma}_{12}}\int[d\tilde{\gamma}]_R[dl]_R\Gamma[\tilde{\gamma}_{34}]P_{34}^{-\tilde{\gamma}_{34}}\int dP_0\prod_{i=1}^4\Gamma[l_i]P_{0i}^{-l_i}\;.
\end{align}
Here the integration measure $[d\tilde{\gamma}]_L[dl]_l$ satisfies  
\begin{align}
l_1+l_2=h+c\;, \quad l_1+\tilde{\gamma}_{12}=\Delta_1\;,\quad l_2+\tilde{\gamma}_{12}=\Delta_2\;,
\end{align}
and the integration measure $[d\tilde{\gamma}]_R[dl]_R$ satisfies 
\begin{align}
l_3+l_4=h-c\;, \quad l_3+\tilde{\gamma}_{34}=\Delta_3\;,\quad l_4+\tilde{\gamma}_{34}=\Delta_4\;.
\end{align}
The integral over the boundary is conformal because $\sum_{i=1}^4l_i=d$ and can be evaluated through the Symanzik formula \cite{Symanzik:1972wj}
\begin{align}
\int dP_0\prod_{i=1}^4\Gamma[l_{i}]P_{0i}^{-l_{i}}=\pi^h\int[d\gamma_{ij}]\left(\prod_{1\leq i<j\leq 4}\Gamma[\gamma_{ij}]P_{ij}^{-\gamma_{ij}}\right)\;,
\end{align}
where the measure is constrained by
\begin{align}
\sum_{\substack{j=1\\j\neq i}}^4\gamma_{ij}=l_i\;,\qquad i=1,2,3,4\;.
\end{align}
After that, one can shift $\gamma_{ij}$ by $\gamma_{ij}\rightarrow \gamma_{ij}-\tilde{\gamma}_{ij}$ for $1\leq i<j\leq 2$ as well as $3\leq i<j\leq 4$. This gives the correct coordinate dependence factor in the definition (\ref{defMellinnpt}) and allows one to easily read off the Mellin amplitudes $\mathcal{M}_{\circ\circ\circ\circ}$, which is given by
\begin{align}
\mathcal{M}_{\circ\circ\circ\circ}=&\frac{\pi^{h}}{8}\prod_{i=1}^4\frac{1}{\Gamma[\Delta_i]}\int\frac{dc}{2\pi i}\frac{1}{(\Delta-h)^2-c^2}\frac{\Gamma[\frac{\Delta_1+\Delta_2-h+c}{2}]\Gamma[\frac{\Delta_3+\Delta_4-h-c}{2}]}{\Gamma[c]\Gamma[-c]}\\\nonumber
&\times\int[d\tilde{\gamma}]_L[dl]_L\frac{\Gamma[\tilde{\gamma}_{12}]\Gamma[\gamma_{12}-\tilde{\gamma}_{12}]}{\Gamma[\gamma_{12}]}\int[d\tilde{\gamma}]_R[dl]_R\frac{\Gamma[\tilde{\gamma}_{34}]\Gamma[\gamma_{34}-\tilde{\gamma}_{34}]}{\Gamma[\gamma_{34}]}\;.
\end{align}
By solving the constraints, we compute the integral over $[d\tilde{\gamma}]_L[dl]_L$ and $[d\tilde{\gamma}]_R[dl]_R$, leading to
\begin{align}
\mathcal{M}_{\circ\circ\circ\circ}=&\frac{\pi^{h}}{8}\prod_{i=1}^4\frac{1}{\Gamma[\Delta_i]}\int\frac{dc}{2\pi i}\frac{1}{(\Delta-h)^2-c^2}\frac{U(s,c)U(s,-c)}{\Gamma[\frac{\Delta_1+\Delta_2-s}{2}]\Gamma[\frac{\Delta_3+\Delta_4-s}{2}]}\;,
\end{align}
where we defined 
\begin{align}
s=\Delta_1+\Delta_2-2\gamma_{12}=\Delta_3+\Delta_4-2\gamma_{34}\;,
\end{align}
and 
\begin{align}
U(s,c)=\frac{\Gamma[\frac{\Delta_1+\Delta_2-h+c}{2}]\Gamma[\frac{\Delta_3+\Delta_4-h+c}{2}]\Gamma[\frac{h+c-s}{2}]}{\Gamma[c]}\;.
\end{align}
By pinching the $c$-contour between two colliding poles in $c$, we can find poles in $s$, given by
\begin{align}
s=\Delta+2m\;,\qquad m\in \mathbb{Z}_{\geq0}\;.
\end{align}
As a result, $\mathcal{M}_{\circ\circ\circ\circ}$ can be written as
\begin{align}
\label{eq:Mcccc}\mathcal{M}_{\circ\circ\circ\circ}=\sum_{m=0}^{\infty}\frac{C^{(0)}_m}{s-\Delta-2m}\;,
\end{align}
and we find the same residues $C^{(0)}_m$ as given in \eqref{C0}.

\section{Four-point function with one bound state}\label{Sec:Wbccc}
\begin{figure}[h]
\centering
\includegraphics[width=0.25\textwidth]{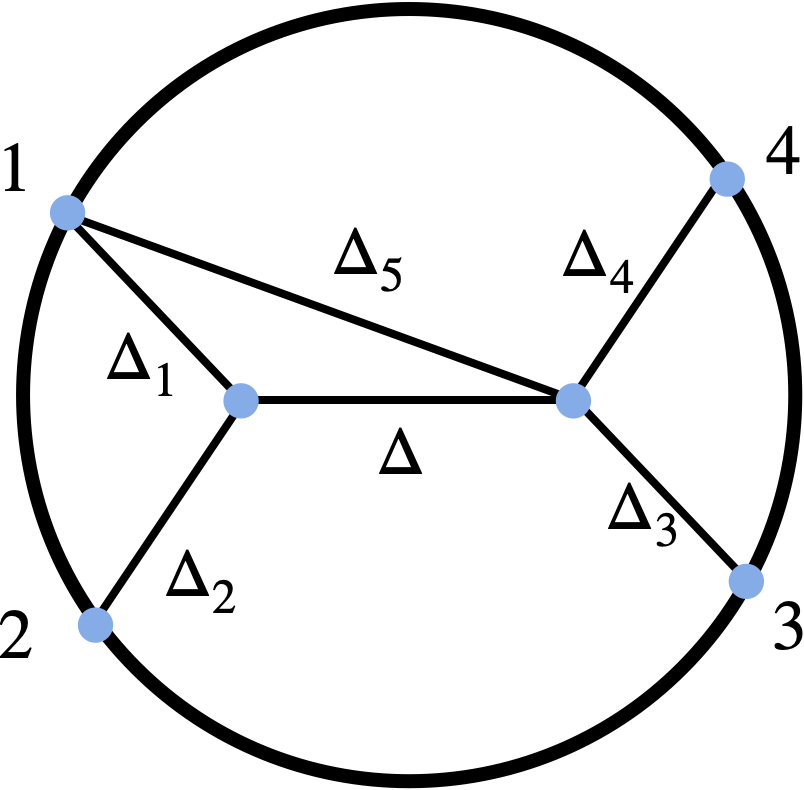}
\caption{The exchange Witten diagram with one bound state. }
    \label{fig:bccc}
\end{figure}

We now proceed to compute the diagram with one bound state, as is depicted in Fig. \ref{fig:bccc}.\footnote{One may wonder if one can get other bound state Witten diagrams with the same set of vertices and propagators. For example, one may consider a diagram where the propagator with $\Delta_5$ starts from the same bulk point but ends on 4. However, this case is trivial because the bulk-to-boundary propagators satisfy the relation $G^{\Delta_i}_{B\partial}(P,Z)G^{\Delta_j}_{B\partial}(P,Z)=G^{\Delta_i+\Delta_j}_{B\partial}(P,Z)$ and the diagram reduces to the diagram (\ref{eq:Wcccc}) without bound states. Therefore, the only nontrivial bound state Witten diagram is Fig. \ref{fig:bccc} up to permutations.} We will use two approaches. The first approach (Section \ref{Subsec:Wbcccivi}) uses the integrated vertex identity and is a straightforward generalization of the calculation presented in Section \ref{Sec:Wcccc}. The second approach (Section \ref{CoincidenceLimit}) obtains the bound state diagram from taking a coincidence limit of a five-point diagram with single-particle external states. 

\subsection{Using the integrated vertex identity}\label{Subsec:Wbcccivi}
The computation of this diagram is similar to that of Section \ref{Subsec:Wccccivi}. We apply the integrated vertex identity to the cubic vertex integral involving the three propagators with dimensions $\Delta_1$, $\Delta_2$ and $\Delta$. This again turns the diagram into infinite sums of $D$-functions
\begin{equation}\label{WbcccinDfun}
\begin{split}
W_{\bullet\circ\circ\circ}={}&\sum_{i=0}^\infty (x_{12}^2)^i T_i D_{\Delta_1+\Delta_5+i,\Delta_2+i,\Delta_3,\Delta_4}\\
{}&+\sum_{i=0}^\infty (x_{12}^2)^{\frac{\Delta-\Delta_1-\Delta_2+2i}{2}} Q_i D_{\frac{\Delta+\Delta_1-\Delta_2}{2}+\Delta_5+i,\frac{\Delta-\Delta_1+\Delta_2}{2}+i,\Delta_3,\Delta_4}\;.
\end{split}
\end{equation}
Translating this result into Mellin space, we get an expression similar to (\ref{MellinMccccinterm}). The expression has poles at $s=\Delta+\Delta_5+2m$ for $m\in \mathbb{Z}_{\geq 0}$. Therefore, we can rewrite the amplitude as a sum over these simple poles. However, we can no longer use the equation of motion identity to rule out additional regular terms because the bi-particle Casimir operator for $x_{1,2}$ necessarily acts on the bulk-to-boundary propagator with dimension $\Delta_5$ as well. On the other hand, for special dimensions satisfying $\Delta_1+\Delta_2-\Delta\in 2\mathbb{Z}_{\geq 0}$, we can verify that regular terms from the two sums cancel and there is no regular part in the Mellin amplitude.\footnote{For example, for $\Delta_1=\Delta_2=\Delta=2$, terms with $i\geq 0$ in the first sum and terms with $i\geq 1$ in the second sum are regular. However, the $(i+1)$-th term in the second sum precisely cancel the $i$-th term in the first sum.} Therefore, we will assume in this subsection that the absence of the regular term is a general feature and the Mellin amplitude has the form
\begin{equation}\label{MellinMbccc}
\mathcal{M}_{\bullet\circ\circ\circ}(s,t)=\sum_{m=0}^\infty \frac{C^{(1)}_m}{s-\Delta-\Delta_5-2m}\;.
\end{equation}
But in the ensuing subsection, we will reproduce this result using a complementary method which also allows us to prove that the regular term is absent. Having determined the analytic structure (\ref{MellinMbccc}), computing $C^{(1)}_m$ is straightforward. These coefficients can be extracted from the residues of (\ref{WbcccinDfun}) in Mellin space and read
\begin{equation}
\label{eq:C1}
\begin{split}
C_m^{(1)}={}&\mathcal{N}^{(1)}{}_3F_2\left(\left.\begin{array}{c}-m, \frac{d}{2}-m-\Delta, 1-m-\frac{\Delta+\Delta_1-\Delta_2}{2}-\Delta_5  \\1-m-\frac{\Delta+\Delta_1-\Delta_2}{2},1-m+\frac{d-\Delta-\Delta_3-\Delta_4-\Delta_5}{2} \end{array}\right.\bigg|1\right)\\
{}&\times \frac{(-1)^m \left(\frac{\Delta -\Delta_1-\Delta_2+2}{2}\right)_m \left(\frac{\Delta +\Delta_1-\Delta_2}{2}\right)_m \left(\frac{-d+\Delta +\Delta_3+\Delta_4+\Delta_5}{2}\right)_m}{m! \left(-\frac{d}{2}+\Delta +1\right)_m \left(\frac{\Delta +\Delta_1-\Delta_2+2 \Delta_5}{2}\right)_m }\;,
\end{split}
\end{equation}
where 
\begin{equation}
\mathcal{N}^{(1)}=-\frac{\pi^{\frac{d}{2}}\Gamma[\frac{\Delta+\Delta_1+\Delta_2-d}{2}]\Gamma[\frac{\Delta+\Delta_3+\Delta_4+\Delta_5-d}{2}]\Gamma[\frac{\Delta-\Delta_2+\Delta_1}{2}]}{4\Gamma[\Delta_1]\Gamma[\Delta_2]\Gamma[\Delta_3]\Gamma[\Delta_4]\Gamma[1-\frac{d}{2}+\Delta]\Gamma[\frac{\Delta-\Delta_2+\Delta_1}{2}+\Delta_5]}\;.
\end{equation}
Note that setting $\Delta_5=0$ reduces the Witten diagram to the exchange Witten diagram in Section \ref{Sec:Wcccc}. We find 
\begin{equation}
C^{(1)}_m\big|_{\Delta_5=0}=C^{(0)}_m\;,
\end{equation}
reproducing the expression (\ref{MellinMcccc}).

\subsection{From the coincidence limit}\label{CoincidenceLimit}

\begin{figure}[h]
\centering
\includegraphics[width=0.25\textwidth]{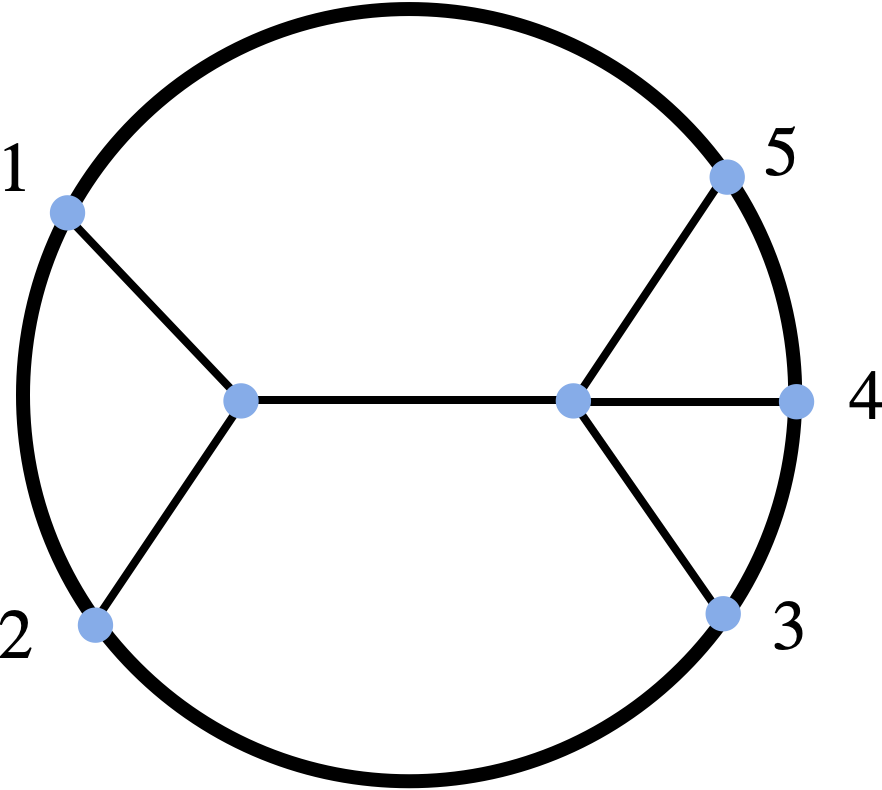}
\caption{The exchange Witten diagram with one bound state can be obtained from a five-point exchange Witten diagram by taking a coincidence limit where $x_5\to x_1$.}
    \label{fig:5pt}
\end{figure}

In this subsection, we will rederive the Mellin amplitude $\mathcal{M}_{\bullet\circ\circ\circ}$ by taking the coincidence limit of $P_1\to P_5$ in Fig. \ref{fig:5pt}. Specifically, the four-point diagram $W_{\bullet\circ\circ\circ}(P_i)$ can be obtained from the five-point diagram $W_{\circ\circ\circ\circ\circ}(P_i)$ through 
\begin{equation}
\begin{split}
W_{\bullet\circ\circ\circ}={}&\lim_{P_5\rightarrow P_1}W_{\circ\circ\circ\circ\circ}(P_i)\\
={}&\lim_{P_5\rightarrow P_1}\int[d\gamma_{ij}]_5\mathcal{M}_{\circ\circ\circ\circ\circ}(s-\Delta_5,t)\prod_{1\leq i<j\leq 5}\Gamma[\gamma_{ij}]P_{ij}^{-\gamma_{ij}}\;,
\end{split}
\end{equation}
where the integration measure $[d\gamma_{ij}]_5$ is constrained by
\begin{align}
\sum_{\substack{j=1\\j\neq i}}^5\gamma_{ij}=\Delta_i
\end{align}
for $1\leq i\leq 5$, and we defined $s$ by 
\begin{equation}
s=\Delta_1+\Delta_2+\Delta_5-2\gamma_{12}\;.
\end{equation}
The first step of the calculation is to compute the five-point Mellin amplitude $\mathcal{M}_{\circ\circ\circ\circ\circ}$, which can be obtained by using the split representation. Because the calculation is very similar to the four-point case presented in Section \ref{SplitM4}, we will omit the details and just write down the result. The Mellin amplitude reads
\begin{equation}
\label{eq:M5}\mathcal{M}_{\circ\circ\circ\circ\circ}(s,t)=\prod_{i=1}^5\frac{1}{\Gamma[\Delta_i]}\sum_{m=0}^{\infty}\frac{K^{(0)}_m}{s-\Delta-2m}\;,
\end{equation}
where
\begin{equation}
\label{eq:Km}K^{(0)}_m=\frac{-\pi^h\Gamma[\frac{\sum_{i=1}^2\Delta_i+\Delta-d}{2}]\Gamma[\frac{\sum_{i=3}^5\Delta_i+\Delta-d}{2}](\frac{2+\Delta-\sum_{i=1}^2\Delta_i}{2})_m(\frac{2+\Delta-\sum_{i=3}^5\Delta_i}{2})_m}{4m!\Gamma[\Delta-h+1+m]}\;.
\end{equation}
We now start to take the coincidence limit. Let us first perform a shift $\gamma_{1j}$ for $2\leq j\leq 4$ by $\gamma_{1j}\rightarrow\gamma_{1j}-\gamma_{j5}$, which gives
\begin{align}
\begin{split}
W_{\bullet\circ\circ\circ}=&\lim_{P_5\rightarrow P_1}\int[d\gamma_{ij}]_5^{\prime}\mathcal{M}_{\circ\circ\circ\circ\circ}(s-\Delta_5+2\gamma_{25},t)P_{15}^{-\gamma_{15}}\prod_{1\leq i<j\leq 4}\Gamma[\gamma_{ij}]P_{ij}^{-\gamma_{ij}}\\
&\times\frac{\Gamma[\gamma_{15}]\Gamma[\gamma_{25}]\Gamma[\gamma_{35}]\Gamma[\gamma_{45}]\Gamma[\gamma_{12}-\gamma_{25}]\Gamma[\gamma_{13}-\gamma_{35}]\Gamma[\gamma_{14}-\gamma_{45}]}{\Gamma[\gamma_{12}]\Gamma[\gamma_{13}]\Gamma[\gamma_{14}]}\;.
\end{split}
\end{align}
Here the integration measure $[d\gamma_{ij}]_5^{\prime}$ is constrained by
\begin{align}
&\sum_{j=2}^4\gamma_{1j}=\Delta_1+\Delta_5-2\gamma_{15},\qquad\sum_{\substack{j=1\\j\neq i}}^4\gamma_{ij}=\Delta_i,\quad 2\leq i\leq5\;.
\end{align}
In the limit that $P_5$ approaches $P_1$, we can close the integration contour of $\gamma_{15}$ to the left in the $\gamma_{15}$-complex plane. Due to the existence of $\Gamma[\gamma_{15}]$, the leading contribution is given by the residue of pole at $\gamma_{15}=0$ and it is the only contribution which we need to keep. Physically, this corresponds to the fact that the limit $x_{15}^2\to 0$ is regular. Thus evaluating the integral over $\gamma_{15}$ leads to
\begin{align}
\begin{split}
W_{\bullet\circ\circ\circ}=&\int[d\gamma_{ij}]_4^{\prime}\prod_{i=2}^3\frac{d\gamma_{i5}}{2\pi i}\mathcal{M}_{\circ\circ\circ\circ\circ}(s-\Delta_5+2\gamma_{25},t)\Gamma[\gamma_{25}]\Gamma[\gamma_{35}]\prod_{1\leq i<j\leq 4}\Gamma[\gamma_{ij}]P_{ij}^{-\gamma_{ij}}\\
&\times\frac{\Gamma[\Delta_5-\gamma_{25}-\gamma_{35}]\Gamma[\gamma_{12}-\gamma_{25}]\Gamma[\gamma_{13}-\gamma_{35}]\Gamma[-\Delta_5+\gamma_{14}+\gamma_{25}+\gamma_{35}]}{\Gamma[\gamma_{12}]\Gamma[\gamma_{13}]\Gamma[\gamma_{14}]}\;,
\end{split}
\end{align}
where the integration measure $[d\gamma_{ij}]_4^{\prime}$ is constrained by
\begin{align}
&\sum_{j=2}^4\gamma_{1j}=\Delta_1+\Delta_5\;,\qquad\sum_{\substack{j=1\\j\neq i}}^4\gamma_{ij}=\Delta_i,\quad 2\leq i\leq4\;.
\end{align}
The Mellin amplitude $\mathcal{M}_{\bullet\circ\circ\circ}$ is thus given by
\begin{align}
\label{eq:gamma}
\begin{split}
\mathcal{M}_{\bullet\circ\circ\circ}=&\prod_{i=2}^3\int\frac{d\gamma_{i5}}{2\pi i}\mathcal{M}_{\circ\circ\circ\circ\circ}(s-\Delta_5+2\gamma_{25},t)\Gamma[\gamma_{25}]\Gamma[\gamma_{12}-\gamma_{25}]\\
&\times\frac{\Gamma[\gamma_{35}]\Gamma[\Delta_5-\gamma_{25}-\gamma_{35}]\Gamma[\gamma_{13}-\gamma_{35}]\Gamma[-\Delta_5+\gamma_{14}+\gamma_{25}+\gamma_{35}]}{\Gamma[\gamma_{12}]\Gamma[\gamma_{13}]\Gamma[\gamma_{14}]}\;.
\end{split}
\end{align}
Performing the integrals over $\gamma_{35}$ and $\gamma_{25}$ leads to an expression for  $\mathcal{M}_{\bullet\circ\circ\circ}$, which agrees with \eqref{MellinMbccc}. We will show this explicitly in Appendix \ref{Mbccc}. Moreover, the approach of taking the coincidence limit also enables us to prove that the regular term vanishes when $\Delta_1>0$. For the sake of readability, we will only outline the proof here and leave the details to Appendix \ref{Mbccc}. The starting point of the proof is to write  $\mathcal{M}_{\bullet\circ\circ\circ}$ as a sum of two parts by performing the remaining integral. Each part can be rewritten as a sum over poles, up to a regular term which we wish to show to be absent. The sum over all the poles can be performed and leads to a generalized hypergeometric function ${}_3F_2$. Thanks to a hypergeometric function identity, which is valid when $\Delta_1>0$, the summation over the poles turns out to be already the same as the original expression. This leads to the conclusion that the regular term must be absent. 

\section{Four-point function with two bound states: Type I}
\begin{figure}[h]
\centering
\includegraphics[width=0.28\textwidth]{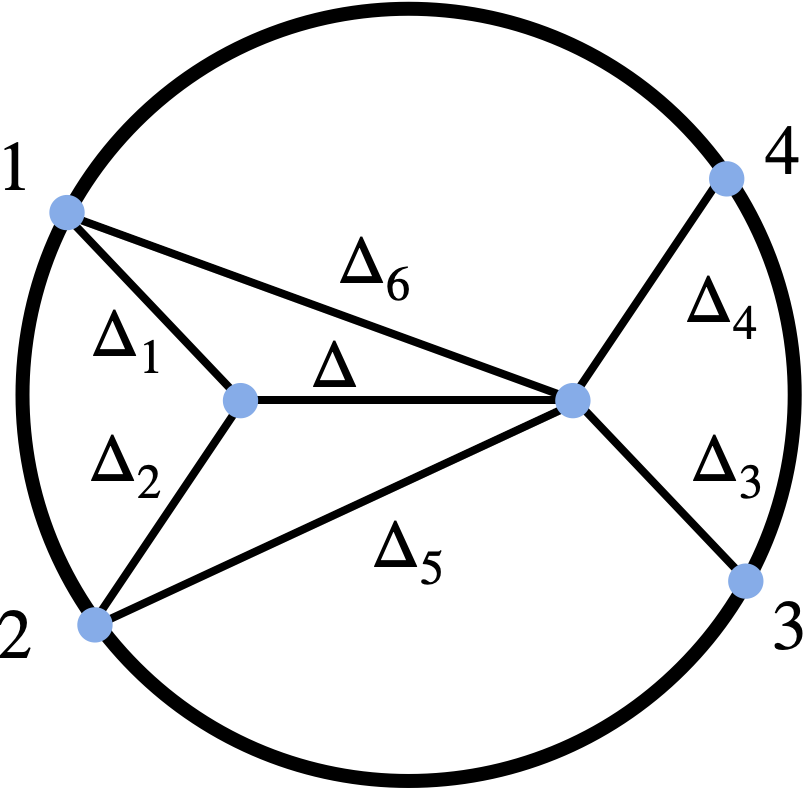}
\caption{Exchange Witten diagram with two bound states (Type I). }
    \label{fig:bbcctypeI}
\end{figure}

We now consider the Witten diagram with two bound states of Type I (Fig. \ref{fig:bbcctypeI}). As in the previous section, we can also evaluate the diagram using two methods and we find the result is structurally similar to the one bound state case.

\label{Sec:WbbcctypeI}


\subsection{Using the integrated vertex identity}
Because the diagram contains a cubic vertex, the method based on the integrated vertex identity can also be applied here. Using the identity on the cubic vertex connecting propagators with dimensions $\Delta_1$, $\Delta_2$, $\Delta$, the diagram is reduced to $D$-functions
\begin{equation}
\begin{split}
W_{\bullet\bullet\circ\circ}^{\rm I}={}&\sum_{i=0}^\infty (x_{12}^2)^i T_i D_{\Delta_1+\Delta_5+i,\Delta_2+\Delta_6+i,\Delta_3,\Delta_4}\\
{}&+\sum_{i=0}^\infty (x_{12}^2)^{\frac{\Delta-\Delta_1-\Delta_2+2i}{2}} Q_i D_{\frac{\Delta+\Delta_1-\Delta_2}{2}+\Delta_5+i,\frac{\Delta-\Delta_1+\Delta_2}{2}+\Delta_6+i,\Delta_3,\Delta_4}\;.
\end{split}
\end{equation}
With the help of (\ref{MellinofDn}) we can translate the result into  Mellin space and obtain an expression for its Mellin amplitude which is similar to (\ref{MellinMccccinterm}). It is not difficult to see that in Mellin space this diagram has poles at $s=\Delta+\Delta_5+\Delta_6+2m$ for $m\in \mathbb{Z}_{\geq 0}$. Like the diagram considered in Section \ref{Sec:Wbccc}, we also cannot use the equation of motion identity argument to rule out the regular term. But based on the explicit results with $\Delta_1+\Delta_2-\Delta\in 2\mathbb{Z}_{\geq 0}$, we will assume that the regular term vanishes in general and the Mellin amplitude has the form 
\begin{equation}\label{MellinMbbccI}
\mathcal{M}^{\rm I}_{\bullet\bullet\circ\circ}(s,t)=\sum_{m=0}^\infty \frac{C^{(2)}_m}{s-\Delta-\Delta_5-\Delta_6-2m}\;.
\end{equation}
Later in Section \ref{Subsec:WbbccIcoinlim}, we will explain how the regular term can be shown to be absent by using the other method based on the coincidence limit. We can compute the residues and find 
\begin{equation}
\begin{split}
\label{eq:C2}
C_m^{(2)}={}&\mathcal{N}^{(2)}{}_4F_3\left(\left.\begin{array}{c}-m, \frac{d}{2}-m-\Delta, 1-m-\frac{\Delta+\Delta_1-\Delta_2}{2}-\Delta_6, 1-m-\frac{\Delta-\Delta_1+\Delta_2}{2}-\Delta_5  \\1-m-\frac{\Delta+\Delta_1-\Delta_2}{2},1-m-\frac{\Delta-\Delta_1+\Delta_2}{2},1-m+\frac{d-\Delta-\Delta_3-\Delta_4-\Delta_5-\Delta_6}{2} \end{array}\right.\bigg|1\right)\\
{}&\times \frac{(-1)^m \left(\frac{\Delta -\Delta_1-\Delta_2+2}{2}\right)_m \left(\frac{\Delta -\Delta_1+\Delta_2}{2}\right)_m \left(\frac{\Delta +\Delta_1-\Delta_2}{2}\right)_m \left(\frac{-d+\Delta +\Delta_3+\Delta_4+\Delta_5+\Delta_6}{2}\right)_m}{m! \left(-\frac{d}{2}+\Delta +1\right)_m \left(\frac{\Delta +\Delta_1-\Delta_2+2 \Delta_6}{2}\right)_m \left(\frac{\Delta -\Delta_1+\Delta_2+2 \Delta_5}{2}\right)_m}\;,
\end{split}
\end{equation}
where 
\begin{equation}
\mathcal{N}^{(2)}=-\frac{\pi^{\frac{d}{2}}\Gamma[\frac{\Delta+\Delta_1+\Delta_2-d}{2}]\Gamma[\frac{\Delta+\Delta_3+\Delta_4+\Delta_5+\Delta_6-d}{2}]\Gamma[\frac{\Delta-\Delta_1+\Delta_2}{2}]\Gamma[\frac{\Delta-\Delta_2+\Delta_1}{2}]}{4\Gamma[\Delta_1]\Gamma[\Delta_2]\Gamma[\Delta_3]\Gamma[\Delta_4]\Gamma[1-\frac{d}{2}+\Delta]\Gamma[\frac{\Delta-\Delta_1+\Delta_2}{2}+\Delta_5]\Gamma[\frac{\Delta-\Delta_2+\Delta_1}{2}+\Delta_6]}\;.
\end{equation}
Note that setting $\Delta_6=0$ reduces to the case considered in Section \ref{Sec:Wbccc}. Furthermore, there is a symmetry of exchanging $(\Delta_1,\Delta_6)$ with $(\Delta_2,\Delta_5)$, which is manifest in Fig. \ref{fig:bbcctypeI}.

\subsection{From the coincidence limit}\label{Subsec:WbbccIcoinlim}

\begin{figure}[h]
\centering
\includegraphics[width=0.28\textwidth]{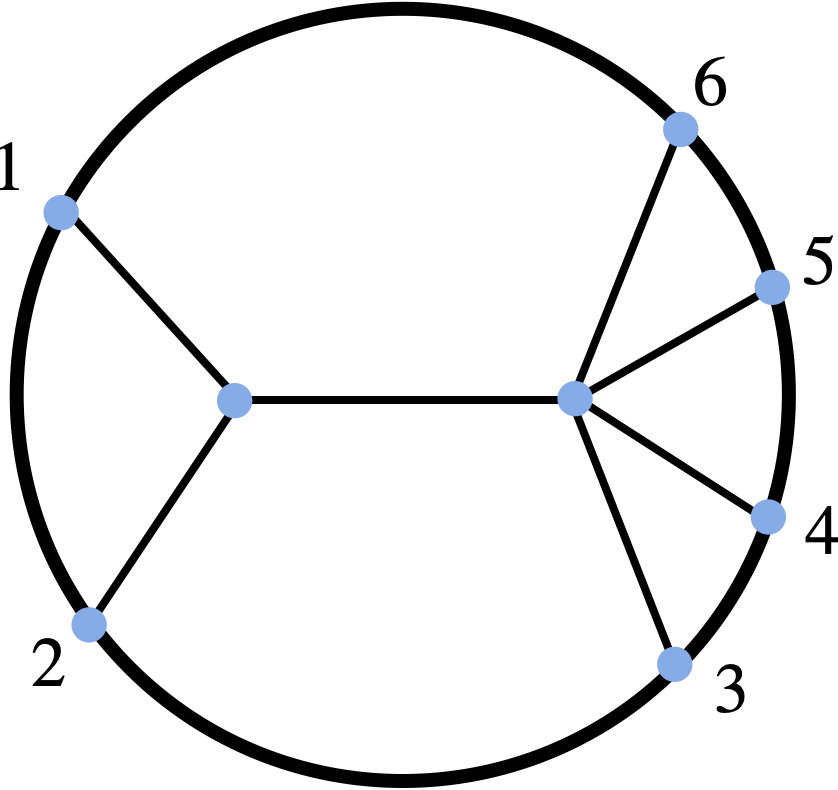}
\caption{The exchange Witten diagram Fig. \ref{fig:bbcctypeI} with two bound states can be obtained from a six-point exchange Witten diagram (denoted by $W^I_{\circ\circ\circ\circ\circ\circ}$) by taking a double coincidence limit where $x_6\to x_1$, $x_5\to x_2$. }
    \label{fig:6ptI}
\end{figure}

In addition to using the integrated vertex identity, we can also obtain the Mellin amplitude $\mathcal{M}^{\rm I}_{\bullet\bullet\circ\circ}$ from the six-point exchange diagram $W^{\rm I}_{\circ\circ\circ\circ\circ\circ}$ in Fig. \ref{fig:6ptI} by taking a coincidence limit where both $P_{6}\to P_1$ and $P_5\to P_2$. The Mellin amplitude  $\mathcal{M}^{\rm I}_{\circ\circ\circ\circ\circ\circ}$ of the  six-point exchange Witten diagram can be deduced by using the split representation. The result is given by
\begin{align}
\label{eq:M6typeI}\mathcal{M}^{\rm I}_{\circ\circ\circ\circ\circ\circ}=\prod_{i=1}^6\frac{1}{\Gamma[\Delta_i]}\sum_{m=0}^{\infty}\frac{R^{(0)}_m}{s^{\prime}-\Delta-2m}\;.
\end{align}
Here $s^{\prime}$ is defined as
\begin{equation}
s^{\prime}=\Delta_1+\Delta_2-2\gamma_{12}\;,
\end{equation}
and the residues are constants
\begin{align}
\label{eq:Rm}R^{(0)}_m=\frac{-\pi^h\Gamma[\frac{\sum_{i=1}^2\Delta_i+\Delta-d}{2}]\Gamma[\frac{\sum_{i=3}^6\Delta_i+\Delta-d}{2}](\frac{2+\Delta-\sum_{i=1}^2\Delta_i}{2})_m(\frac{2+\Delta-\sum_{i=3}^6\Delta_i}{2})_m}{4m!\Gamma[\Delta-h+1+m]}\;.
\end{align}
With $\mathcal{M}^{\rm I}_{\circ\circ\circ\circ\circ\circ}$ in hand, we can proceed with taking the coincidence limit. Let us first identify $P_6$ and $P_1$. This gives a five-point Witten diagram with one bound state
\begin{align}
W^{\rm I}_{\bullet\circ\circ\circ\circ}&=\lim_{P_6\rightarrow P_1}\int[d\gamma_{ij}]_6\mathcal{M}^{\rm I}_{\circ\circ\circ\circ\circ\circ}(s^{\prime\prime}-\Delta_6,t)\prod_{1\leq i<j\leq 6}\Gamma[\gamma_{ij}]P_{ij}^{-\gamma_{ij}}\;,
\end{align}
where the integration measure $[d\gamma_{ij}]_6$ is constrained by
\begin{align}\label{eq:6ptmeasure}
\sum_{\substack{j=1\\j\neq i}}^6\gamma_{ij}=\Delta_i\;,
\end{align}
for $1\leq i\leq 6$ and we defined $s^{\prime\prime}$ by
\begin{align}
s^{\prime\prime}=\Delta_1+\Delta_2+\Delta_6-2\gamma_{12}\;.
\end{align}
Following the same steps as in Section \ref{CoincidenceLimit}, we can shift the Mellin parameters and evaluate the integral over $\gamma_{16}$. After that, we compute the integral over $\gamma_{36}$ and $\gamma_{46}$ by using the first Barnes' lemma, leading to
\begin{align}
\begin{split}
\mathcal{M}^{\rm I}_{\bullet\circ\circ\circ\circ}=&\int\frac{d\gamma_{26}}{2\pi i}\mathcal{M}_6(s^{\prime\prime}-\Delta_6+2\gamma_{26},t)\Gamma[\gamma_{26}]\\
&\times\frac{\Gamma[\Delta_6-\gamma_{26}]\Gamma[\frac{\Delta_1-\Delta_2-\Delta_6+s^{\prime\prime}}{2}+\gamma_{26}]\Gamma[\frac{\Delta_1+\Delta_2+\Delta_6-s^{\prime\prime}}{2}-\gamma_{26}]}{\Gamma[\frac{\Delta_1+\Delta_2+\Delta_6-s^{\prime\prime}}{2}]\Gamma[\frac{\Delta_1-\Delta_2+\Delta_6+s^{\prime\prime}}{2}]}\;.
\end{split}
\end{align}
As in Section \ref{CoincidenceLimit} (which was further detailed in Appendix \ref{Mbccc}), we can perform the integral over $\gamma_{26}$ to obtain an expression for $\mathcal{M}^{\rm I}_{\bullet\circ\circ\circ\circ}$. The result is given by
\begin{align}
\label{eq:Mbcccc}\mathcal{M}^{\rm I}_{\bullet\circ\circ\circ\circ}(s,t)=&\prod_{i=1}^5\frac{1}{\Gamma[\Delta_i]}\sum_{m=0}^{\infty}\frac{K^{(1)}_{m}}{s^{\prime\prime}-\Delta-\Delta_6-2m}\;,
\end{align}
where we have assumed $\Delta_1>0$ and the residue $K^{(1)}_m$ is given by
\begin{align}
\begin{split}
K^{(1)}_{m}=&\frac{R^{(0)}_m\Gamma[\frac{\Delta_1-\Delta_2+\Delta+2m}{2}]}{\Gamma[\frac{\Delta_1-\Delta_2+2\Delta_6+\Delta+2m}{2}]}{}_3F_2\left(\left.\begin{array}{c}-m, \Delta_6,h-\Delta-m \\1-\frac{\Delta_1-\Delta_2+\Delta+2m}{2},\frac{\sum_{i=3}^6\Delta_i-\Delta-2m}{2}\end{array}\right.\bigg|1\right)\;.
\end{split}
\end{align}
To get $\mathcal{M}^{\rm I}_{\bullet\bullet\circ\circ}$, we need to further identify $P_5$ and $P_2$, \textit{i.e.},
\begin{align}
W^{\rm I}_{\bullet\bullet\circ\circ}=&\lim_{P_5\rightarrow P_2}\int[d\gamma_{ij}]_5\mathcal{M}_{\bullet\circ\circ\circ\circ}^{\rm I}(s-\Delta_5,t)\prod_{1\leq i<j\leq 5}\Gamma(\gamma_{ij})P_{ij}^{-\gamma_{ij}}\;,
\end{align}
where we redefined $s$ by
\begin{align}
s=\Delta_1+\Delta_2+\Delta_5+\Delta_6-2\gamma_{12}\;.
\end{align}
Repeating the steps in Section \ref{CoincidenceLimit}, we arrive at an integral representation for  $\mathcal{M}^{\rm I}_{\bullet\bullet\circ\circ}$, which is given by
\begin{align}\label{gamma15typeI}
\begin{split}
\mathcal{M}^{\rm I}_{\bullet\bullet\circ\circ}=&\int\frac{d\gamma_{15}}{2\pi i}\mathcal{M}_{\bullet\circ\circ\circ\circ}^{\rm I}(s-\Delta_5+2\gamma_{15},t)\Gamma[\gamma_{15}]\Gamma[\Delta_5-\gamma_{15}]\\
&\times\frac{\Gamma[\frac{\Delta_1+\Delta_2+\Delta_5+\Delta_6-s}{2}-\gamma_{15}]\Gamma[\frac{\Delta_2-\Delta_1-\Delta_5-\Delta_6+s}{2}+\gamma_{15}]}{\Gamma[\frac{\Delta_1+\Delta_2+\Delta_5+\Delta_6-s}{2}]\Gamma[\frac{-\Delta_1+\Delta_2+\Delta_5-\Delta_6+s}{2}]}\;.
\end{split}
\end{align}
Performing the integral over $\gamma_{15}$ finally leads to an expression for $\mathcal{M}^{\rm I}_{\bullet\bullet\circ\circ}$
\begin{align}
\label{eq:MbbcctypeI}\mathcal{M}^{\rm I}_{\bullet\bullet\circ\circ}(s,t)=&\sum_{m=0}^{\infty}\frac{C^{(2)}_{m}}{s-\Delta-\Delta_5-\Delta_6-2m}\;,
\end{align}
where we assumed $\Delta_1, \Delta_2>0$ and the residue $C^{(2)}$ is given by
\begin{align}\label{eq: C21}
C^{(2)}_{m}=&\sum_{n=0}^{\infty}\frac{(-1)^nK^{(1)}_{m-n}}{n!}\frac{\Gamma[\Delta_5+n]\Gamma[\frac{\Delta_1+\Delta_2-\Delta}{2}-m+n]\Gamma[\frac{\Delta_2-\Delta_1+\Delta}{2}+m-n]}{\Gamma[\Delta_1]\Gamma[\Delta_2]\Gamma[\Delta_3]\Gamma[\Delta_4]\Gamma[\frac{\Delta_1+\Delta_2-\Delta}{2}-m]\Gamma[\frac{-\Delta_1+\Delta_2+2\Delta_5+\Delta}{2}+m]}\;.
\end{align}
In this case, one can also show that the regular term vanishes by using a similar argument based on the identity \eqref{3F2Identity0}. Details of the computation can be found in Appendix \ref{App:MbbcctypeI}. Note that to compare the above $C^{(2)}_m$ with \eqref{eq:C2} a re-summation has to be performed.

\section{Four-point function with two bound states: Type II}\label{Sec:WbbcctypeII}

\subsection{Mellin amplitude}

\begin{figure}[h]
\centering
\includegraphics[width=0.25\textwidth]{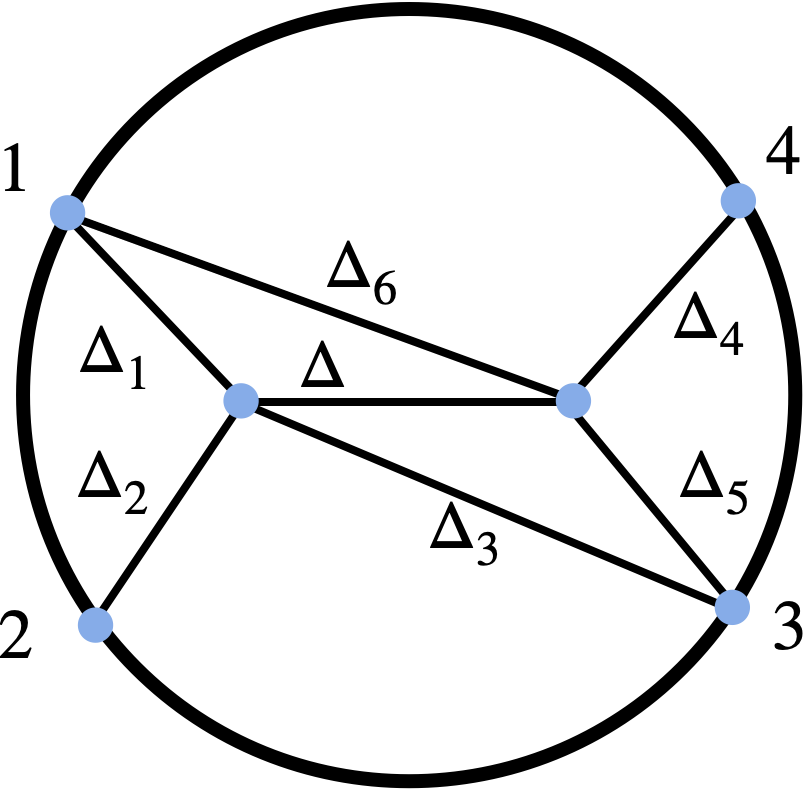}
\caption{Exchange Witten diagram with two bound states (type II). }
    \label{fig:bbcctypeII}
\end{figure}

\begin{figure}[h]
\centering
\includegraphics[width=0.28\textwidth]{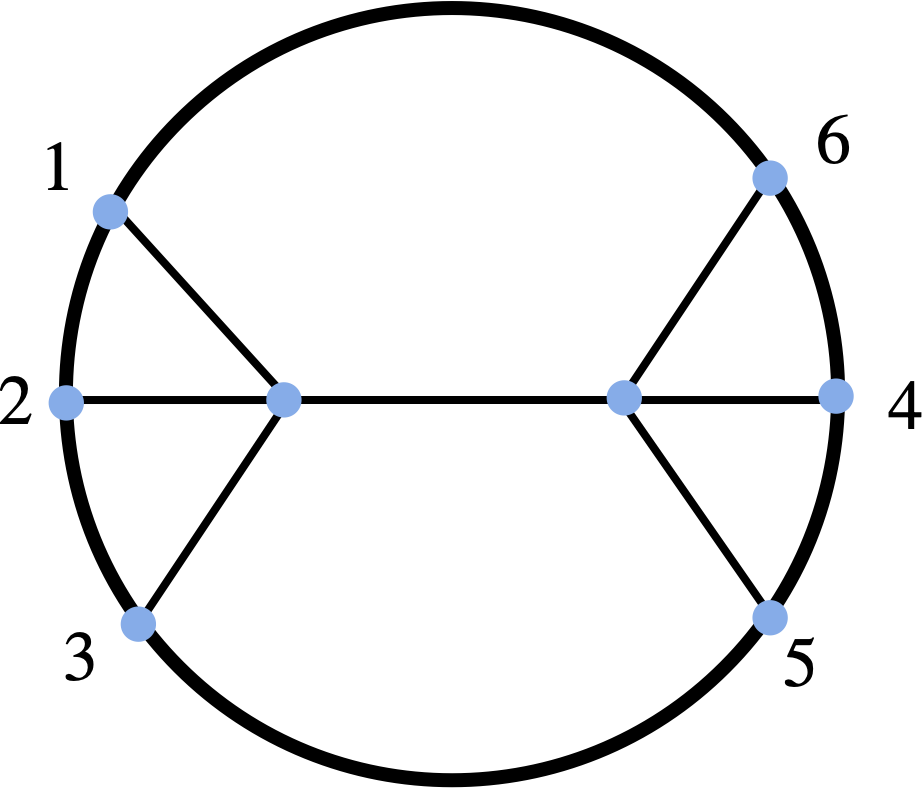}
\caption{The exchange Witten diagram with two bound states can be obtained from a six-point exchange Witten diagram by taking a double coincidence limit where $x_6\to x_1$, $x_5\to x_3$.}
    \label{fig:6pt}
\end{figure}

The two bound state diagram of Type II (Fig. \ref{fig:bbcctypeII}) no longer contains cubic vertices and therefore can not be computed using the method based on the integrated vertex identity. However, the method using the  coincidence limit can still be applied to this case. In this section, we will obtain its Mellin amplitude $\mathcal{M}^{\rm II}_{\bullet\bullet\circ\circ}$ from the six-point Mellin amplitudes $\mathcal{M}^{\rm II}_{\circ\circ\circ\circ\circ\circ}$, depicted in Fig. \ref{fig:6pt}, 
by taking the coincidence limit $P_6\rightarrow P_1$ together with $P_5\rightarrow P_3$.  The six-point Mellin amplitudes $\mathcal{M}^{\rm II}_{\circ\circ\circ\circ\circ\circ}$ can be computed by using the split representation, giving
\begin{align}
\label{eq:M6typeII}\mathcal{M}^{\rm II}_{\circ\circ\circ\circ\circ\circ}(s,t)=\prod_{i=1}^6\frac{1}{\Gamma[\Delta_i]}\sum_{m=0}^{\infty}\frac{\widetilde{R}^{(0)}_m}{s^{\prime}-\Delta-2m}\;.
\end{align}
Here $s^{\prime}$ is defined as
\begin{equation}
s^{\prime}=\Delta_1+\Delta_2+\Delta_3-2\gamma_{12}-2\gamma_{13}-2\gamma_{23}\;,
\end{equation}
and the residues are constants
\begin{align}
\label{eq:Rtm}\widetilde{R}^{(0)}_m=\frac{-\pi^h\Gamma[\frac{\sum_{i=1}^3\Delta_i+\Delta-d}{2}]\Gamma[\frac{\sum_{i=4}^6\Delta_i+\Delta-d}{2}](\frac{2+\Delta-\sum_{i=1}^3\Delta_i}{2})_m(\frac{2+\Delta-\sum_{i=4}^6\Delta_i}{2})_m}{4m!\Gamma[\Delta-h+1+m]}\;.
\end{align}
\begin{figure}[h]
\centering
\includegraphics[width=0.3\textwidth]{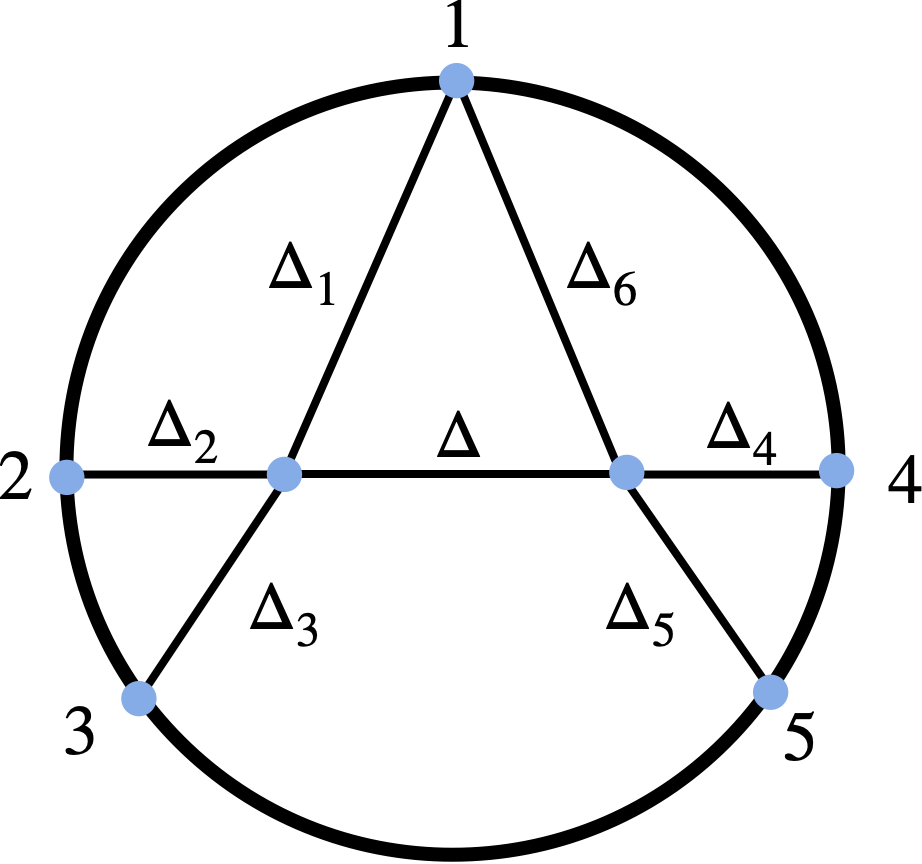}
\caption{A five-point Witten diagram with one bound state. }
    \label{fig:5ptbcccc}
\end{figure}
After taking $P_6\rightarrow P_1$ in $\mathcal{M}^{\rm II}_{\circ\circ\circ\circ\circ\circ}$, a five-point function with one bound state in Fig. \ref{fig:5ptbcccc} can be obtained
\begin{align}
W^{\rm II}_{\bullet\circ\circ\circ\circ}&=\lim_{P_6\rightarrow P_1}\int[d\gamma_{ij}]_6\mathcal{M}^{\rm II}_{\circ\circ\circ\circ\circ\circ}(s^{\prime\prime}-\Delta_6,t)\prod_{1\leq i<j\leq 6}\Gamma[\gamma_{ij}]P_{ij}^{-\gamma_{ij}}\;,
\end{align}
where the integration measure $[d\gamma_{ij}]_6$ is again constrained by \eqref{eq:6ptmeasure} and we defined $s^{\prime\prime}$ by
\begin{align*}
s^{\prime\prime}=\Delta_1+\Delta_2+\Delta_3+\Delta_6-2\gamma_{12}-2\gamma_{13}-2\gamma_{23}\;.
\end{align*}
Similar to the previous sections, one can change variables and evaluate the integral over $\gamma_{36}$, $\gamma_{46}$ and $\gamma_{56}$, leading to
\begin{align}\label{eq:gamma26typeII}
\begin{split}
\mathcal{M}^{\rm II}_{\bullet\circ\circ\circ\circ}=&\int\frac{d\gamma_{26}}{2\pi i}\mathcal{M}^{\rm II}_{\circ\circ\circ\circ\circ\circ}(s^{\prime\prime}-\Delta_6+2\gamma_{26},t^{\prime\prime})\\
&\times\frac{\Gamma[\Delta_6-\gamma_{26}]\Gamma[\frac{s^{\prime\prime}-t^{\prime\prime}+\Delta_1-\Delta_6+2\gamma_{26}}{2}]\Gamma[\frac{t^{\prime\prime}-s^{\prime\prime}+\Delta_1+\Delta_6-2\gamma_{26}}{2}]\Gamma[\gamma_{26}]}{\Gamma[\frac{\Delta_1+\Delta_6-s^{\prime\prime}+t^{\prime\prime}}{2}]\Gamma[\frac{\Delta_1+\Delta_6+s^{\prime\prime}-t^{\prime\prime}}{2}]}\;,
\end{split}
\end{align}
where we defined another Mellin-Mandelstam variable
\begin{align}
t^{\prime\prime}=\Delta_1+\Delta_4+\Delta_5+\Delta_6-2\gamma_{14}-2\gamma_{15}-2\gamma_{45}\;.
\end{align}
Although now $\mathcal{M}^{\rm II}_{\bullet\circ\circ\circ\circ}$ depends on both $s^{\prime\prime}$ and $t^{\prime\prime}$, one can still expand $\mathcal{M}^{\rm II}_{\bullet\circ\circ\circ\circ}$ around its poles and show that the regular term vanishes by using \eqref{3F2Identity0}. We leave the details to Appendix \ref{App:MbbcctypeII} and just write down the results here 
\begin{align}
\label{eq:MbcccctypeII}\mathcal{M}^{\rm II}_{\bullet\circ\circ\circ\circ}(s^{\prime\prime},t^{\prime\prime})=\mathcal{M}^{1}_{\bullet\circ\circ\circ\circ}(s^{\prime\prime},t^{\prime\prime})+\delta_{\Delta_6,0}\mathcal{M}^{2}_{\bullet\circ\circ\circ\circ}(s^{\prime\prime},t^{\prime\prime})+\delta_{\Delta_1,0}\mathcal{M}^{3}_{\bullet\circ\circ\circ\circ}(s^{\prime\prime},t^{\prime\prime})\;.
\end{align}
Here $\mathcal{M}^{2}_{\bullet\circ\circ\circ\circ}(s^{\prime\prime},t^{\prime\prime})$ and $\mathcal{M}^{3}_{\bullet\circ\circ\circ\circ}(s^{\prime\prime},t^{\prime\prime})$ contain only single poles and are given by
\begin{align}
&\mathcal{M}^{2}_{\bullet\circ\circ\circ\circ}(s^{\prime\prime},t^{\prime\prime})=\prod_{i=1}^5\frac{1}{\Gamma[\Delta_i]}\sum_{m=0}^{\infty}\frac{\widetilde{R}^{(0)}_m|_{\Delta_6=0}}{s^{\prime\prime}-\Delta-2m}\;,\\
&\mathcal{M}^{3}_{\bullet\circ\circ\circ\circ}(s^{\prime\prime},t^{\prime\prime})=\prod_{i=2}^6\frac{1}{\Gamma[\Delta_i]}\sum_{m=0}^{\infty}\frac{\widetilde{R}^{(0)}_m|_{\Delta_1=0}}{t^{\prime\prime}-\Delta-2m}\;.
\end{align}
By contrast, $\mathcal{M}^{1}_{\bullet\circ\circ\circ\circ}(s^{\prime\prime},t^{\prime\prime})$ contains simultaneous poles in $s^{\prime\prime}$ and $t^{\prime\prime}$, and is given by 
\begin{align}
&\mathcal{M}^{1}_{\bullet\circ\circ\circ\circ}(s^{\prime\prime},t^{\prime\prime})=\prod_{i=1}^6\frac{1}{\Gamma[\Delta_i]}\sum_{m_1,m_2=0}^{\infty}\frac{\widetilde{K}^{(1)}_{m_1m_2}}{(s^{\prime\prime}-\Delta_6-\Delta-2m_1)(t^{\prime\prime}-\Delta_1-\Delta-2m_2)}\;,
\end{align}
with the residue
\begin{align}\label{K1tildem1m2}
\begin{split}
\widetilde{K}^{(1)}_{m_1m_2}=\frac{\pi^h\Gamma[\frac{\sum_{i=1}^3\Delta_i+\Delta-d}{2}]\Gamma[\frac{\sum_{i=4}^6\Delta_i+\Delta-d}{2}](1-\Delta_6-m_1)_{m_2}(1-\Delta_1-m_2)_{m_1}}{2m_1!m_2!\Gamma[\Delta-h+1]}\\
\times{}_4F_3\left(\left.\begin{array}{c}-m_1, -m_2, 1-\frac{\sum_{i=1}^3\Delta_i-\Delta}{2}, 1-\frac{\sum_{i=4}^6\Delta_i-\Delta}{2} \\ 1-\Delta_1-m_2, 1-\Delta_6-m_1, 1+\Delta-h \end{array}\right.\bigg|1\right)\;.
\end{split}
\end{align}
Setting $\Delta_1=0$ or $\Delta_6=0$, one finds that $\mathcal{M}^{1}_{\bullet\circ\circ\circ\circ}(s^{\prime\prime},t^{\prime\prime})=0$ and the expression \eqref{eq:MbcccctypeII} directly reduces to the desired five-point Mellin amplitudes.

The four-point Mellin amplitudes in Fig. \ref{fig:bbcctypeII} can be obtained by further identifying $P_5$ and $P_3$ in Fig. \ref{fig:5ptbcccc}
\begin{align}\label{eq:WbbcctypeII}
W^{\rm II}_{\bullet\bullet\circ\circ}&=\lim_{P_5\rightarrow P_2}\int[d\gamma_{ij}]_5\mathcal{M}^{\rm II}_{\bullet\circ\circ\circ\circ}(s^{\prime\prime},t^{\prime\prime})\prod_{1\leq i<j\leq 5}\Gamma[\gamma_{ij}]P_{ij}^{-\gamma_{ij}}\;.
\end{align}
Due to the fact that the expression \eqref{eq:MbcccctypeII} for $\mathcal{M}^{\rm II}_{\bullet\circ\circ\circ\circ}$ contains three parts, $\mathcal{M}^{\rm II}_{\bullet\bullet\circ\circ}$ can be correspondingly written as
\begin{align}\label{eq:MbbcctypeII}
\mathcal{M}^{\rm II}_{\bullet\bullet\circ\circ}&=\mathcal{M}^{1}_{\bullet\bullet\circ\circ}(s,t)+\mathcal{M}^{2}_{\bullet\bullet\circ\circ}(s,t)+\mathcal{M}^{3}_{\bullet\bullet\circ\circ}(s,t)\;,
\end{align}
where $\mathcal{M}^{i}_{\bullet\bullet\circ\circ}$ for $i=1, 2, 3$ are obtained by substituting the corresponding $\mathcal{M}^{i}_{\bullet\circ\circ\circ\circ}$ into \eqref{eq:WbbcctypeII}. Moreover, $\mathcal{M}^{2}_{\bullet\circ\circ\circ\circ}$ and $\mathcal{M}^{3}_{\bullet\circ\circ\circ\circ}$ represent two five-point Mellin amplitudes, the resulting $\mathcal{M}^{2}_{\bullet\bullet\circ\circ}$ and $\mathcal{M}^{3}_{\bullet\bullet\circ\circ}$ are just four-point Mellin amplitudes with one bound state. In other words, we have
\begin{align}\label{eq:M23bbcc}
\begin{split}
\mathcal{M}^{2}_{\bullet\bullet\circ\circ}(s,t)=\delta_{\Delta_6,0}\mathcal{M}_{\bullet\circ\circ\circ}\bigg[\begin{array}{c} 12345\\ 45123\end{array}\bigg]\;,\qquad\mathcal{M}^{3}_{\bullet\bullet\circ\circ}(s,t)=\delta_{\Delta_1,0}\mathcal{M}_{\bullet\circ\circ\circ}\bigg[\begin{array}{c} 12345\\ 32465\end{array}\bigg]\;,
\end{split}
\end{align}
where $\mathcal{M}_{\bullet\circ\circ\circ}\bigg[\begin{array}{c} 12345\\ abcde\end{array}\bigg]$ means that we relabel $1, 2, 3, 4, 5$ in \eqref{MellinMbccc} by $a, b, c, d, e$, respectively. To compute $\mathcal{M}^{1}_{\bullet\bullet\circ\circ}(s,t)$, we follow the steps in the previous sections and reach an integral representation for $\mathcal{M}^{1}_{\bullet\bullet\circ\circ}$, given by
\begin{align}\label{eq:integral}
\begin{split}
\mathcal{M}^{1}_{\bullet\bullet\circ\circ}=\int\frac{d\gamma_{25}}{2\pi i}\frac{d\gamma_{45}}{2\pi i}\frac{\mathcal{M}_{\bullet\circ\circ\circ\circ}^{1}(\Delta_4+\Delta_5-2\gamma_{45},t-\Delta_5+2\gamma_{25})\Gamma[\Delta_5-\gamma_{45}-\gamma_{25}]}{\Gamma[\frac{s+t-\Delta_2-\Delta_4}{2}]}\\
\frac{\Gamma[\gamma_{25}]\Gamma[\gamma_{45}]\Gamma[\frac{\sum_{i=3}^5\Delta_i-s}{2}-\gamma_{45}]\Gamma[\frac{\Delta_2+\Delta_3+\Delta_5-t}{2}-\gamma_{25}]\Gamma[\frac{s+t-\Delta_2-\Delta_4-2\Delta_{5}}{2}+\gamma_{25}+\gamma_{45}]}{\Gamma[\frac{\Delta_2+\Delta_3+\Delta_5-t}{2}]\Gamma[\frac{\Delta_3+\Delta_4+\Delta_5-s}{2}]}\;,
\end{split}
\end{align}
where 
we have defined $s$ and $t$ as
\begin{align}
s=\Delta_1+\Delta_2+\Delta_6-2\gamma_{12}\;,\quad t=\Delta_1+\Delta_4+\Delta_6-2\gamma_{14}\;.
\end{align}
Performing the remaining integral finally leads an expression for $\mathcal{M}^{1}_{\bullet\bullet\circ\circ}(s,t)$. The computation is technical and tedious. We leave the details in the Appendix \ref{App:MbbcctypeII} and only write down the final result here
\begin{align}\label{eq:M1bbcc}
\begin{split}
\mathcal{M}^{1}_{\bullet\bullet\circ\circ}(s,t)=&\delta_{\Delta_5,0}\mathcal{M}_{\bullet\circ\circ\circ}\bigg[\begin{array}{c} 12345\\ 64231\end{array}\bigg]+\delta_{\Delta_3,0}\mathcal{M}_{\bullet\circ\circ\circ}\bigg[\begin{array}{c} 12345\\ 12456\end{array}\bigg]\\
&+\sum_{m_1,m_2=0}^{\infty}\frac{\widetilde{C}^{(2)}_{m_1m_2}}{(s-\Delta_3-\Delta_6-\Delta-2m_1)(t-\Delta_1-\Delta_5-\Delta-2m_2)}\;.
\end{split}
\end{align}
The coefficients $\widetilde{C}^{(2)}_{m_1m_2}$ in the second line are constants given by
\begin{align}\label{eq:C2typeII}
\begin{split}
\widetilde{C}^{(2)}_{m_1m_2}=\sum_{n_1,n_2=0}^{\infty}\frac{(1-\frac{\Delta_4+\Delta_5-\Delta_6-\Delta}{2}+n_1)_{m_1-n_1}(1-\frac{-\Delta_1+\Delta_2+\Delta_3-\Delta}{2}+n_2)_{m_2-n_2}}{\Gamma[\Delta_1]\Gamma[\Delta_2]\Gamma[\Delta_3]\Gamma[\Delta_4]\Gamma[\Delta_5]\Gamma[\Delta_6](m_1-n_1)!(m_2-n_2)!}\\
\frac{\Gamma[\frac{-\Delta_4+\Delta_5+\Delta_6+\Delta}{2}+n_1-n_2+m_2]\Gamma[\frac{\Delta_1-\Delta_2+\Delta_3+\Delta}{2}+m_1-n_1+n_2]}{\Gamma[\frac{\Delta_1-\Delta_2+\Delta_3-\Delta_4+\Delta_5+\Delta_6+2\Delta}{2}+m_1+m_2]}\widetilde{K}^{(1)}_{n_1n_2}\;,
\end{split}
\end{align}
where $\widetilde{K}^{(1)}_{n_1n_2}$ has already been defined in (\ref{K1tildem1m2}). Substituting \eqref{eq:M1bbcc} and \eqref{eq:M23bbcc} into \eqref{eq:MbbcctypeII} leads to the final expression for $\mathcal{M}^{\rm II}_{\bullet\bullet\circ\circ}(s,t)$.


\subsection{Relation with AdS one-loop box diagrams}\label{Sec:Wbbbccand1loop}

The appearance of simultaneous poles in \eqref{eq:M1bbcc} is reminiscent of the Mellin amplitudes for one-loop box diagrams in AdS \cite{Alday:2018kkw,Alday:2019nin,Alday:2021ajh}. However, a direct comparison that pinpoints the precise AdS diagram is difficult. On the one hand, a closed form expression for one-loop box diagrams with generic conformal dimensions in any spacetime dimension is still absent. On the other, in the supersymmetric cases where there are explicit results \cite{Alday:2018kkw,Alday:2019nin,Alday:2021ajh} one works with the {\it reduced} correlator\footnote{This is analogous to stripping off a factor supercharge delta functions in flat-space super amplitudes. See, {\it e.g.}, the review \cite{Alday:2021ajh} for more details.}. The one-loop correction of the reduced correlator does not admit a clear interpretation as a collection of one-loop Witten diagrams when the AdS radius is finite. The one-loop correction sees not only the AdS factor of the background but the internal space as well. Therefore, these diagrams are extended into the internal dimensions and are not pure AdS diagrams. In this subsection we will not attempt to find the exact AdS loop diagrams. Instead,  we will content ourselves with confirming that the bound state amplitudes $\mathcal{M}_{\bullet\bullet\circ\circ}^{\rm II}(s,t)$ with certain conformal dimensions reduce to the flat-space massless box diagrams in the flat-space limit \cite{Penedones:2010ue}. We leave the precise identification with AdS loop diagrams at a finite radius for the future. 

We consider a special class of bound state amplitudes $\mathcal{M}_{\bullet\bullet\circ\circ}^{\rm II}(s,t)$ with conformal dimensions $\Delta_1=\Delta_3=\Delta_5=\Delta_6=1$ and $\Delta_2=\Delta_4=\Delta$. The four-point function therefore has external dimensions $2$, $2$, $\Delta$, $\Delta$. In this case, we find that the Mellin amplitude $\mathcal{M}_{\bullet\bullet\circ\circ}^{\rm II}(s,t)$ reads
\begin{align}\label{eq:flatspacelimt}
\begin{split}
\mathcal{M}^{\rm II}_{\bullet\bullet\circ\circ}(s,t)=&\sum_{m_1,m_2=0}^{\infty}\frac{\widetilde{C}^{(2)}_{m_1m_2}}{(s-2-\Delta-2m_1)(t-2-\Delta-2m_2)}\;,
\end{split}
\end{align}
with residues
\begin{align}
\begin{split}
\widetilde{C}^{(2)}_{m_1m_2}=\frac{\pi^h\Gamma[\Delta+1-h]}{2\Gamma[\Delta]^2(m_1+m_2+1)}\;.
\end{split}
\end{align}
Note that the poles of the Mellin amplitude are precisely those corresponding to the double-trace operators. This is the same situation as in the one-loop case \cite{Alday:2018kkw,Alday:2019nin,Alday:2021ajh}. Let us also mention that the sums in (\ref{eq:flatspacelimt}) can be performed and gives 
\begin{align}
\begin{split}
\mathcal{M}^{\rm II}_{\bullet\bullet\circ\circ}(s,t)=&\frac{\pi^h\Gamma[\Delta+1-h]}{8\Gamma^2[\Delta](s+t-2-2\Delta)}\bigg((\gamma-H_{\frac{\Delta-t}{2}})(\gamma-2H_{\frac{\Delta-s}{2}}+H_{\frac{\Delta-t}{2}})+\psi^{(1)}[\frac{s-\Delta}{2}]\\
&+\psi^{(1)}[\frac{2-t+\Delta}{2}]+\psi^{(0)2}[\frac{s-\Delta}{2}]-2\psi^{(0)}[\frac{s-\Delta}{2}]\psi^{(0)}[\frac{2-s+\Delta}{2}]\bigg)\;.
\end{split}
\end{align}
Here $\gamma$ is the Euler constant and $H_x$ and $\psi^{(n)}[x]$ are the Harmonic number and polygamma function with order $n$ respectively. Let us now examine the flat-space limit of this bound state Mellin amplitude. The flat-space limit corresponds to the high energy limit where both $s$, $t$ become large \cite{Penedones:2010ue}. The leading contribution in the sum (\ref{eq:flatspacelimt}) arises from the region with large $m_1,m_2\sim s,t$. From the explicit expression, we find that $\widetilde{C}^{(2)}_{m_1m_2}$ has the following large $m_1$, $m_2$ behavior 
\begin{align}\label{eq:C flatspacelimit}
\widetilde{C}^{(2)}_{m_1m_2}=\frac{1}{m_1+m_2}+\cdots.
\end{align}
This is a special case of the one-loop diagrams considered in \cite{Alday:2021ajh} where the Mellin amplitudes have the form 
\begin{equation}
\mathcal{M}(s,t)=\sum_{m_1m_2}\frac{c_{m_1m_2}}{(s-2m_1)(t-2m_2)}\;.
\end{equation}
The coefficients $c_{m_1m_2}$ are assumed to have the asymptotic behavior
\begin{equation}
c_{m_1m_2}=\frac{(m_1m_2)^{\frac{D}{2}-3}}{(m_1+m_2)^{\frac{D}{2}-2}}+\cdots\;,
\end{equation}
in the large $m_1$, $m_2$ limit. It was shown in \cite{Alday:2021ajh} that in the flat-space limit the Mellin amplitude reduces to the massless one-loop box diagram in a $D$-dimensional flat spacetime. The behavior \eqref{eq:C flatspacelimit} implies $D=6$. Therefore, we find that the bound state Mellin amplitude \eqref{eq:flatspacelimt} 
becomes the 6D one-loop box diagram in the flat-space limit.


\subsection{A special case: The two-loop four-mass ladder diagram}
\begin{figure}[h]
  \centering
\begin{subfigure}{0.45\textwidth}
 \centering
  \includegraphics[width=0.7\linewidth]{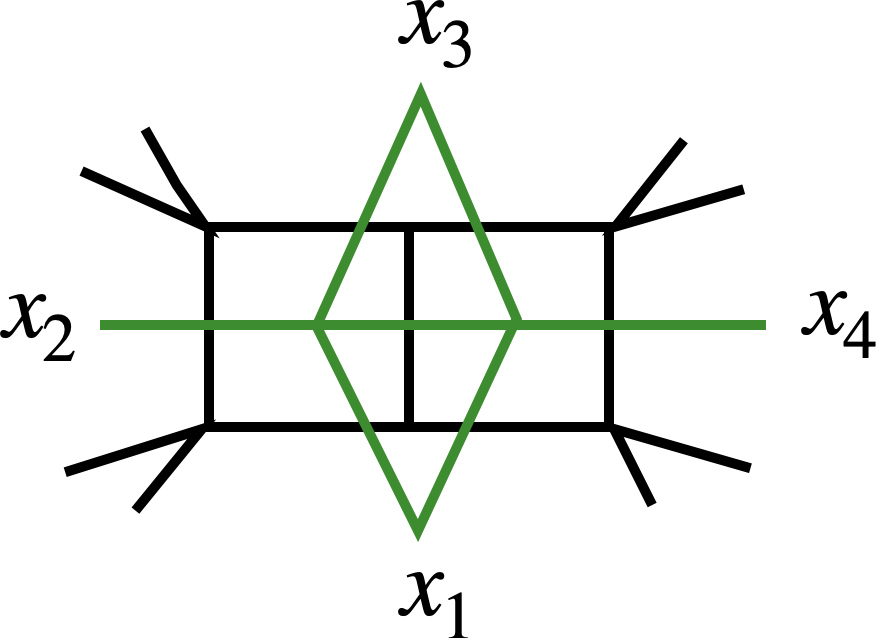}
  \caption{The four-mass diagram}
  \label{subfig:laddermassless}
\end{subfigure}
\begin{subfigure}{0.45\textwidth}
  \centering
  \includegraphics[width=0.7\linewidth]{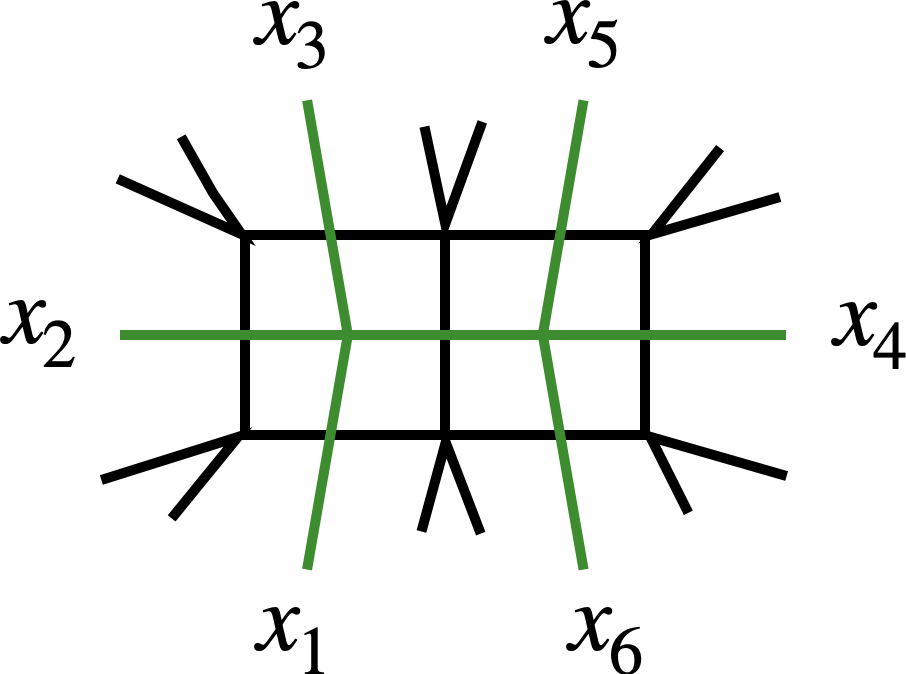}
  \caption{The fully massive diagram}
  \label{subfig:laddermassive}
\end{subfigure}
\caption{Flat-space two-loop ladder diagrams in momentum space (in black) and their dual diagrams (in green). The diagram (a) is a special limit of (b) obtained by taking $x_1\to x_6$ and $x_3\to x_5$.}
\label{fig:ladderdiagrams}
\end{figure}

Finally, as a consistency check of our results, let us consider a special case of (\ref{eq:flatspacelimt}) with $\Delta=1$ and reproduce the result of an important conformal integral in the literature. This special case should correspond to the two-loop four-mass diagram in flat space which is depicted in Fig. \ref{subfig:laddermassless}. In terms of the dual coordinates, the diagram  \ref{subfig:laddermassless} is defined by the integral 
\begin{equation}
A_{\rm 2-mass}(x_i)=\int \frac{d^4xd^4y}{(x_1-x)^2(x_2-x)^2(x_3-x)^2(x-y)^2(x_1-y)^2(x_3-y)^2(x_4-y)^2}\;,
\end{equation}
and can be obtained from the fully massive diagram \ref{subfig:laddermassive}
\begin{equation}
A_{\text{fully massive}}(x_i)=\int \frac{d^4xd^4y}{(x_1-x)^2(x_2-x)^2(x_3-x)^2(x-y)^2(x_4-y)^2(x_5-y)^2(x_6-y)^2}\;,
\end{equation}
by taking the massless limit $x_1\to x_6$, $x_3\to x_5$. To see why the diagram \ref{subfig:laddermassless} should match $\mathcal{M}^{\rm II}_{\bullet\bullet\circ\circ}(s,t)\big|_{\Delta_i=\Delta=1}$, let us note that the Mellin amplitude of the fully massive diagram was computed in \cite{Paulos:2012nu} and turned out to be 
\begin{equation}
\mathcal{M}_{\text{fully massive}}\propto \frac{1}{s'-1}\;.
\end{equation}
This is precisely the six-point exchange Witten diagram Fig. \ref{fig:6pt} with $\Delta_i=\Delta=1$.\footnote{Note that this identification is valid for all spacetime dimensions $d$. This is because the Mellin amplitude contains only one pole and $d$ only appears in the numerator as an overall factor.} Therefore, by further taking the coincidence limit the four-mass diagram \ref{subfig:laddermassless} should be identical to the two bound state Witten diagram $W^{\rm II}_{\bullet\bullet\circ\circ}$.

The result for the four-mass two-loop ladder diagram is well known in the literature and is given by \cite{Usyukina:1992wz,Usyukina:1993ch}
\begin{equation}
A_{\rm 2-mass}(x_i)=-\frac{\pi^4}{x_{13}^4x_{24}^2}\Phi^{(2)}(U,V)\;,
\end{equation}
where 
\begin{align}\label{Phi2}
\begin{split}
\Phi^{(2)}(U,V)=&\frac{1}{\lambda}\bigg(6(\text{Li}_4(-\rho U)+\text{Li}_4(-\rho V))+3\text{log}\frac{V}{U}(\text{Li}_3(-\rho U)-\text{Li}_3(-\rho V))\\
&\qquad+\frac{1}{2}\text{log}^2\frac{V}{U}(\text{Li}_2(-\rho U)+\text{Li}_2(-\rho V))+\frac{1}{4}\text{log}^2(\rho U)\text{log}^2(\rho V)\\
&\qquad+\frac{\pi^2}{2}\text{log}(\rho U)\text{ln}(\rho V)+\frac{\pi^2}{12}\text{log}^2\frac{V}{U}+\frac{7\pi^2}{60}\bigg)\;,
\end{split}
\end{align}
with $\lambda$ and $\rho$ given by
\begin{align}
\lambda=\sqrt{(1-U-V)^2-4UV},\hspace{1cm}\rho=\frac{2}{1-U-V+\lambda}\;.
\end{align}
Here $U$ and $V$ are the conformal cross ratios defined in \eqref{eq:ConformalCrossRatios}. Plugging the Mellin amplitude (\ref{eq:flatspacelimt})  with $\Delta=1$  in the Mellin representation (\ref{defMellin4pt}) and closing the contours for $s$ and $t$ to pick up the residues, we obtain an expansion in small $U$ and $V$. It is not difficult to verify that up to an overall normalization the expansion matches precisely with the small $U$, $V$ expansion of (\ref{Phi2}). Therefore, we can conclude that we have reproduced the four-mass two-loop ladder diagram in flat space as a special case of bound state Witten diagrams.


\section{More general diagrams}\label{Sec:morediagrams}
From the basic diagrams we computed in the previous sections, we can construct a bevy of tree-level diagrams with one or two bound states by using the integrated vertex identity. In this section, we briefly explain how this works. 

\begin{figure}[h]
\centering
\includegraphics[width=0.8\textwidth]{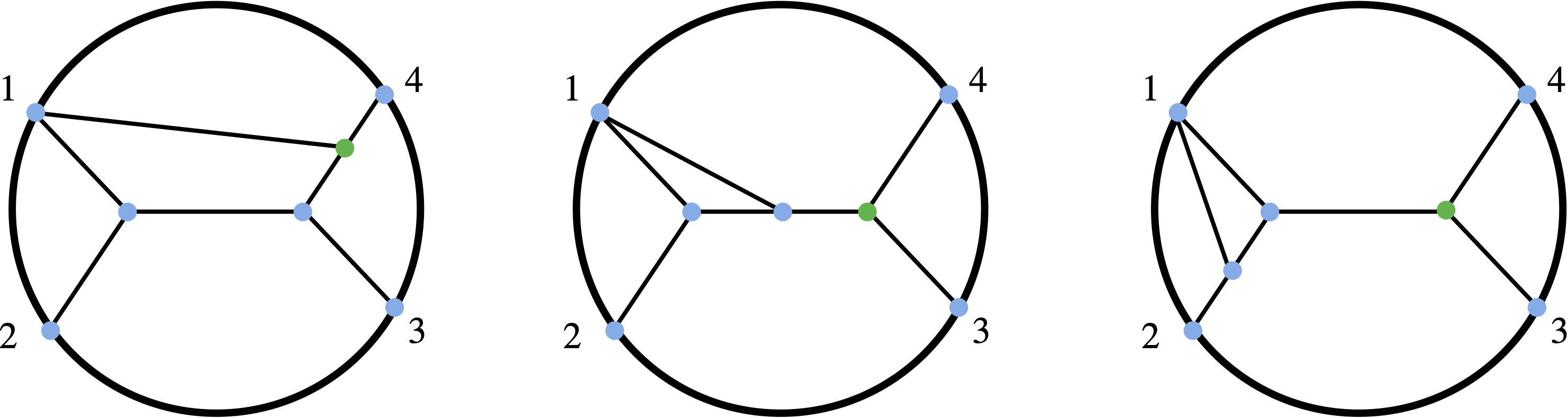}
\caption{Diagrams with one bound state and only cubic vertices.}
    \label{fig:MGDbccc}
\end{figure}

Let us start with case with only one bound state. In Fig. \ref{fig:bccc} there is only one bulk-to-bulk propagator. We can consider the more complicated diagrams with two bulk-to-bulk propagators by moving the bulk point of the bulk-to-boundary propagator with dimension $\Delta_5$ away from the quartic vertex to end on other propagators. The new diagrams contain only cubic vertices. There are three inequivalent possibilities which are depicted in Fig. \ref{fig:MGDbccc}. Using the integrated vertex identity on the green vertices eliminates the bulk-to-bulk propagator,  and we reduce the three diagrams to the basic diagram $W_{\bullet\circ\circ\circ}$. 

\begin{figure}[h]
\centering
\includegraphics[width=\textwidth]{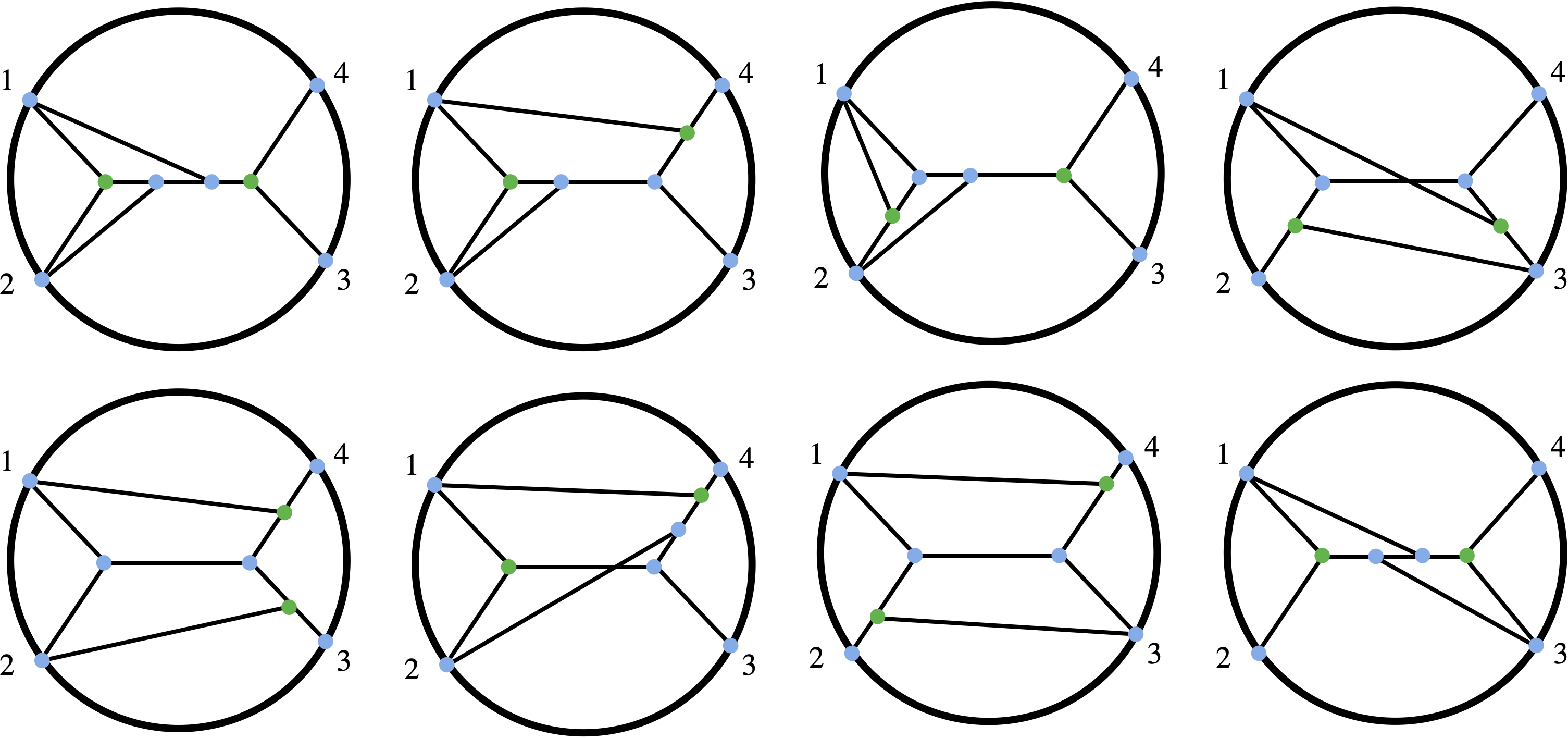}
\caption{Examples of diagrams with two bound states and only cubic vertices. Using the integrated vertex identity at the green indices reduces the diagrams to the two basic diagrams $W^{\rm I}_{\bullet\bullet\circ\circ}$ and $W^{\rm II}_{\bullet\bullet\circ\circ}$.}
    \label{fig:MGDbbcc}
\end{figure}

We now move on to the tree-level diagrams with two bound states. Let us first consider the case where the diagrams only have cubic vertices. These diagrams have three bulk-to-bulk propagators. They can be obtained from the one bound state diagrams in Fig. \ref{fig:MGDbccc} by further attaching another bulk-to-boundary propagator which starts from 2, 3 or 4 and terminates on the existing propagators. There are now many more diagrams. However, one can show that using integrated vertex identities twice allows us to reduce all these diagrams to $W^{\rm I}_{\bullet\bullet\circ\circ}$ and $W^{\rm II}_{\bullet\bullet\circ\circ}$. Some examples of these diagrams are included in Fig. \ref{fig:MGDbbcc}, and the integrated vertex identity is applied to the vertices in green. Similarly, one can consider tree-level diagrams with two bound states and two bulk-to-bulk propagators. This requires the diagrams to have one quartic vertex.  One can use the integrated vertex identity to show that these diagrams reduce to the basic diagrams considered in this paper. In fact, they arise from the aforementioned case with three bulk-to-bulk propagators after using once the integrated vertex identity. 

\section{Outlook}\label{Sec:outlook}
In this paper, we initiated the study of tree-level Witten diagrams with external bound states. We considered the case with only scalar fields and focused on three basic diagrams using which we can construct an array of more complicated diagrams. We showed that these diagrams have simple analytic structures in Mellin space and obtained explicit expressions for their Mellin amplitudes for generic conformal dimensions. These explorations lead to many interesting directions for future research. 

\begin{itemize}
\item One immediate generalization is to include diagrams in which the internal propagators carry Lorentz spins. Such diagrams with spins up to two appear in correlators in full-fledged supergravity theories. The techniques which we developed in this paper are also useful in this more complicated scenario. In particular, the method based on the integrated vertex identity generalize straightforwardly to the spinning case. We expect that the Mellin amplitudes of these diagrams will be structurally similar to the ones studied here.  
\item Once these diagrams with internal spinning operators have been computed, we can use them in the bootstrap calculation of four-point functions in 4d $\mathcal{N}=4$ SYM at strong coupling with two $\frac{1}{4}$-BPS operators and two $\frac{1}{2}$-BPS operators. The $\frac{1}{2}$-BPS operators are dual to single-particle states while the $\frac{1}{4}$-BPS operators are dual to bi-particle bound states. The starting point of such a calculation is an ansatz for the four-point function in terms of all possible diagrams with unfixed coefficients. We then use superconformal constraints to solve these unknowns. The superconformal kinematics of such ``bound state'' four-point functions has recently been analyzed in \cite{Bissi:2021hjk}. Such a bootstrap strategy is similar in spirit to the one first devised in \cite{Rastelli:2016nze,Rastelli:2017udc}.
\item Relatedly, it would be interesting to see if our results in Mellin space, for special operator spectra appearing in top-down holographic models, can be re-expressed in position space in terms of the generalized Bloch-Wigner-Ramakrishnan functions found in the $AdS_3\times S^3$ case \cite{Ceplak:2021wzz}. A good starting point is to consider the double-discontinuities \cite{Caron-Huot:2017vep,Alday:2017vkk}, which are simpler objects but contain all the essential information. Being able to find an efficient algorithm to rewrite the Mellin results in terms of these building block functions will be useful for implementing the bootstrap strategy in position space. 
\item Our analysis of the type II tree-level bound state diagram in Fig. \ref{fig:bbcctypeII} showed that it has intimate connections with one-loop diagrams in AdS. The  resemblance between the two is manifested in the Mellin representation where they can be both represented as a sum of simultaneous poles. Moreover, we showed that the flat-space limits of certain bound state diagrams coincide with those of the AdS one-loop diagrams. It would be very interesting to develop a more systematic understanding in the future to establish a precise connection that remains valid at finite AdS radius. For example, a possible route is to consider generalizations of the integrated vertex identities \cite{Jepsen:2019svc}. It might be possible to use these identities to transform the diagrams and directly prove the equivalence between the bound state tree diagrams and certain one-loop diagrams. 
\item It would also be of great interest to explore other methods for computing bound state processes in AdS. For example, a powerful technique is the AdS unitarity method initiated in \cite{Aharony:2016dwx}, which mirrors the unitarity method in flat space. One can cut an AdS diagram into tree-level diagrams and ``glue'' them together in a sense that can be rigorously defined in the CFT language. For example, both diagrams Fig. \ref{subfig:bbcctypeI} and \ref{subfig:bbcctypeII} with two bound states can be viewed as the results of gluing together two five-point functions. The application of this method to AdS loop diagrams has already been streamlined in the literature. It would be very interesting to extend this technique to calculate bound state Witten diagrams and reproduce our results. Moreover, this perspective might also offer a more intuitive explanation of why these two diagrams have drastically different analytic structures. Another exciting direction to explore is to make connections with the Schwinger-Dyson equation in AdS, which will allow us to resum the $1/N$ corrections. Related works include \cite{Carmi:2018qzm,Carmi:2021dsn,Fichet:2021pbn}.
\item Note that in this paper we only considered correlators with at most two  bi-particle bound states. Clearly, an important generalization of the analysis in the future is  to include more bound state operators. It would also be interesting to look at correlators with multi-particle bound states which are obtained from taking the OPE limit of more than two single-particle operators. 
\item Finally, it would also be interesting to consider bound state Witten diagrams in other backgrounds such as those providing holographic duals for boundary CFTs (or interface CFTs). The simplest example for holographic interface CFTs is the so called probe brane setup where there are localized degrees of freedom living on an $AdS_d$ subspace inside $AdS_{d+1}$ \cite{DeWolfe:2001pq,Aharony:2003qf}. Witten diagrams in this background with single-particle external states were systematically studied in  \cite{Rastelli:2017ecj,Mazac:2018biw}, and the Mellin formalism for BCFT correlators was developed in \cite{Rastelli:2017ecj}. The techniques developed in this paper will be useful for studying bound state scattering in these systems.
\end{itemize}

\acknowledgments
We thank Fernando Alday, Agnese Bissi, Giulia Fardelli, Vasco Goncalves and Andrea Manenti for interesting discussions and useful comments on the draft. The work of X.Z. is supported by funds from University of Chinese Academy of Sciences (UCAS) and funds from the Kavli Institute for Theoretical Sciences (KITS). The work of X.Z. is also supported by the Fundamental Research Funds for the Central Universities.

\appendix

\section{More details of $\mathcal{M}_{\bullet\circ\circ\circ}$}

\label{Mbccc}

In this appendix, we will show how to obtain \eqref{MellinMbccc} from \eqref{eq:gamma}. We first note that the $\gamma_{35}$-integral can be computed by using the first Barnes' lemma:
\begin{align}
\int\frac{ds}{2\pi i}\Gamma[a+s]\Gamma[b+s]\Gamma[c-s]\Gamma[d-s]=\frac{\Gamma[a+c]\Gamma[a+d]\Gamma[b+c]\Gamma[b+d]}{\Gamma[a+b+c+d]},
\end{align}
giving
\begin{align}
\begin{split}
\mathcal{M}_{\bullet\circ\circ\circ}=&\int\frac{d\gamma_{25}}{2\pi i}\mathcal{M}_{\circ\circ\circ\circ\circ}(s-\Delta_5+2\gamma_{25},t)\\
&\times\frac{\Gamma[\gamma_{25}]\Gamma[\frac{\Delta_1+\Delta_2+\Delta_5-s}{2}-\gamma_{25}]\Gamma[\Delta_5-\gamma_{25}]\Gamma[\frac{\Delta_1-\Delta_2-\Delta_5+s}{2}+\gamma_{25}]}{\Gamma[\frac{\Delta_1+\Delta_2+\Delta_5-s}{2}]\Gamma[\frac{\Delta_1-\Delta_2+\Delta_5+s}{2}]}\;,
\end{split}
\end{align}
where the following identities have been used
\begin{align}
\gamma_{12}=\frac{\Delta_1+\Delta_2+\Delta_5-s}{2}\;,\qquad\gamma_{13}+\gamma_{14}=\frac{\Delta_1-\Delta_2+\Delta_5+s}{2}\;.
\end{align}
We then perform the integral over $\gamma_{25}$ through enclosing the contour to the left, leading to
\begin{align}
\label{eq:M=A1+A2}\mathcal{M}_{\bullet\circ\circ\circ}=&\sum_{m,n=0}^{\infty}\bigg(\frac{K^{(0)}_{m}A^1_{n}}{s-\Delta-\Delta_5-2m-2n}+\frac{K^{(0)}_{m}A^2_{n}}{-\Delta_1+\Delta_2-\Delta-2m-2n}\bigg)\;,
\end{align}
with 
\begin{align}
A^1_{n}=&\bigg(\prod_{i=1}^5\frac{1}{\Gamma[\Delta_i]}\bigg)\frac{(-1)^n\Gamma[\Delta_5+n]\Gamma[\frac{\Delta_1+\Delta_2+\Delta_5-s}{2}+n]\Gamma[\frac{\Delta_1-\Delta_2-\Delta_5+s}{2}-n]}{n!\Gamma[\frac{\Delta_1+\Delta_2+\Delta_5-s}{2}]\Gamma[\frac{\Delta_1-\Delta_2+\Delta_5+s}{2}]}\;,
\end{align}
and
\begin{align}
A^2_{n}=&\bigg(\prod_{i=1}^5\frac{1}{\Gamma[\Delta_i]}\bigg)\frac{(-1)^n\Gamma[\frac{-\Delta_1+\Delta_2+\Delta_5-s}{2}-n]\Gamma[\frac{\Delta_1-\Delta_2+\Delta_5+s}{2}+n]\Gamma[\Delta_1+n]}{n!\Gamma[\frac{\Delta_1+\Delta_2+\Delta_5-s}{2}]\Gamma[\frac{\Delta_1-\Delta_2+\Delta_5+s}{2}]}\;.
\end{align}
We first focus on $A^1_{n}$. When $\Delta_5=0$, $\Gamma[\Delta_5]$ in the denominator forces $n$ in $\Gamma[\Delta_5+n]$ to be zero. Therefore, $A^1_n$ reduces to 
\begin{align}
A^1_{n}=&\delta_{n,0}\bigg(\prod_{i=1}^4\frac{1}{\Gamma[\Delta_i]}\bigg)\;.
\end{align}
When $\Delta_5>0$, as a function of $s$, $A^1_{n}$ can be expanded as
\begin{align}
A^1_{n}=&\bigg(\prod_{i=1}^5\frac{1}{\Gamma[\Delta_i]}\bigg)\sum_{r=0}^{\infty}\frac{\mathcal{A}_{nr}}{s+\Delta_1-\Delta_2-\Delta_5-2n+2r}+F_{n}(s)\;,
\end{align}
where $F_{n}(s)$ is the regular term and
\begin{align}
\mathcal{A}_{nr}=&\frac{2(-1)^{n+r}\Gamma[\Delta_5+n]\Gamma[\Delta_1+r]}{n!r!\Gamma[\Delta_1+r-n]\Gamma[\Delta_5+n-r]}\;.
\end{align}
On the other hand, we note that
\begin{align}
\begin{split}
&\bigg(\prod_{i=1}^5\frac{1}{\Gamma[\Delta_i]}\bigg)\sum_{r=0}^{\infty}\frac{\mathcal{A}_{nr}}{s+\Delta_1-\Delta_2-\Delta_5-2n+2r}\\
=&\bigg(\prod_{i=1}^5\frac{1}{\Gamma[\Delta_i]}\bigg)\frac{(-1)^{n}\Gamma[\frac{s+\Delta_1-\Delta_2-\Delta_5}{2}-n]\Gamma[\Delta_1]}{n!\Gamma[\frac{s+\Delta_1-\Delta_2-\Delta_5}{2}-n+1]\Gamma[\Delta_1-n]\Gamma[\Delta_5+n-r]}\\
&\qquad\times{}_3F_2\left(\left.\begin{array}{c} \Delta_1,\frac{s+\Delta_1-\Delta_2-\Delta_5}{2}-n, 1-\Delta_5-n \\ \Delta_1-n,\frac{s+\Delta_1-\Delta_2-\Delta_5}{2}-n+1\end{array}\right.\bigg|1\right)\;,
\end{split}
\end{align}
where we wrote the sum over $r$ as a generalized hypergeometric function ${}_3F_2$. Using the identity \eqref{3F2Identity0} for generalized hypergeometric function, we can replace the above ${}_3F_2$ function by
\begin{align}
\begin{split}
&{}_3F_2\left(\left.\begin{array}{c} \Delta_1,\frac{s+\Delta_1-\Delta_2-\Delta_5}{2}-n, 1-\Delta_5-n \\ \Delta_1-n,\frac{s+\Delta_1-\Delta_2-\Delta_5}{2}-n+1\end{array}\right.\bigg|1\right)\\
&\qquad=\frac{\Gamma[\Delta_5+n]\Gamma[\frac{s+\Delta_1-\Delta_2-\Delta_5}{2}-n+1](\frac{s-\Delta_1-\Delta_2-\Delta_5}{2}-n+1)_n}{\Gamma[\frac{s+\Delta_1-\Delta_2+\Delta_5}{2}](1-\Delta_1)_n}\;,
\end{split}
\end{align}
leading to
\begin{align}
&\bigg(\prod_{i=1}^5\frac{1}{\Gamma[\Delta_i]}\bigg)\sum_{r=0}^{\infty}\frac{\mathcal{A}_{nr}}{s+\Delta_1-\Delta_2-\Delta_5-2n+2r}=A^1_{n}\;.
\end{align}
We thus conclude that the regular term $F_n(s)$ vanishes and $A_n^1$ can be written as
\begin{align}
\label{eq:A1}A^1_{n}=&\bigg(\prod_{i=1}^5\frac{1}{\Gamma[\Delta_i]}\bigg)\sum_{r=0}^{\infty}\frac{\mathcal{A}_{nr}}{s+\Delta_1-\Delta_2-\Delta_5-2n+2r}+\delta_{\Delta_5,0}\delta_{n,0}\bigg(\prod_{i=1}^4\frac{1}{\Gamma[\Delta_i]}\bigg)\;.
\end{align}
$A^2_{n}$ can be obtained analogously. Actually, one can directly deduce $A^2_{n}$ by switching $\Delta_1$ and $\Delta_5$ as well as $\Delta_2$ and $s$ in $A^1_{n}$. This gives
\begin{align}
\label{eq:A2}A^2_{n}=&\bigg(\prod_{i=1}^5\frac{1}{\Gamma[\Delta_i]}\bigg)\sum_{r=0}^{\infty}\frac{\mathcal{A}_{rn}}{-s-\Delta_1+\Delta_2+\Delta_5-2n+2r}+\delta_{\Delta_1,0}\delta_{n,0}\bigg(\prod_{i=2}^5\frac{1}{\Gamma[\Delta_i]}\bigg)\;,
\end{align}
where we used the fact that $\mathcal{A}_{nr}\big|_{\Delta_1\leftrightarrow\Delta_5, s\leftrightarrow\Delta_2}=\mathcal{A}_{rn}$. Substituting \eqref{eq:A1} and \eqref{eq:A2} into \eqref{eq:M=A1+A2} thus leads to
\begin{align}
\begin{split}
\mathcal{M}_{\bullet\circ\circ\circ}=&\bigg(\prod_{i=1}^5\frac{1}{\Gamma(\Delta_i)}\bigg)\sum_{m,n,r=0}^{\infty}\frac{K^{(0)}_{m}\mathcal{A}_{nr}}{(s-\Delta-\Delta_5-2m-2n)(\Delta_1-\Delta_2+\Delta+2m+2r)}\\
&\qquad\qquad+\delta_{\Delta_5,0}\bigg(\prod_{i=1}^4\frac{1}{\Gamma[\Delta_i]}\bigg)\sum_{m=0}^{\infty}\frac{K^{(0)}_m|_{\Delta_5=0}}{s-\Delta-2m}\;.
\end{split}
\end{align}
where we assumed $\Delta_1>0$. The sum over $r$ in the first term can be computed by virtue of \eqref{3F2Identity0}. This gives
\begin{align}
\label{eq:Mbccc}\mathcal{M}_{\bullet\circ\circ\circ}=&\sum_{m=0}^{\infty}\frac{C^{(1)}_{m}}{s-\Delta-\Delta_5-2m}+\delta_{\Delta_5,0}\bigg(\prod_{i=1}^4\frac{1}{\Gamma[\Delta_i]}\bigg)\sum_{m=0}^{\infty}\frac{K^{(0)}_m|_{\Delta_5=0}}{s-\Delta-2m}\;,
\end{align}
where the residue $C^{(1)}_m$ is given by
\begin{align}
\begin{split}
C^{(1)}_m=&\bigg(\prod_{i=1}^4\frac{1}{\Gamma[\Delta_i]}\bigg)\frac{\Gamma[\frac{\Delta_1-\Delta_2+\Delta}{2}+m]}{\Gamma[\frac{\Delta_1-\Delta_2+2\Delta_5+\Delta}{2}+m]}K^{(0)}_m\\
&\qquad\times{}_3F_2\left(\left.\begin{array}{c}-m, \Delta_5,h-\Delta-m \\1-m-\frac{\Delta_1-\Delta_2+\Delta}{2},\frac{\sum_{i=3}^5\Delta_i-\Delta}{2}-m\end{array}\right.\bigg|1\right)\;.
\end{split}
\end{align}
We note the second term in the above equality can be absorbed into the first term. After that, we use the identity \eqref{3F2Identity1} for generalized hypergeometric function, translating $C^{(1)}_m$ into a form in \eqref{eq:C1}.

 
\section{More details of $\mathcal{M}_{\bullet\bullet\circ\circ}$}

In this appendix, we will show the details of the computation for $\mathcal{M}^{\rm I}_{\bullet\bullet\circ\circ}$ and $\mathcal{M}^{\rm II}_{\bullet\bullet\circ\circ}$.

\subsection{Type I}\label{App:MbbcctypeI}

We start with the integral representation \eqref{gamma15typeI}. After performing the $\gamma_{15}$-integral, $\mathcal{M}^{\rm I}_{\bullet\bullet\circ\circ}$ becomes
\begin{align}\label{Ap1+Ap2}
\begin{split}
\mathcal{M}^{\rm I}_{\bullet\bullet\circ\circ}=\sum_{m,n=0}^{\infty}\bigg(\frac{K^{(1)}_{m}A^{\prime 1}_n}{s-\Delta-\Delta_5-\Delta_6-2m-2n}+\frac{K^{(1)}_{m}A^{\prime 2}_n}{\Delta_1-\Delta_2-\Delta-2n-2m}\bigg)\;,
\end{split}
\end{align}
where we define $A^{\prime 1}_n$ and $A^{\prime 2}_n$ as
\begin{align}
\begin{split}
A^{\prime 1}_n=&\prod_{i=1}^5\frac{1}{\Gamma[\Delta_i]}\frac{(-1)^n\Gamma[\Delta_5+n]\Gamma[\frac{\Delta_1+\Delta_2+\Delta_5+\Delta_6-s}{2}+n]\Gamma[\frac{\Delta_2-\Delta_1-\Delta_5-\Delta_6+s}{2}-n]}{n!\Gamma[\frac{\Delta_1+\Delta_2+\Delta_5+\Delta_6-s}{2}]\Gamma[\frac{-\Delta_1+\Delta_2+\Delta_5-\Delta_6+s}{2}]}\;,
\end{split}
\end{align}
and
\begin{align}
\begin{split}
A^{\prime 2}_n=\prod_{i=1}^5\frac{1}{\Gamma[\Delta_i]}\frac{(-1)^n\Gamma[\frac{\Delta_1-\Delta_2+\Delta_5+\Delta_6-s}{2}-n]\Gamma[\frac{\Delta_2-\Delta_1+\Delta_5-\Delta_6+s}{2}+n]\Gamma[\Delta_2+n]}{n!\Gamma[\frac{\Delta_1+\Delta_2+\Delta_5+\Delta_6-s}{2}]\Gamma[\frac{-\Delta_1+\Delta_2+\Delta_5-\Delta_6+s}{2}]}\;.
\end{split}
\end{align}
By noting that $A^{\prime 1}=A^{1}|_{s-\rightarrow s-\Delta_6,\Delta_1\leftrightarrow\Delta_2}$ and $A^{\prime 2}=A^{2}|_{s-\rightarrow s-\Delta_6,\Delta_1\leftrightarrow\Delta_2}$, one can immediately write down expressions for $A^{\prime 1}$ and $A^{\prime 2}$ from \eqref{eq:A1} and \eqref{eq:A2}. Substituting $A^{\prime 1}_n$ and $A^{\prime 2}_n$ into \eqref{Ap1+Ap2} then reproduces \eqref{eq:MbbcctypeI}.


\subsection{Type II}\label{App:MbbcctypeII}

To derive the expressions \eqref{eq:MbcccctypeII} for $\mathcal{M}^{\rm II}_{\bullet\circ\circ\circ\circ}$ and \eqref{eq:M1bbcc} for $\mathcal{M}^1_{\bullet\bullet\circ\circ}$, we start with \eqref{eq:gamma26typeII}. Performing the integral over $\gamma_{26}$ in \eqref{eq:gamma26typeII} leads to
\begin{align}\label{M=B1+B2}
\mathcal{M}^{\rm II}_{\bullet\circ\circ\circ\circ}=\sum_{m,n=0}^{\infty}\bigg(\frac{\widetilde{R}^{(0)}_mB^1_n}{s^{\prime\prime}-\Delta_6-\Delta-2m-2n}+\frac{\widetilde{R}^{(0)}_mB^2_n}{t^{\prime\prime}-\Delta_1-\Delta-2m-2n}\bigg)\;,
\end{align}
with
\begin{align}
B^1_n=&\bigg(\prod_{i=1}^6\frac{1}{\Gamma[\Delta_i]}\bigg)\frac{(-1)^n\Gamma[\Delta_6+n]\Gamma[\frac{\Delta_1-\Delta_6+s^{\prime\prime}-t^{\prime\prime}}{2}-n]\Gamma[\frac{\Delta_1+\Delta_6-s^{\prime\prime}+t^{\prime\prime}}{2}+n]}{n!\Gamma[\frac{\Delta_1+\Delta_6-s^{\prime\prime}+t^{\prime\prime}}{2}]\Gamma[\frac{\Delta_1+\Delta_6+s^{\prime\prime}-t^{\prime\prime}}{2}]}\;,
\end{align}
and
\begin{align}
B^2_n=&\bigg(\prod_{i=1}^6\frac{1}{\Gamma[\Delta_i]}\bigg)\frac{(-1)^n\Gamma[\frac{\Delta_1+\Delta_6+s^{\prime\prime}-t^{\prime\prime}}{2}+n]\Gamma[\Delta_1+n]\Gamma[\frac{-\Delta_1+\Delta_6-s^{\prime\prime}+t^{\prime\prime}}{2}-n]}{n!\Gamma[\frac{\Delta_1+\Delta_6-s^{\prime\prime}+t^{\prime\prime}}{2}]\Gamma[\frac{\Delta_1+\Delta_6+s^{\prime\prime}-t^{\prime\prime}}{2}]}\;.
\end{align}
$B^1_n$ and $B^2_n$ can be obtained by following the steps in the Appendix \ref{Mbccc}. Specifically, for $\Delta_6=0$ ($\Delta_1=0$) we find that $B^1_n=\delta_{n,0}\prod_{i=1}^5\frac{1}{\Gamma[\Delta_i]}$ ($B^2_n=\delta_{n,0}\prod_{i=2}^6\frac{1}{\Gamma[\Delta_i]}$), while for $\Delta_6>0$ ($\Delta_1>0$) one can expand $B^1_n$ ($B^2_n$) around its poles and show the regular term vanishes by using \eqref{3F2Identity0}. Substituting the expression for $B^1_n$ and $B^2_n$ into \eqref{M=B1+B2} and shifting $m$ by $m\rightarrow m-n$ then reproduce the expression \eqref{eq:MbcccctypeII}.

To get  \eqref{eq:M1bbcc} for $\mathcal{M}^1_{\bullet\bullet\circ\circ}$, we evaluate the $\gamma_{45}$- and $\gamma_{25}$-integral in \eqref{eq:integral} by using the residue theorem, leading to
\begin{align}\label{I1+I2+I3}
\mathcal{M}^1_{\bullet\bullet\circ\circ}=I^1+I^2+I^3\;,
\end{align}
where $I_1$, $I_2$ and $I_3$ are given by
\begin{align}
\begin{split}
I_{1}=\bigg(\prod_{i=1}^6\frac{1}{\Gamma[\Delta_i]}\bigg)\sum_{n_1,n_2,k_1,k_2=0}^{\infty}\frac{(-1)^{k_1+k_2}\widetilde{K}^{(1)}_{n_1n_2}\Gamma[\Delta_5+k_1+k_2]}{(t-\Delta_1-\Delta_5-\Delta-2n_2-2k_2)\Gamma[\frac{\Delta_3+\Delta_4+\Delta_5-s}{2}]}\\
\frac{\Gamma[\frac{\Delta_3+\Delta_4-\Delta_5-s}{2}-k_1-k_2]\Gamma[\frac{\Delta_2+\Delta_3+\Delta_5-t}{2}+k_2]\Gamma[\frac{s+t-\Delta_2-\Delta_4}{2}+k_1]}{k_1!k_2!(\Delta_4-\Delta_5-\Delta_6-\Delta-2n_1-2k_1-2k_2)\Gamma[\frac{s+t-\Delta_2-\Delta_4}{2}]\Gamma[\frac{\Delta_2+\Delta_3+\Delta_5-t}{2}]}\;,
\end{split}
\end{align}
\begin{align}
\begin{split}
I_{2}=\bigg(\prod_{i=1}^6\frac{1}{\Gamma[\Delta_i]}\bigg)\sum_{n_1,n_2,k_1,k_2=0}^{\infty}\frac{(-1)^{k_1+k_2}\widetilde{K}^{(1)}_{n_1n_2}\Gamma[\frac{\sum_{i=3}^5\Delta_i-s}{2}+k_1]}{(s-\Delta_3-\Delta_6-\Delta-2k_1-2n_1)\Gamma[\frac{\Delta_3+\Delta_4+\Delta_5-s}{2}]}\\
\frac{\Gamma[\frac{-t+\Delta_2-\Delta_3+\Delta_{5}}{2}-k_1-k_2]\Gamma[\frac{s+t-\Delta_2-\Delta_4}{2}+k_2]\Gamma[\Delta_3+k_1+k_2]}{k_1!k_2!(-\Delta_1+\Delta_2-\Delta_3-\Delta-2n_2-2k_1-2k_2)\Gamma[\frac{s+t-\Delta_2-\Delta_4}{2}]\Gamma[\frac{\Delta_2+\Delta_3+\Delta_5-t}{2}]}\;,
\end{split}
\end{align}
and
\begin{align}
\begin{split}
I_{3}=\bigg(\prod_{i=1}^6\frac{1}{\Gamma[\Delta_i]}\bigg)\sum_{n_1,n_2,k_1,k_2=0}^{\infty}\frac{(-1)^{k_1+k_2}\widetilde{K}^{(1)}_{n_1n_2}\Gamma[\frac{\sum_{i=3}^5\Delta_i-s}{2}+k_1]}{(t-\Delta_1-\Delta_5-\Delta-2n_2-2k_2)\Gamma[\frac{\Delta_3+\Delta_4+\Delta_5-s}{2}]}\\
\frac{\Gamma[\frac{s-\Delta_3-\Delta_4+\Delta_5}{2}-k_1+k_2]\Gamma[\frac{\Delta_2+\Delta_3+\Delta_5-t}{2}+k_2]\Gamma[\frac{t-\Delta_2+\Delta_3-\Delta_{5}}{2}+k_1-k_2]}{k_1!k_2!(s-\Delta_3-\Delta_6-\Delta-2k_1-2n_1)\Gamma[\frac{s+t-\Delta_2-\Delta_4}{2}]\Gamma[\frac{\Delta_2+\Delta_3+\Delta_5-t}{2}]}\;,
\end{split}
\end{align}
respectively. Let us focus on $I^1$ first. After defining $\widetilde{A}_{k_1k_2}$ as
\begin{align}
\widetilde{A}_{k_1k_2}=&\frac{\Gamma[\Delta_5+k_1+k_2]\Gamma[\frac{\Delta_3+\Delta_4-\Delta_5-s-2k_1-2k_2}{2}]\Gamma[\frac{\Delta_2+\Delta_3+\Delta_5-t+2k_2}{2}]\Gamma[\frac{s+t-\Delta_2-\Delta_4+2k_1}{2}]}{(-1)^{k_1+k_2}k_1!k_2!\Pi_6(\Delta_i)\Gamma[\frac{s+t-\Delta_2-\Delta_4}{2}]\Gamma[\frac{\Delta_2+\Delta_3+\Delta_5-t}{2}]\Gamma[\frac{\Delta_3+\Delta_4+\Delta_5-s}{2}]}\;,
\end{align}
with $\Pi_a(\Delta_i)$ defined as
\begin{align}
\Pi_a(\Delta_i)=\prod_{i=1}^a\Gamma[\Delta_i]\;,
\end{align}
$I^1$ can be written as
\begin{align}
I^1=&\sum_{n_1,n_2,k_1,k_2=0}^{\infty}\frac{\widetilde{K}^{(1)}_{n_1n_2}\widetilde{A}_{k_1k_2}}{(\Delta_4-\Delta_5-\Delta_6-\Delta-2n_1-2k_1-2k_2)(t-\Delta_1-\Delta_5-\Delta-2n_2-2k_2)}\;.
\end{align}

Following the steps in the Appendix \ref{Mbccc}, $\widetilde{A}_{k_1k_2}$ can be expanded around its poles at $s=\Delta_3+\Delta_4-\Delta_5-2k_1-2k_2+2m$. The absence of regular term can be verified by using \eqref{3F2Identity0}. As a result, $I_1$ is expressible as
\begin{align}\label{eq:I1}
\begin{split}
I^1=I^1|_{\Delta_5>0}+\delta_{\Delta_5,0}\mathcal{M}_{\bullet\circ\circ\circ}\bigg[\begin{array}{c} 12345\\ 64231\end{array}\bigg]\;.
\end{split}
\end{align}
Here $I^1|_{\Delta_5>0}$ is expressible as
\begin{align}
I^1|_{\Delta_5>0}=\sum_{n_1,n_2,k_1,k_2=0}^{\infty}\frac{\widetilde{K}^{(1)}_{n_1n_2}\widetilde{\mathcal{A}}_{k_1k_2m}}{(t-\Delta_1-\Delta_5-\Delta-2n_2-2k_2)(\Delta_3+\Delta_4-\Delta_5-s-2k_1-2k_2+2m)}\;,
\end{align}
where $\widetilde{\mathcal{A}}_{k_1k_2n_1m}$ is the residue of $\widetilde{A}_{k_1k_2}$ at poles $s=\Delta_3+\Delta_4-\Delta_5-2k_1-2k_2+2m$ divided by $(\Delta_4-\Delta_5-\Delta_6-\Delta-2n_1-2k_1-2k_2)$ 
\begin{align}
\widetilde{\mathcal{A}}_{k_1k_2n_1m}=&\frac{2(-1)^{k_2}(1-\Delta_5-k_1-k_2)_m(\frac{\Delta_2+\Delta_3+\Delta_5-t}{2})_{k_2}(1-\frac{t-\Delta_2+\Delta_3-\Delta_5-2k_2+2m}{2})_{k_1}}{k_1!k_2!m!\Pi_6(\Delta_i)(\Delta_4-\Delta_5-\Delta_6-\Delta-2n_1-2k_1-2k_2)}\;.
\end{align}
In a similar way, one can derive an expression for $I^2$, given by 
\begin{align}\label{eq:I2}
\begin{split}
I^2=I^2|_{\Delta_3>0}+\delta_{\Delta_3,0}\mathcal{M}_{\bullet\circ\circ\circ}\bigg[\begin{array}{c} 12345\\ 12456\end{array}\bigg]\;,
\end{split}
\end{align}
where $I^2|_{\Delta_3>0}$ is
\begin{align}
I^2|_{\Delta_3>0}=\sum\frac{\widetilde{K}^{(1)}_{n_1n_2}\widetilde{\mathcal{A}}_{k_1k_2n_1m}|_{s\leftrightarrow t,\Delta_1\leftrightarrow\Delta_6,\Delta_2\leftrightarrow\Delta_4,\Delta_3\leftrightarrow\Delta_5}}{(s-\Delta_6-\Delta_3-\Delta-2n_2-2k_2)(\Delta_5+\Delta_2-\Delta_3-t-2k_1-2k_2+2m)}\;,
\end{align}
where all of the summation variables $n_1$, $n_2$, $k_1$ and $k_2$ run from zero to infinity. The computation of $I^3$ is more involved. We first note that $I^3$ vanishes when $\Delta_5=0$. Thus we only need to deal with the case when $\Delta_5>0$. In that case, we can expand $I^3$ around its poles at $s=\Delta_3+\Delta_6+\Delta+2k_1+2n_1$ and $s=\Delta_3+\Delta_4-\Delta_5+2k_1-2k_2-2r$. One can directly check that the term obtained by the expansion of $I^3$ around $s=\Delta_3+\Delta_4-\Delta_5+2k_1-2k_2-2r$ is exactly $I^1|_{\Delta_5>0}$ up to a minus sign. This leads to
\begin{align}\label{I3}
I^3=\sum_{n_1,n_2,k_1,k_2=0}^{\infty}\frac{\widetilde{B}^{1}_{n_1n_2k_1k_2}}{s-\Delta_3-\Delta_6-\Delta-2k_1-2n_1}-I^1|_{\Delta_5>0}+F(s,t)\;,
\end{align}
where $F(s,t)$ is a possible regular term and the residue $\widetilde{B}^{1}_{n_1n_2k_1k_2}$ of poles at $s=\Delta_3+\Delta_6+\Delta+2k_1+2n_1$ is given by
\begin{align}
\begin{split}
\widetilde{B}^{1}_{n_1n_2k_1k_2}=\frac{(-1)^{k_1+k_2}\widetilde{K}^{(1)}_{n_1n_2}\Gamma[\frac{\Delta_4+\Delta_5-\Delta_6-\Delta}{2}-n_1]\Gamma[\frac{-\Delta_4+\Delta_5+\Delta_6+\Delta}{2}+n_1+k_2]}{k_1!k_2!(t-\Delta_1-\Delta_5-\Delta-2n_2-2k_2)\Gamma[\frac{\Delta_2+\Delta_3+\Delta_5-t}{2}]}\\
\frac{\Gamma[\frac{\Delta_2+\Delta_3+\Delta_5-t}{2}+k_2]\Gamma[\frac{t-\Delta_2+\Delta_3-\Delta_{5}}{2}+k_1-k_2]}{\Pi_5(\Delta_i)\Gamma[\frac{t-\Delta_2+\Delta_3-\Delta_4+\Delta_6+\Delta}{2}+k_1+n_1]\Gamma[\frac{\Delta_4+\Delta_5-\Delta_6-\Delta}{2}-k_1-n_1]}\;.
\end{split}
\end{align}
In the following, we will show that the regular term $F(s,t)$ vanishes. Equivalently, we will prove that
\begin{align}\label{I31}
\sum_{n_1,n_2,k_1,k_2=0}^{\infty}\frac{\widetilde{B}^{1}_{n_1n_2k_1k_2}}{s-\Delta_3-\Delta_6-\Delta-2k_1-2n_1}=I_3+I^1|_{\Delta_5>0}\;.
\end{align}
To do this, we first rewrite the sum over $k_1$ as a generalized hypergeometric function
\begin{align}
\begin{split}
&\sum_{n_1,n_2,k_1,k_2=0}^{\infty}\frac{\widetilde{B}^{1}_{n_1n_2k_1k_2}}{s-\Delta_3-\Delta_6-\Delta-2k_1-2n_1}=\sum_{n_1,n_2,k_2=0}^{\infty}\widehat{B}_{n_1n_2k_2}\\
&\qquad\times{}_3F_2\left(\left.\begin{array}{c}\frac{\Delta_3+\Delta_6+\Delta-s}{2}+n_1,\frac{t-\Delta_2+\Delta_3-\Delta_{5}}{2}-k_{2}, 1-\frac{\Delta_4+\Delta_5-\Delta_6-\Delta}{2}+n_1 \\ \frac{\Delta_3+\Delta_6+\Delta-s}{2}+n_1+1,\frac{t-\Delta_2+\Delta_3-\Delta_4+\Delta_6+\Delta}{2}+n_1\end{array}\right.\bigg|1\right)\;,
\end{split}
\end{align}
where we defined $\widehat{B}_{n_1n_2k_2}$ as
\begin{align}
\begin{split}
\widehat{B}_{n_1n_2k_2}=\frac{(-1)^{k_2}\widetilde{K}^{(1)}_{n_1n_2}\Gamma[\frac{\Delta_3+\Delta_6+\Delta-s}{2}+n_1]}{-2k_2!(t-\Delta_1-\Delta_5-\Delta-2n_2-2k_2)\Gamma[\frac{\Delta_3+\Delta_6+\Delta-s}{2}+n_1+1]}\\
\frac{\Gamma[\frac{-\Delta_4+\Delta_5+\Delta_6+\Delta}{2}+n_1+k_2]\Gamma[\frac{\Delta_2+\Delta_3+\Delta_5-t}{2}+k_2]\Gamma[\frac{t-\Delta_2+\Delta_3-\Delta_{5}}{2}-k_2]}{\Pi_5(\Delta_i)\Gamma[\frac{t-\Delta_2+\Delta_3-\Delta_4+\Delta_6+\Delta}{2}+n_1]\Gamma[\frac{\Delta_2+\Delta_3+\Delta_5-t}{2}]}\;.
\end{split}
\end{align}
Using the identity \eqref{3F2Identity2}, we can rewrite the generalized hypergeometric function, leading to
\begin{align}
&\sum_{n_1,n_2,k_1,k_2=0}^{\infty}\frac{\widetilde{B}^{1}_{n_1n_2k_1k_2}}{s-\Delta_3-\Delta_6-\Delta-2k_1-2n_1}=\widehat{A}^1+\widehat{A}^2\;,
\end{align}
where $\widehat{A}^1$ and $\widehat{A}^2$ are given by
\begin{align}
\begin{split}
\widehat{A}^1=&\sum_{n_1,n_2,k_2=0}^{\infty}\frac{(-1)^{k_1+k_2}\widetilde{K}_{n_1n_2}\Gamma[\frac{\Delta_3+\Delta_6+\Delta-s}{2}+n_1]\Gamma[\frac{\Delta_2-\Delta_3-\Delta_5-t}{2}+2]}{-2k_2!\Pi_5(\Delta_i)(t-\Delta_1-\Delta_5-\Delta-2n_2-2k_2)}\\
&\frac{\Gamma[\frac{s-\Delta_3-\Delta_4+\Delta_5}{2}+k_2]\Gamma[\frac{\Delta_2+\Delta_3+\Delta_5-t}{2}+k_2]\Gamma[\frac{t-\Delta_2+\Delta_3-\Delta_{5}}{2}-k_2]}{\Gamma[\frac{s+t-\Delta_2-\Delta_4}{2}]\Gamma[\frac{\Delta_2+\Delta_3+\Delta_5-t}{2}]\Gamma[\frac{\Delta_2-\Delta_5+\Delta_6+\Delta-s-t}{2}+n_1+2]}\\
&\qquad\times{}_3F_2\left(\left.\begin{array}{c}\frac{\Delta_3+\Delta_6+\Delta-s}{2}+n_1,\frac{\Delta_2+\Delta_4-s-t}{2}+1, 1-\Delta_5-k_2 \\ \frac{\Delta_3+\Delta_4-\Delta_5-s}{2}-k_2+1,\frac{\Delta_2-\Delta_5+\Delta_6+\Delta-s-t}{2}+n_1+2\end{array}\right.\bigg|1\right)\;,
\end{split}
\end{align}
and
\begin{align}
\begin{split}
\widehat{A}^2&=\sum_{n_1,n_2,k_2=0}^{\infty}\frac{(-1)^{k_1+k_2}\widetilde{K}^{(1)}_{n_1n_2}\Gamma[\frac{\Delta_3+\Delta_4-\Delta_5-s}{2}-k_2]\Gamma[\frac{\Delta_2-\Delta_3-\Delta_5-t}{2}+2]}{-2k_2!\Pi_5(\Delta_i)(t-\Delta_1-\Delta_5-\Delta-2n_2-2k_2)\Gamma[\frac{\Delta_3+\Delta_4+\Delta_5-s}{2}]}\\
&\frac{\Gamma[\frac{s-\Delta_3-\Delta_4+\Delta_5}{2}+k_2+1]\Gamma[\frac{-\Delta_4+\Delta_5+\Delta_6+\Delta}{2}+n_1+k_2]\Gamma[\frac{\Delta_2+\Delta_3+\Delta_5-t}{2}+k_2]}{\Gamma[\frac{\Delta_2+\Delta_3+\Delta_5-t}{2}]\Gamma[\frac{s-\Delta_3-\Delta_4-\Delta_5}{2}+2]\Gamma[\frac{\Delta_2-\Delta_3-\Delta_4+\Delta_6+\Delta-t}{2}+n_1+k_2+2]}\\
&\qquad\times{}_3F_2\left(\left.\begin{array}{c}1-\frac{\Delta_4+\Delta_5-\Delta_6-\Delta}{2}+n_1, \frac{\Delta_2-\Delta_3-\Delta_{5}-t}{2}+2, \frac{s-\Delta_3-\Delta_4-\Delta_5}{2}+1 \\ \frac{s-\Delta_3-\Delta_4-\Delta_5}{2}+2,\frac{\Delta_2-\Delta_3-\Delta_4+\Delta_6+\Delta-t}{2}+n_1+k_2+2\end{array}\right.\bigg|1\right)\;.
\end{split}
\end{align}
To prove \eqref{I31}, we need to show that  $\widehat{A}^1+\widehat{A}^2$ is equal to the right-hand side of \eqref{I31}. Thanks to the identities \eqref{3F2Identity3} and \eqref{3F2Identity4}, we can show that 
\begin{align}
I^3=\widehat{A}^1\;,\qquad I^{1}|_{\Delta_5>0}=\widehat{A}^2\;.
\end{align}
This completes our proof. We conclude that 
\begin{align}\label{eq:0}
I^3=\sum_{n_1,n_2,k_1,k_2=0}^{\infty}\frac{\widetilde{B}^{1}_{n_1n_2k_1k_2}}{s-\Delta_3-\Delta_6-\Delta-2k_1-2n_1}-I^1|_{\Delta_5>0}\;.
\end{align}
Now we further expand $\widetilde{B}^1_{n_1n_2k_1k_2}$ as 
\begin{align}\label{eq:B}
\widetilde{B}^1_{n_1n_2k_1k_2}=&\frac{(-1)^{k_1+k_2}\widetilde{K}^{(1)}_{n_1,n_2}}{k_1!k_2!(t-\Delta_1-\Delta_5-\Delta-2n_2-2k_2)}\sum_{r=0}^{\infty}\frac{\widetilde{\mathcal{B}}_{n_1k_1k_2r}}{t-\Delta_2+\Delta_3-\Delta_5+2k_1-2k_2+2r}\;,
\end{align}
with
\begin{align}
\widetilde{\mathcal{B}}_{n_1k_1k_2r}=&\frac{2(-1)^r\Gamma[\frac{\Delta_4+\Delta_5-\Delta_6-\Delta}{2}-n_1]\Gamma[\frac{-\Delta_4+\Delta_5+\Delta_6+\Delta+2n_1+2k_2}{2}](\Delta_3+k_1+r-k_2)_{k_2}}{r!\Pi_5(\Delta_i)\Gamma[\frac{-\Delta_4+\Delta_5+\Delta_6+\Delta+2k_2+2n_1-2r}{2}]\Gamma[\frac{\Delta_4+\Delta_5-\Delta_6-\Delta-2k_1-2n_1}{2}]}\;.
\end{align}
The identity \eqref{3F2Identity0} again enables us to show the regular term in the above expansion vanishes when $-\Delta_4+\Delta_5+\Delta_6+\Delta>0$. Plugging \eqref{eq:B} into \eqref{eq:0} and expanding $I^3$ further around poles at $t=\Delta_1+\Delta_5+\Delta+2k_2+2n_2$ and $t=\Delta_2-\Delta_3+\Delta_5-2k_1+2k_2-2r$ give
\begin{align}\label{eq:I3}
I^3=&\sum_{m_1,m_2=0}^{\infty}\frac{\widetilde{C}^{(2)}_{m_1m_2}}{(s-\Delta_3-\Delta_6-\Delta-2m_1)(t-\Delta_1-\Delta_5-\Delta-2m_2)}-I^1|_{\Delta_5>0}-I^2|_{\Delta_3>0}\;,
\end{align}
with the residue $\widetilde{C}^{(2)}_{m_1m_2}$ given by \eqref{eq:C2typeII}. Substituting \eqref{eq:I1}, \eqref{eq:I2} and \eqref{eq:I3} into \eqref{I1+I2+I3} reproduces \eqref{eq:M1bbcc}.

\section{Hypergeometric function identities}
In this appendix, we collect some useful identities for generalized hypergeometric functions ${}_3F_2$ which were used in this paper. A useful identity, which has been used to translate \eqref{eq: C21} into \eqref{eq:C2}, is  
\begin{align}
\label{3F2Identity1}{}_3F_2\left(\left.\begin{array}{c}-n, b, c \\ d,e\end{array}\right.\bigg|1\right)=\frac{(d-b)_n}{(d)_n}{}_3F_2\left(\left.\begin{array}{c}-n, b, e-c \\ b-d-n+1,e\end{array}\right.\bigg|1\right),\hspace{1cm}n\in\mathbb{Z}_{\geq0}\;.
\end{align}
Another useful identity, given by
\begin{align}
\label{3F2Identity0}{}_3F_2\left(\left.\begin{array}{c}a, b, c \\ a-n,b+1\end{array}\right.\bigg|1\right)=\frac{\Gamma[1-c]\Gamma[b+1](b-a+1)_n}{\Gamma[b-c+1](1-a)_n},\hspace{1cm}\text{Re}(c)<1-n\;,
\end{align}
with $n$ a non-negative integer, enables us to rewrite certain generalized hypergeometric functions ${}_3F_2$ as Gamma functions \cite{O.Marichev1}. This identity was used to show that the regular terms in $\mathcal{M}_{\bullet\circ\circ\circ}(s,t)$ and $\mathcal{M}^{\rm I}_{\bullet\bullet\circ\circ}(s,t)$ vanish. 

To prove the absence of regular terms in $\mathcal{M}_{\bullet\bullet\circ\circ}^{\rm II}(s,t)$ we need to combine the above identity with the following three identities. The first two identities are \cite{olver2010nist}
\begin{align}\label{3F2Identity3}
{}_3F_2\left(\left.\begin{array}{c}a_1, a_2, a_3 \\ b_1, b_2\end{array}\right.\bigg|1\right)=&\frac{\Gamma[b_1]\Gamma[c]}{\Gamma[b_1-a_1]\Gamma[c+a_1]}{}_3F_2\left(\left.\begin{array}{c}a_1, b_2-a_2, b_2-a_3 \\ b_2, c+a_1\end{array}\right.\bigg|1\right)\hspace{0.2cm}\text{Re}(b_1-a_1)>0\;,
\end{align}
and
\begin{align}\label{3F2Identity4}
\begin{split}
{}_3F_2\left(\left.\begin{array}{c}a_1, a_2, a_3 \\ b_1, b_2\end{array}\right.\bigg|1\right)=&\frac{\Gamma[b_1]\Gamma[b_2]\Gamma[c]}{\Gamma[a_1]\Gamma[c+a_2]\Gamma[c+a_3]}{}_3F_2\left(\left.\begin{array}{c}b_1-a_1, b_2-a_1, c \\ c+a_2, c+a_3\end{array}\right.\bigg|1\right)\hspace{0.2cm}\text{Re}(a_1)>0\;,
\end{split}
\end{align}
where we have defined 
\begin{align}
c\equiv b_1+b_2-a_1-a_2-a_3\;.
\end{align}
Let us mention that an analytic continuation from the region $\text{Re}(c)>0$ to the region $c\neq 0, -1, \cdots$ has to be implemented to make these two identities hold. To get the third identity, let us start with the following identity \cite{O.Marichev2}
\begin{align}\label{3F2Identity5}
\begin{split}
{}_3F_2\left(\left.\begin{array}{c}a_1, a_2, a_3 \\ b_1, b_2\end{array}\right.\bigg|1\right)=F_1+F_2\;,
\end{split}
\end{align}
where $F_1$ and $F_2$ are given by
\begin{align}
\begin{split}
F_1=&\frac{\Gamma[b_1-a_1-a_2]\Gamma[b_1]}{\Gamma[b_1-a_1]\Gamma[b_1-a_2]}{}_3F_2\left(\left.\begin{array}{c}a_1, a_2, b_2-a_3 \\ a_1+a_2-b_1+1, b_2\end{array}\right.\bigg|1\right)\;,
\end{split}
\end{align}
and
\begin{align}
\begin{split}
F_2=&\frac{\Gamma[b_1]\Gamma[b_2]\Gamma[a_1+a_2-b_1]\Gamma[c]}{\Gamma[a_1]\Gamma[a_2]\Gamma[c+a_3]\Gamma[b_2-a_3]}{}_3F_2\left(\left.\begin{array}{c}b_1-a_1, b_1-a_2, c \\ b_1-a_1-a_2+1, b_1+b_2-a_1-a_2\end{array}\right.\bigg|1\right)\;.
\end{split}
\end{align}
This identity is valid when $\text{Re}(c)>0$ and $\text{Re}(a_3-b_1+1)>0$. Combining \eqref{3F2Identity3}, \eqref{3F2Identity4}, and \eqref{3F2Identity5} leads to the last identity
\begin{align}\label{3F2Identity2}
\begin{split}
{}_3F_2\left(\left.\begin{array}{c}a_1, a_2, a_3 \\ b_1, b_2\end{array}\right.\bigg|1\right)=F_3+F_4\;,
\end{split}
\end{align}
where $F_3$ and $F_4$ are given by
\begin{align}
\begin{split}
F_3=&\frac{\Gamma[b_1]\Gamma[b_2]\Gamma[a_3-b_1+1]\Gamma[b_1-a_1-a_2]}{\Gamma[b_1-a_1]\Gamma[b_2-a_1]\Gamma[b_1-a_2]\Gamma[a_1+a_3-b_1+1]}\\
&\qquad\times{}_3F_2\left(\left.\begin{array}{c}a_1, a_1-b_1+1, 1-c \\ a_1+a_2-b_1+1, a_1+a_3-b_1+1\end{array}\right.\bigg|1\right)\;,
\end{split}
\end{align}
and
\begin{align}
\begin{split}
F_4=&\frac{\Gamma[b_1]\Gamma[b_2]\Gamma[a_1+a_2-b_1]\Gamma[a_3-b_1+1]\Gamma[b_1-a_1-a_2+1]}{\Gamma[a_1]\Gamma[a_2]\Gamma[a_3-a_1+1]\Gamma[a_3-a_2+1]\Gamma[b_2-a_3]}\\
&\qquad\times{}_3F_2\left(\left.\begin{array}{c}a_3, a_3-b_1+1, a_3-b_2+1 \\ a_3-a_1+1, a_3-a_2+1\end{array}\right.\bigg|1\right)\;,
\end{split}
\end{align}
respectively. This identity holds when $\text{Re}(c)>0$ and $\text{Re}(b_2-a_1)>0$. 
\bibliography{refs} 

\providecommand{\href}[2]{#2}\begingroup\raggedright\begin{thebibliography}{10}

\bibitem{Rastelli:2016nze}
L.~Rastelli and X.~Zhou, ``{Mellin amplitudes for $AdS_5\times S^5$},''
  \href{http://dx.doi.org/10.1103/PhysRevLett.118.091602}{{\em Phys. Rev.
  Lett.} {\bfseries 118} no.~9, (2017) 091602},
\href{http://arxiv.org/abs/1608.06624}{{\ttfamily arXiv:1608.06624 [hep-th]}}.

\bibitem{Rastelli:2017udc}
L.~Rastelli and X.~Zhou, ``{How to Succeed at Holographic Correlators Without
  Really Trying},'' \href{http://dx.doi.org/10.1007/JHEP04(2018)014}{{\em JHEP}
  {\bfseries 04} (2018) 014},
\href{http://arxiv.org/abs/1710.05923}{{\ttfamily arXiv:1710.05923 [hep-th]}}.

\bibitem{Rastelli:2019gtj}
L.~Rastelli, K.~Roumpedakis, and X.~Zhou, ``{$\mathbf{AdS_3\times S^3}$
  Tree-Level Correlators: Hidden Six-Dimensional Conformal Symmetry},''
  \href{http://dx.doi.org/10.1007/JHEP10(2019)140}{{\em JHEP} {\bfseries 10}
  (2019) 140},
\href{http://arxiv.org/abs/1905.11983}{{\ttfamily arXiv:1905.11983 [hep-th]}}.

\bibitem{Alday:2020lbp}
L.~F. Alday and X.~Zhou, ``{All Tree-Level Correlators for M-theory on $AdS_7
  \times S^4$},'' \href{http://dx.doi.org/10.1103/PhysRevLett.125.131604}{{\em
  Phys. Rev. Lett.} {\bfseries 125} no.~13, (2020) 131604},
  \href{http://arxiv.org/abs/2006.06653}{{\ttfamily arXiv:2006.06653
  [hep-th]}}.

\bibitem{Alday:2020dtb}
L.~F. Alday and X.~Zhou, ``{All Holographic Four-Point Functions in All
  Maximally Supersymmetric CFTs},''
  \href{http://dx.doi.org/10.1103/PhysRevX.11.011056}{{\em Phys. Rev. X}
  {\bfseries 11} no.~1, (2021) 011056},
  \href{http://arxiv.org/abs/2006.12505}{{\ttfamily arXiv:2006.12505
  [hep-th]}}.

\bibitem{Alday:2021odx}
L.~F. Alday, C.~Behan, P.~Ferrero, and X.~Zhou, ``{Gluon Scattering in AdS from
  CFT},'' \href{http://dx.doi.org/10.1007/JHEP06(2021)020}{{\em JHEP}
  {\bfseries 06} (2021) 020}, \href{http://arxiv.org/abs/2103.15830}{{\ttfamily
  arXiv:2103.15830 [hep-th]}}.

\bibitem{Bissi:2022mrs}
A.~Bissi, A.~Sinha, and X.~Zhou, ``{Selected Topics in Analytic Conformal
  Bootstrap: A Guided Journey},''
  \href{http://arxiv.org/abs/2202.08475}{{\ttfamily arXiv:2202.08475
  [hep-th]}}.

\bibitem{Arutyunov:1999en}
G.~Arutyunov and S.~Frolov, ``{Some cubic couplings in type IIB supergravity on
  AdS(5) x S**5 and three point functions in SYM(4) at large N},''
  \href{http://dx.doi.org/10.1103/PhysRevD.61.064009}{{\em Phys. Rev. D}
  {\bfseries 61} (2000) 064009},
  \href{http://arxiv.org/abs/hep-th/9907085}{{\ttfamily arXiv:hep-th/9907085}}.

\bibitem{Arutyunov:2000ima}
G.~Arutyunov and S.~Frolov, ``{On the correspondence between gravity fields and
  CFT operators},'' \href{http://dx.doi.org/10.1088/1126-6708/2000/04/017}{{\em
  JHEP} {\bfseries 04} (2000) 017},
  \href{http://arxiv.org/abs/hep-th/0003038}{{\ttfamily arXiv:hep-th/0003038}}.

\bibitem{Aprile:2018efk}
F.~Aprile, J.~Drummond, P.~Heslop, and H.~Paul, ``{The double-trace spectrum of
  $N=4$ SYM at strong coupling},''
\href{http://arxiv.org/abs/1802.06889}{{\ttfamily arXiv:1802.06889 [hep-th]}}.

\bibitem{Aprile:2019rep}
F.~Aprile, J.~Drummond, P.~Heslop, and H.~Paul, ``{One-loop amplitudes in
  $AdS_5 \times S^5$ supergravity from $ \mathcal{N} $ = 4 SYM at strong
  coupling},'' \href{http://dx.doi.org/10.1007/JHEP03(2020)190}{{\em JHEP}
  {\bfseries 03} (2020) 190}, \href{http://arxiv.org/abs/1912.01047}{{\ttfamily
  arXiv:1912.01047 [hep-th]}}.

\bibitem{Alday:2019nin}
L.~F. Alday and X.~Zhou, ``{Simplicity of AdS Supergravity at One Loop},''
  \href{http://dx.doi.org/10.1007/JHEP09(2020)008}{{\em JHEP} {\bfseries 09}
  (2020) 008}, \href{http://arxiv.org/abs/1912.02663}{{\ttfamily
  arXiv:1912.02663 [hep-th]}}.

\bibitem{Aprile:2020uxk}
F.~Aprile, J.~M. Drummond, P.~Heslop, H.~Paul, F.~Sanfilippo, M.~Santagata, and
  A.~Stewart, ``{Single particle operators and their correlators in free $
  \mathcal{N} $ = 4 SYM},''
  \href{http://dx.doi.org/10.1007/JHEP11(2020)072}{{\em JHEP} {\bfseries 11}
  (2020) 072}, \href{http://arxiv.org/abs/2007.09395}{{\ttfamily
  arXiv:2007.09395 [hep-th]}}.

\bibitem{Goncalves:2019znr}
V.~Gon{\c c}alves, R.~Pereira, and X.~Zhou, ``{$20'$ Five-Point Function from
  $AdS_5\times S^5$ Supergravity},''
  \href{http://dx.doi.org/10.1007/JHEP10(2019)247}{{\em JHEP} {\bfseries 10}
  (2019) 247},
\href{http://arxiv.org/abs/1906.05305}{{\ttfamily arXiv:1906.05305 [hep-th]}}.

\bibitem{Alday:2022lkk}
L.~F. Alday, V.~Gon\c{c}alves, and X.~Zhou, ``{Supersymmetric Five-Point Gluon
  Amplitudes in AdS Space},''
  \href{http://dx.doi.org/10.1103/PhysRevLett.128.161601}{{\em Phys. Rev.
  Lett.} {\bfseries 128} no.~16, (2022) 161601},
  \href{http://arxiv.org/abs/2201.04422}{{\ttfamily arXiv:2201.04422
  [hep-th]}}.

\bibitem{Giombi:2018qox}
S.~Giombi and S.~Komatsu, ``{Exact Correlators on the Wilson Loop in
  $\mathcal{N}=4$ SYM: Localization, Defect CFT, and Integrability},''
  \href{http://dx.doi.org/10.1007/JHEP05(2018)109}{{\em JHEP} {\bfseries 05}
  (2018) 109}, \href{http://arxiv.org/abs/1802.05201}{{\ttfamily
  arXiv:1802.05201 [hep-th]}}. [Erratum: JHEP 11, 123 (2018)].

\bibitem{Antunes:2021abs}
A.~Antunes, M.~S. Costa, J.~a. Penedones, A.~Salgarkar, and B.~C. van Rees,
  ``{Towards bootstrapping RG flows: sine-Gordon in AdS},''
  \href{http://dx.doi.org/10.1007/JHEP12(2021)094}{{\em JHEP} {\bfseries 12}
  (2021) 094}, \href{http://arxiv.org/abs/2109.13261}{{\ttfamily
  arXiv:2109.13261 [hep-th]}}.

\bibitem{Ceplak:2021wzz}
N.~Ceplak, S.~Giusto, M.~R.~R. Hughes, and R.~Russo, ``{Holographic correlators
  with multi-particle states},''
  \href{http://dx.doi.org/10.1007/JHEP09(2021)204}{{\em JHEP} {\bfseries 09}
  (2021) 204}, \href{http://arxiv.org/abs/2105.04670}{{\ttfamily
  arXiv:2105.04670 [hep-th]}}.

\bibitem{Giusto:2019pxc}
S.~Giusto, R.~Russo, A.~Tyukov, and C.~Wen, ``{Holographic correlators in
  AdS$_3$ without Witten diagrams},''
  \href{http://dx.doi.org/10.1007/JHEP09(2019)030}{{\em JHEP} {\bfseries 09}
  (2019) 030}, \href{http://arxiv.org/abs/1905.12314}{{\ttfamily
  arXiv:1905.12314 [hep-th]}}.

\bibitem{Giusto:2018ovt}
S.~Giusto, R.~Russo, and C.~Wen, ``{Holographic correlators in AdS$_{3}$},''
  \href{http://dx.doi.org/10.1007/JHEP03(2019)096}{{\em JHEP} {\bfseries 03}
  (2019) 096}, \href{http://arxiv.org/abs/1812.06479}{{\ttfamily
  arXiv:1812.06479 [hep-th]}}.

\bibitem{Galliani:2017jlg}
A.~Galliani, S.~Giusto, and R.~Russo, ``{Holographic 4-point correlators with
  heavy states},'' \href{http://dx.doi.org/10.1007/JHEP10(2017)040}{{\em JHEP}
  {\bfseries 10} (2017) 040}, \href{http://arxiv.org/abs/1705.09250}{{\ttfamily
  arXiv:1705.09250 [hep-th]}}.

\bibitem{Bombini:2017sge}
A.~Bombini, A.~Galliani, S.~Giusto, E.~Moscato, and R.~Russo, ``{Unitary
  4-point correlators from classical geometries},''
  \href{http://dx.doi.org/10.1140/epjc/s10052-017-5492-3}{{\em Eur. Phys. J. C}
  {\bfseries 78} no.~1, (2018) 8},
  \href{http://arxiv.org/abs/1710.06820}{{\ttfamily arXiv:1710.06820
  [hep-th]}}.

\bibitem{Giusto:2020neo}
S.~Giusto, R.~Russo, A.~Tyukov, and C.~Wen, ``{The CFT$_6$ origin of all
  tree-level 4-point correlators in AdS$_3 \times S^3$},''
  \href{http://dx.doi.org/10.1140/epjc/s10052-020-8300-4}{{\em Eur. Phys. J. C}
  {\bfseries 80} no.~8, (2020) 736},
  \href{http://arxiv.org/abs/2005.08560}{{\ttfamily arXiv:2005.08560
  [hep-th]}}.

\bibitem{Aprile:2017bgs}
F.~Aprile, J.~M. Drummond, P.~Heslop, and H.~Paul, ``{Quantum Gravity from
  Conformal Field Theory},''
  \href{http://dx.doi.org/10.1007/JHEP01(2018)035}{{\em JHEP} {\bfseries 01}
  (2018) 035},
\href{http://arxiv.org/abs/1706.02822}{{\ttfamily arXiv:1706.02822 [hep-th]}}.

\bibitem{Aprile:2017qoy}
F.~Aprile, J.~M. Drummond, P.~Heslop, and H.~Paul, ``{Loop corrections for
  Kaluza-Klein AdS amplitudes},''
  \href{http://dx.doi.org/10.1007/JHEP05(2018)056}{{\em JHEP} {\bfseries 05}
  (2018) 056},
\href{http://arxiv.org/abs/1711.03903}{{\ttfamily arXiv:1711.03903 [hep-th]}}.

\bibitem{Bissi:2020wtv}
A.~Bissi, G.~Fardelli, and A.~Georgoudis, ``{Towards all loop supergravity
  amplitudes on $AdS_5\times S^5$},''
  \href{http://dx.doi.org/10.1103/PhysRevD.104.L041901}{{\em Phys. Rev. D}
  {\bfseries 104} no.~4, (2021) L041901},
  \href{http://arxiv.org/abs/2002.04604}{{\ttfamily arXiv:2002.04604
  [hep-th]}}.

\bibitem{Bissi:2020woe}
A.~Bissi, G.~Fardelli, and A.~Georgoudis, ``{All loop structures in
  supergravity amplitudes on $AdS_5\times S^5$ from CFT},''
  \href{http://dx.doi.org/10.1088/1751-8121/ac0ebf}{{\em J. Phys. A} {\bfseries
  54} no.~32, (2021) 324002}, \href{http://arxiv.org/abs/2010.12557}{{\ttfamily
  arXiv:2010.12557 [hep-th]}}.

\bibitem{Huang:2021xws}
Z.~Huang and E.~Y. Yuan, ``{Graviton Scattering in
  $\mathrm{AdS}_5\times\mathrm{S}^5$ at Two Loops},''
  \href{http://arxiv.org/abs/2112.15174}{{\ttfamily arXiv:2112.15174
  [hep-th]}}.

\bibitem{Drummond:2022dxw}
J.~M. Drummond and H.~Paul, ``{Two-loop supergravity on AdS$_5\times$S$^5$ from
  CFT},'' \href{http://arxiv.org/abs/2204.01829}{{\ttfamily arXiv:2204.01829
  [hep-th]}}.

\bibitem{Alday:2018kkw}
L.~F. Alday, ``{On genus-one string amplitudes on $AdS_5 \times S^5$},''
  \href{http://dx.doi.org/10.1007/JHEP04(2021)005}{{\em JHEP} {\bfseries 04}
  (2021) 005}, \href{http://arxiv.org/abs/1812.11783}{{\ttfamily
  arXiv:1812.11783 [hep-th]}}.

\bibitem{Alday:2021ajh}
L.~F. Alday, A.~Bissi, and X.~Zhou, ``{One-loop gluon amplitudes in AdS},''
  \href{http://dx.doi.org/10.1007/JHEP02(2022)105}{{\em JHEP} {\bfseries 02}
  (2022) 105}, \href{http://arxiv.org/abs/2110.09861}{{\ttfamily
  arXiv:2110.09861 [hep-th]}}.

\bibitem{Mack:2009mi}
G.~Mack, ``{D-independent representation of Conformal Field Theories in D
  dimensions via transformation to auxiliary Dual Resonance Models. Scalar
  amplitudes},''
\href{http://arxiv.org/abs/0907.2407}{{\ttfamily arXiv:0907.2407 [hep-th]}}.

\bibitem{Penedones:2010ue}
J.~Penedones, ``{Writing CFT correlation functions as AdS scattering
  amplitudes},'' \href{http://dx.doi.org/10.1007/JHEP03(2011)025}{{\em JHEP}
  {\bfseries 03} (2011) 025},
\href{http://arxiv.org/abs/1011.1485}{{\ttfamily arXiv:1011.1485 [hep-th]}}.

\bibitem{DHoker:1999mqo}
E.~D'Hoker, D.~Z. Freedman, and L.~Rastelli, ``{AdS / CFT four point functions:
  How to succeed at z integrals without really trying},''
  \href{http://dx.doi.org/10.1016/S0550-3213(99)00526-X}{{\em Nucl. Phys.}
  {\bfseries B562} (1999) 395--411},
\href{http://arxiv.org/abs/hep-th/9905049}{{\ttfamily arXiv:hep-th/9905049
  [hep-th]}}.

\bibitem{Zhou:2018sfz}
X.~Zhou, ``{Recursion Relations in Witten Diagrams and Conformal Partial
  Waves},'' \href{http://dx.doi.org/10.1007/JHEP05(2019)006}{{\em JHEP}
  {\bfseries 05} (2019) 006}, \href{http://arxiv.org/abs/1812.01006}{{\ttfamily
  arXiv:1812.01006 [hep-th]}}.

\bibitem{Costa:2014kfa}
M.~S. Costa, V.~Gon{\c c}alves, and J.~Penedones, ``{Spinning AdS
  Propagators},'' \href{http://dx.doi.org/10.1007/JHEP09(2014)064}{{\em JHEP}
  {\bfseries 09} (2014) 064},
\href{http://arxiv.org/abs/1404.5625}{{\ttfamily arXiv:1404.5625 [hep-th]}}.

\bibitem{Symanzik:1972wj}
K.~Symanzik, ``{On Calculations in conformal invariant field theories},''
\href{http://dx.doi.org/10.1007/BF02824349}{{\em Lett. Nuovo Cim.} {\bfseries
  3} (1972) 734--738}.

\bibitem{Paulos:2012nu}
M.~F. Paulos, M.~Spradlin, and A.~Volovich, ``{Mellin Amplitudes for Dual
  Conformal Integrals},'' \href{http://dx.doi.org/10.1007/JHEP08(2012)072}{{\em
  JHEP} {\bfseries 08} (2012) 072},
  \href{http://arxiv.org/abs/1203.6362}{{\ttfamily arXiv:1203.6362 [hep-th]}}.

\bibitem{Usyukina:1992wz}
N.~I. Usyukina and A.~I. Davydychev, ``{Some exact results for two loop
  diagrams with three and four external lines},'' {\em Phys. Atom. Nucl.}
  {\bfseries 56} (1993) 1553--1557,
  \href{http://arxiv.org/abs/hep-ph/9307327}{{\ttfamily arXiv:hep-ph/9307327}}.

\bibitem{Usyukina:1993ch}
N.~I. Usyukina and A.~I. Davydychev, ``{Exact results for three and four point
  ladder diagrams with an arbitrary number of rungs},''
  \href{http://dx.doi.org/10.1016/0370-2693(93)91118-7}{{\em Phys. Lett. B}
  {\bfseries 305} (1993) 136--143}.

\bibitem{Bissi:2021hjk}
A.~Bissi, G.~Fardelli, and A.~Manenti, ``{Rebooting quarter-BPS operators in
  $\mathcal{N}=4$ Super Yang-Mills},''
  \href{http://arxiv.org/abs/2111.06857}{{\ttfamily arXiv:2111.06857
  [hep-th]}}.

\bibitem{Caron-Huot:2017vep}
S.~Caron-Huot, ``{Analyticity in Spin in Conformal Theories},''
  \href{http://dx.doi.org/10.1007/JHEP09(2017)078}{{\em JHEP} {\bfseries 09}
  (2017) 078},
\href{http://arxiv.org/abs/1703.00278}{{\ttfamily arXiv:1703.00278 [hep-th]}}.

\bibitem{Alday:2017vkk}
L.~F. Alday and S.~Caron-Huot, ``{Gravitational S-matrix from CFT dispersion
  relations},''
\href{http://arxiv.org/abs/1711.02031}{{\ttfamily arXiv:1711.02031 [hep-th]}}.

\bibitem{Jepsen:2019svc}
C.~B. Jepsen and S.~Parikh, ``{Propagator identities, holographic conformal
  blocks, and higher-point AdS diagrams},''
  \href{http://dx.doi.org/10.1007/JHEP10(2019)268}{{\em JHEP} {\bfseries 10}
  (2019) 268}, \href{http://arxiv.org/abs/1906.08405}{{\ttfamily
  arXiv:1906.08405 [hep-th]}}.

\bibitem{Aharony:2016dwx}
O.~Aharony, L.~F. Alday, A.~Bissi, and E.~Perlmutter, ``{Loops in AdS from
  Conformal Field Theory},''
  \href{http://dx.doi.org/10.1007/JHEP07(2017)036}{{\em JHEP} {\bfseries 07}
  (2017) 036},
\href{http://arxiv.org/abs/1612.03891}{{\ttfamily arXiv:1612.03891 [hep-th]}}.

\bibitem{Carmi:2018qzm}
D.~Carmi, L.~Di~Pietro, and S.~Komatsu, ``{A Study of Quantum Field Theories in
  AdS at Finite Coupling},''
  \href{http://dx.doi.org/10.1007/JHEP01(2019)200}{{\em JHEP} {\bfseries 01}
  (2019) 200}, \href{http://arxiv.org/abs/1810.04185}{{\ttfamily
  arXiv:1810.04185 [hep-th]}}.

\bibitem{Carmi:2021dsn}
D.~Carmi, ``{Loops in AdS: from the spectral representation to position space.
  Part II},'' \href{http://dx.doi.org/10.1007/JHEP07(2021)186}{{\em JHEP}
  {\bfseries 07} (2021) 186}, \href{http://arxiv.org/abs/2104.10500}{{\ttfamily
  arXiv:2104.10500 [hep-th]}}.

\bibitem{Fichet:2021pbn}
S.~Fichet, ``{Dressing in AdS and a Conformal Bethe-Salpeter Equation},''
  \href{http://arxiv.org/abs/2106.04604}{{\ttfamily arXiv:2106.04604
  [hep-th]}}.

\bibitem{DeWolfe:2001pq}
O.~DeWolfe, D.~Z. Freedman, and H.~Ooguri, ``{Holography and defect conformal
  field theories},'' \href{http://dx.doi.org/10.1103/PhysRevD.66.025009}{{\em
  Phys. Rev. D} {\bfseries 66} (2002) 025009},
  \href{http://arxiv.org/abs/hep-th/0111135}{{\ttfamily arXiv:hep-th/0111135}}.

\bibitem{Aharony:2003qf}
O.~Aharony, O.~DeWolfe, D.~Z. Freedman, and A.~Karch, ``{Defect conformal field
  theory and locally localized gravity},''
  \href{http://dx.doi.org/10.1088/1126-6708/2003/07/030}{{\em JHEP} {\bfseries
  07} (2003) 030},
\href{http://arxiv.org/abs/hep-th/0303249}{{\ttfamily arXiv:hep-th/0303249
  [hep-th]}}.

\bibitem{Rastelli:2017ecj}
L.~Rastelli and X.~Zhou, ``{The Mellin Formalism for Boundary CFT$_d$},''
  \href{http://dx.doi.org/10.1007/JHEP10(2017)146}{{\em JHEP} {\bfseries 10}
  (2017) 146},
\href{http://arxiv.org/abs/1705.05362}{{\ttfamily arXiv:1705.05362 [hep-th]}}.

\bibitem{Mazac:2018biw}
D.~Mazac, L.~Rastelli, and X.~Zhou, ``{An analytic approach to BCFT$_{d}$},''
  \href{http://dx.doi.org/10.1007/JHEP12(2019)004}{{\em JHEP} {\bfseries 12}
  (2019) 004}, \href{http://arxiv.org/abs/1812.09314}{{\ttfamily
  arXiv:1812.09314 [hep-th]}}.

\bibitem{O.Marichev1}
O.~Marichev, M.~Trott, and S.~Wolfram, ``{The Wolfram Functions Site}.''
\newblock
  \url{https://functions.wolfram.com/HypergeometricFunctions/Hypergeometric3F2/03/02/03/0007/}.

\bibitem{olver2010nist}
F.~W. Olver, D.~W. Lozier, R.~F. Boisvert, and C.~W. Clark, {\em {NIST Handbook
  of Mathematical Functions}}.
\newblock Cambridge university press, 2010.

\bibitem{O.Marichev2}
O.~Marichev, M.~Trott, and S.~Wolfram, ``{The Wolfram Functions Site}.''
\newblock
  \url{https://functions.wolfram.com/HypergeometricFunctions/Hypergeometric3F2/17/02/06/0004/}.

\end{thebibliography}\endgroup
\bibliographystyle{utphys}
\end{document}